\documentclass[%
 aip,jcp
 amsmath,amssymb,
 reprint,%
]{revtex4-1}
\usepackage{hyperref}
\usepackage{graphicx}
\usepackage{dcolumn}
\usepackage{bm}
\usepackage{physics}
\usepackage{subfigure}
\usepackage[utf8]{inputenc}
\usepackage[T1]{fontenc}
\usepackage{etoolbox}
\usepackage{makecell}
\usepackage{mathrsfs}
\usepackage{ulem}
\usepackage{booktabs}
\usepackage{multirow}
\makeatletter
\def\@email#1#2{%
 \endgroup
 \patchcmd{\titleblock@produce}
  {\frontmatter@RRAPformat}
  {\frontmatter@RRAPformat{\produce@RRAP{*#1\href{mailto:#2}{#2}}}\frontmatter@RRAPformat}
  {}{}
}%
\makeatother
\begin{document}

\preprint{AIP/123-QED}

\title[]{Variational Quantum Imaginary Time Evolution for Matrix Product State Ansatz with Tests on Transcorrelated Hamiltonians}
\author{Hao-En Li}
\affiliation{ 
Department of Chemistry and Engineering Research Center of Advanced Rare-Earth Materials\\ of Ministry of Education, Tsinghua University, Beijing 100084, China.
}
\author{Xiang Li}
\affiliation{ 
Department of Chemistry and Engineering Research Center of Advanced Rare-Earth Materials\\ of Ministry of Education, Tsinghua University, Beijing 100084, China.
}
\author{Jia-Cheng Huang}
\affiliation{ 
Department of Chemistry and Engineering Research Center of Advanced Rare-Earth Materials\\ of Ministry of Education, Tsinghua University, Beijing 100084, China.
}
\author{Guang-Ze Zhang}
\affiliation{ 
Department of Chemistry and Engineering Research Center of Advanced Rare-Earth Materials\\ of Ministry of Education, Tsinghua University, Beijing 100084, China.
}

\author{Zhu-Ping Shen}
\affiliation{ 
Department of Chemistry and Engineering Research Center of Advanced Rare-Earth Materials\\ of Ministry of Education, Tsinghua University, Beijing 100084, China.
}
\author{Chen Zhao}
\affiliation{ 
Department of Chemistry and Engineering Research Center of Advanced Rare-Earth Materials\\ of Ministry of Education, Tsinghua University, Beijing 100084, China.
}
\author{Jun Li}
\affiliation{ 
Department of Chemistry and Engineering Research Center of Advanced Rare-Earth Materials\\ of Ministry of Education, Tsinghua University, Beijing 100084, China.
}
\affiliation{Department of Chemistry and Guangdong Provincial Key Laboratory of Catalytic Chemistry, Southern University of Science and Technology, Shenzhen 518055, China.}
\affiliation{Fundamental Science Center of Rare Earths, Ganjiang Innovation Academy, Chinese Academy of Sciences, Ganzhou 341000, China.}
\author{Han-Shi Hu}
\thanks{Author to whom correspondence should be addressed:\\
\href{mailto:hshu@mail.tsinghua.edu.cn}{hshu@mail.tsinghua.edu.cn}
}
\affiliation{ 
Department of Chemistry and Engineering Research Center of Advanced Rare-Earth Materials\\ of Ministry of Education, Tsinghua University, Beijing 100084, China.
}

\date{\today}

\begin{abstract}
    The matrix product state (MPS) ansatz offers a promising approach for finding the ground state of molecular Hamiltonians and solving quantum chemistry problems. Building on this concept, the proposed technique of quantum circuit MPS (QCMPS) enables the simulation of chemical systems using a relatively small number of qubits. 
In this study, we enhance the optimization performance of the QCMPS ansatz by employing the variational quantum imaginary time evolution (VarQITE) approach. Guided by McLachlan's variational principle, the VarQITE method provides analytical metrics and gradients, resulting in improved convergence efficiency and robustness of the QCMPS. We validate these improvements numerically through simulations of $\rm H_2$, $\rm H_4$, and $\rm LiH$ molecules. Additionally, given that VarQITE is applicable to non-Hermitian Hamiltonians, we evaluate its effectiveness in preparing the ground state of transcorrelated (TC) Hamiltonians. This approach yields energy estimates comparable to the complete basis set (CBS) limit while using even fewer qubits. Specifically, we perform simulations of the beryllium atom and $\rm LiH$ molecule using only three qubits, maintaining high fidelity with the CBS ground state energy of these systems. This qubit reduction is achieved through the combined advantages of both the QCMPS ansatz and transcorrelation. Our findings demonstrate the potential practicality of this quantum chemistry algorithm on near-term quantum devices.
\end{abstract}

\maketitle

%

\section{\label{sec:Introduction}Introduction}

Finding numerical solutions to the many-electron Schrödinger equation with high precision is a fundamental challenge in quantum chemistry and electronic structure theory. Among existing numerical methods, the full configuration interaction (FCI) method can fully characterize many-electron wave functions within a certain basis set in the second-quantized framework.\cite{fullci2020, fullci2024} Additionally, explicitly correlated methods such as R12, F12,\cite{explicit2012, explicit2017} and transcorrelated (TC)\cite{tc2010, tc2011, tc2012} approaches reduce the number of necessary spin orbitals required to accurately represent the exact ground state, approximating the complete basis set (CBS) energy using a smaller basis set. However, the exponential growth of the Hilbert space for the electronic wave function with system size necessitates the development of wave function methods with compact parameterization to circumvent this ``exponential wall'' problem.  Classical methods like the matrix-product-state (MPS)-based density matrix renormalization group (DMRG) algorithm,\cite{DMRG1992,DMRG2011, DMRG2020, TDDMRG2020, TDDMRG2021} general tensor network (TN) ansatz,\cite{tensor2013,tensor2015,tensor2019,tensor2021} and neural-network quantum state (NQS) ansatz\cite{NQS2017,NQS2020,NQS2023,SCNQS2023,DYNNQS2019} exploit compact parameterization and well-designed variational algorithms to capture ground states, excited states, and dynamic behaviors of chemical systems.

Recent advances in quantum technologies offer additional solutions.\cite{QuantumRev2020} Parameterized quantum circuits (PQCs) can naturally describe the quantum states of chemical systems and search for ground states through variational optimization, as exemplified by the variational quantum eigensolver (VQE)\cite{vqe2014, HEA2017, VQA2021} method and the quantum approximate optimization algorithm (QAOA).\cite{QAOA} These classical-quantum hybrid algorithms employ relatively shallow quantum circuits, garnering significant attention in the noisy intermediate-scale quantum (NISQ) era.

Variational quantum algorithms generally entail high-dimensional, highly nonlinear, and non-convex optimization problems.\cite{mrsqk2020,nonlinear2020,VQA2021} 
Furthermore, the number of available qubits in near-term quantum devices remains limited. Therefore, designing qubit-efficient and easy-to-optimize PQCs is crucial. Chemically inspired unitary coupled cluster (UCC) ansatz,\cite{ucc2017,ucc2018,ucc2022,symUCCSD2022} adaptive derivative-assembled pseudo-Trotter (ADAPT) ansatz,\cite{ADAPT} qubit-ADAPT ansatz (a more economical version than the original ADAPT),\cite{qbitADAPT} hardware-efficient ansatz (HEA),\cite{HEA2017} and other circuit-designing strategies have demonstrated efficacy in accurately describing the ground state of specific molecular systems.\cite{lucj2023,PCHEA2024} Notably, in 2019, Liu et al.\cite{QCMPS2019} illustrated that instead of encoding the degrees of freedom of spin orbital occupation configurations using a 0-1 string in the qubit Hilbert space, isomorphic to $(\mathbb{C}^2)^{\otimes N_{\rm orb}}$, qubits can also represent the degrees of freedom of the virtual bond space in MPS-form wave functions by incorporating mid-circuit measurements in PQCs. This approach enables the preparation of PQCs capable of generating an MPS with an exponentially large bond dimension, allowing VQE to be performed with fewer qubits.\cite{QCMPS2019, QCMPSSU22020,QCMPS2021, QRBM2021,QCMPS2022,QCMPS2023,QCMPSchem2023,QCMPS2024} Moreover, certain quantum circuit architectures like QCMPS, quantum tree tensor networks and quantum multiscale entanglement renormalization ansatz are free from the barren plateau problem, thus these ansatzes may be ideal for VQE and VarQITE.\cite{barren2023} For instance, Liu et al. simulated a $4\times 4$ square lattice spin model using only $6$ qubits.\cite{QCMPS2019} Fan et al. further explored the quantum circuit matrix product state (QCMPS) architecture, simulating the $\rm H_{50}$ molecule system with just $8$ qubits, achieving fidelity comparable to the classical MPS ansatz using the DMRG algorithm.\cite{QCMPSchem2023}

Nevertheless, significant challenges persist in applying QCMPS to real chemical systems, primarily due to optimization difficulties in VQE. Most VQE algorithms provide direct access only to the expectation value of energy measured from the quantum circuit, which serves as the loss function value.\cite{tc_quantum2023,barrenvqe2024}  Consequently, non-gradient or pseudo-gradient optimizers are used, treating the loss function as a ``black box'' and neglecting the intrinsic geometric structure of the parameter space, rendering optimization particularly challenging. Additionally, the expectation values of non-Hermitian observables may not be real numbers, precluding direct optimization using conventional VQE and impeding the extension to scenarios involving TC methods.\cite{tc_quantum2023,tc_quantum2024,adapt_tc2024}

In this article, we address these challenges by incorporating the ansatz-based variational quantum imaginary time evolution (VarQITE)\cite{varqite2019, varqite2023} method into QCMPS optimization, replacing the previous optimizers VQE based on energy expectation values. We design a specialized QCMPS circuit where analytical metrics and gradients of the ansatz space can be conveniently measured on a quantum circuit using the parameter-shift rule, enhancing optimization efficiency and stability.\cite{parameter2019} Furthermore, as shown by Dobrautz et al,\cite{tc_quantum2023,tc_quantum2024} the imaginary time evolution formulation also applies to the non-Hermitian transcorrelated (TC) Hamiltonians. This method allows us to integrate the TC method within the QCMPS framework, further reducing qubit requirements while improving the accuracy of quantum chemical simulations.\cite{hermitian2020} Based on these insights, this method presents a potentially viable solution for further mitigating qubit resource requirements on the near-term quantum devices.

The rest of this paper is organized as follows. In Section \ref{sec:theory}, we provide a succinct overview of the VarQITE-QCMPS algorithm. Detailed descriptions of our theory and methodology are available in Appendices \ref{app:A} and \ref{app:B}. Section \ref{sec:hermite} presents the performance of our algorithmic improvements in simulating Hermitian Hamiltonians.  We also evaluate the cost and accuracy of this method using TC Hamiltonians, including the $\rm Be$ atom and the $\rm LiH$ molecule system.  Finally, we conclude in Section \ref{sec:conclusion}.
\section{\label{sec:theory}Theory and Methods}
\subsection{\label{sec:mps}Quantum circuit matrix product state}

\begin{figure*}
    \centering
    \includegraphics[width=1.0\textwidth]{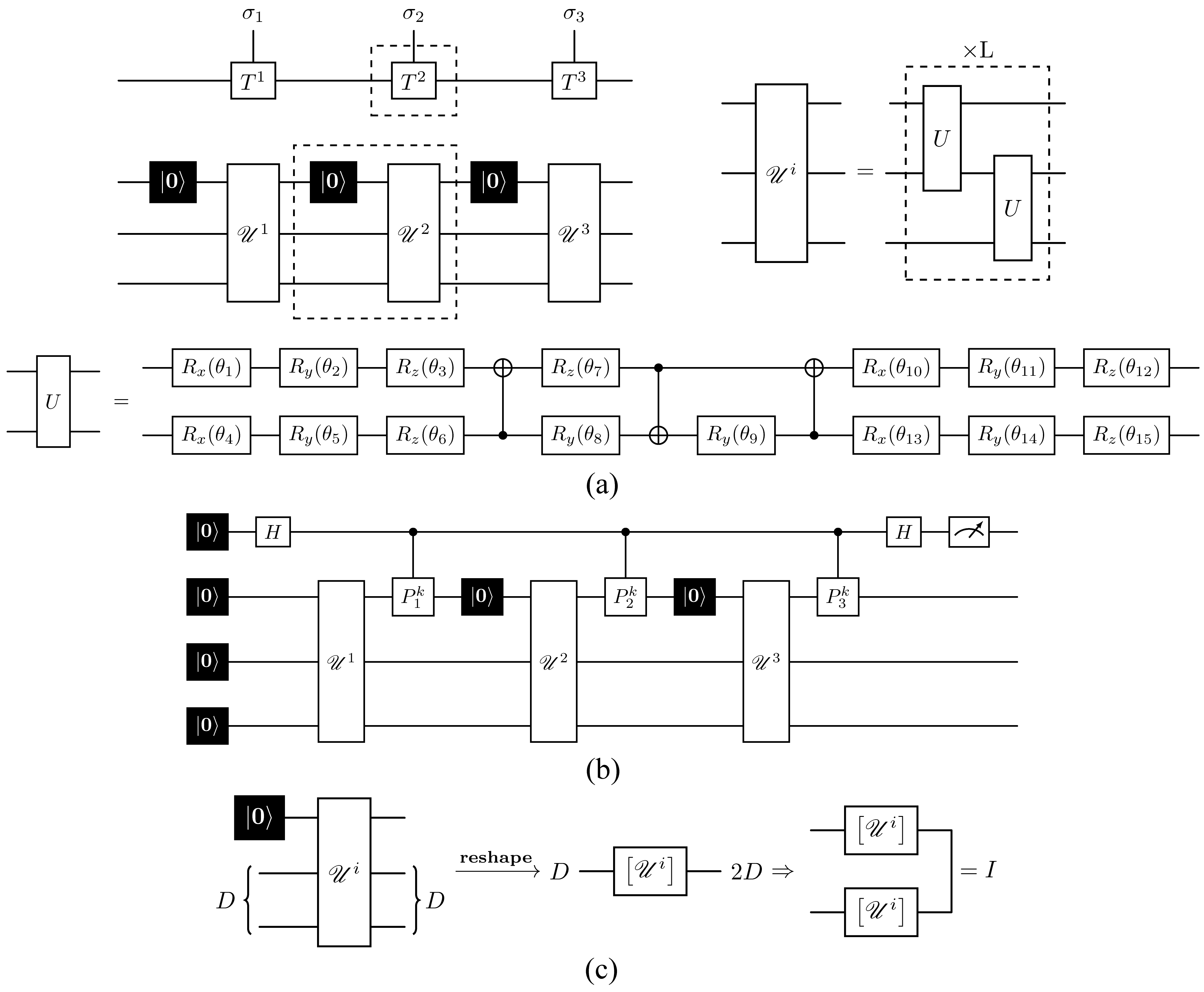}
    \caption{(\textbf{a}) The upper left part depicts the graphical representation of the open-boundary classical MPS alongside the quantum circuit diagram used to prepare a quantum version of the MPS. The unitary evolution $\mathscr{U}^i$, involving $N_b + 1$ qubits (where $N_b = 2$), corresponds to the component tensor $T^i$ in the classical MPS. In the upper right part, a staircase-like quantum circuit is shown, which constructs each component tensor $\mathscr{U}^i$ within the QCMPS ansatz. The parameter degrees of freedom of each $\mathscr{U}^i$ increase linearly with the repetition count (or number of layers) denoted as $L$, thereby enhancing its expressive power. As illustrated in the lower part of the figure, each block approximates a general 2-qubit unitary matrix using a 15-parameter decomposition, transforming all parameterized quantum gates into rotation gates and facilitating their application in VarQITE \cite{QCMPSkagome2024}. (\textbf{b}) The Hadamard's test circuit for measuring the expectation value $\langle \psi_{\mathrm{QCMPS}}|P_1^k\otimes P_2^k\otimes P_3^k|\psi_{\mathrm{QCMPS}}\rangle$. (\textbf{c}) The graphical representation of the right orthogonal condition of the QCMPS ansatz, where $[\mathscr{U}^i]$ represents the matrization of the quantum circuit component tensor.  } 
    \label{fig:1}
\end{figure*}

We begin with a brief overview of the classical matrix product state (MPS) ansatz in quantum physics and chemistry. A many-body electronic wave function can be expressed as a linear combination of electron configurations in Slater determinant form:
\begin{equation}
\ket{\psi} = \sum_{\sigma_1,\cdots,\sigma_N} \psi_{\sigma_1,\cdots,\sigma_N}\ket{\sigma_1,\cdots,\sigma_N},
\end{equation}
where the configuration coefficients $\psi_{\sigma_1,\cdots,\sigma_N}\in \mathbb{C}$ is generally a high-order tensor of $2^N$ dimension. Each $\sigma_i\in \{0,1\}$ indicates the occupation number of each spin orbital. As shown in DMRG method, we can represent many-body wave function in the form of contraction of several low-rank tensors and then use variational methods to find the ground state. Specifically, by factorizing $\{\psi_{\sigma_1,\cdots,\sigma_N}\}$, we obtain the open-boundary MPS ansatz
\begin{equation}
    \begin{aligned}
    \ket{\psi} &= \sum_{\sigma_1,\cdots,\sigma_N}\sum_{\alpha_1,\cdots,\alpha_{N-1}} (T^1)^{\sigma_1}_{\alpha_1} (T^2)^{\sigma_2}_{\alpha_1,\alpha_2}\\
    &\cdots (T^{N-1})^{\sigma_{N-1}}_{\alpha_{N-2},\alpha_{N-1}} (T^N)_{\alpha_{N-1}}^{\sigma_N}\ket{\sigma_1,\cdots,\sigma_N},
    \end{aligned}
\end{equation}
where each $T^{i}(i=1,\cdots,N)$ is a rank-$2$ for $i=1$ and $i=N$, or a rank-$3$ tensor for $i=2,\cdots,N-1$. The indices $\sigma_i$
and $\alpha_i$ are called physical index and virtual index, respectively. The expressiveness of an MPS is determined by the virtual bond dimension $D$, i.e. the dimension of virtual index. As a wave function ansatz, MPS can effectively represent the ground state of physical and chemical systems that comply with the one-dimensional area law when $D$ is sufficiently large, making the DMRG algorithm a practical high-precision method widely used in electronic structure theory. 

Tensor contraction involves multiplying tensor elements and summing them over specific indices, similar to matrix multiplication. A graphical method has been developed to clearly illustrate tensor contractions. A typical MPS and its component tensor are depicted in Figure \ref{fig:1}(a). The legs of the tensor correspond to its indices, and jointed legs indicate that two tensors contract with respect to this index. 

Inspired by this graphical representation, we can envision using $N_b$ qubits on a quantum device to create a virtual bond space with dimension $D=2^{N_b}$. An additional qubit is also introduced to represent a 2-dimensional space for each physical index, simulating spin-1/2 fermions. Corresponding to the component tensors of the classical MPS, a series of $N_{\rm orb}$ unitary matrix blocks $\{\mathscr{U}^i\}_{i=1}^{N_{\rm orb}}$ evolve collectively on these qubits. The physical qubit channel is traced out using mid-circuit measurements and then reset to $\ket{0}$ between each pair of adjacent blocks for reuse,\cite{QCMPS2021, QCMPSchem2023} as shown in Figure \ref{fig:1}(a). Each block $\mathscr{U}^i$ includes many parameterized quantum gates, which are updated during variational optimization.

In this setup, only $\log_2 D+1$ qubits are needed to prepare a PQC equivalent to an MPS with bond dimension $D$. In contrast, traditional ansatz settings map the fermionic Fock space and fermionic creation-annihilation operators to qubit space and qubit operators using the Jordan-Wigner transformation, requiring as many qubits as there are spin orbitals $N_{\rm orb}$. Typically, the MPS bond dimension $D$ needed to accurately describe the ground state is much smaller than $2^{N_{\rm orb}}$. Consequently, $\log_2 D+1$ is usually much less than $N_{\rm orb}$, allowing for large-scale quantum simulations with relatively few qubits.

To employ the variational optimization for the QCMPS-type PQC, the first step is to measure the energy expectation value $\langle \psi|\hat{H}|\psi\rangle$ of QCMPS on the quantum device. This involves converting the electronic (fermionic) Hamiltonian into a qubit Hamiltonian using the Jordan-Wigner mapping, which expresses it as a real linear combination of Pauli strings:\cite{JWT1928, QuantumRev2020}
\begin{equation}
    \hat{H} = \sum_{k} c_k \mathscr{P}_k,\quad c_k\in \mathbb{R},
\end{equation}
where each Pauli string $\mathscr{P}_k$ is given by:
\begin{equation}
    \mathscr{P}_k = \bigotimes_{i=1}^{N_\mathrm{orb}} P^k_i,\quad P^k_i\in \{\mathrm{Id},X,Y,Z\}.
\end{equation}

Here, $\{\mathrm{Id},X,Y,Z\}$ denotes the set of four $2\times 2$ pauli operators. For the conventional ansatzes based on orbital-qubit mapping, we can directly perform measurements with respect to the eigenbasis of $\mathscr{P}_k = \bigotimes_{i=1}^{N_{\rm orb}}P_i^k$, or alternatively employ indirect measurement techniques such as Hadamard's test, to obtain the expectation value $\langle \psi(\boldsymbol{\theta})|\mathscr{P}_k|\psi(\boldsymbol{\theta})\rangle$ of the parametrized quantum state $\ket{\psi(\boldsymbol{\theta})}$ with respect to each Pauli string $\mathscr{P}_k$. The linear combination of them gives the expectation value of the Hamiltonian operator:
\begin{equation}
    \langle \psi(\boldsymbol{\theta})|\hat{H}|\psi(\boldsymbol{\theta})\rangle = \sum_{k}c_k \langle  {\psi(\boldsymbol{\theta})}|\mathscr{P}_k|\psi(\boldsymbol{\theta})\rangle.
\end{equation}

In the context of the QCMPS ansatz, Hadamard's test remains applicable with a slight modification. Due to the placement of physical indices following each block $\mathscr{U}^i$
  in QCMPS, each controlled-Pauli quantum gate C-$P_i^k$ must be applied between $\mathscr{U}^i$ and the reset-to-zero channel on the physical qubit, rather than at the end of the entire parameterized quantum circuit, as traditionally done in other ansatzes. The Hadamard's test circuit is illustrated in Figure \ref{fig:1}(b). For a detailed discussion on measurement techniques, please refer to Appendix \ref{app:A}.

Once we obtain the energy expectation value, VQE can be used to find the ground-state energy and its corresponding eigenstate of $\hat{H}$ by minimizing the cost function\cite{QCMPSchem2023} 
\begin{equation}
    \boldsymbol{\theta}^\ast =\mathrm{argmin}_{\boldsymbol{\theta}}\mathcal{E}(\boldsymbol{\theta})= \mathrm{argmin}_{\boldsymbol{\theta}} \langle \psi(\boldsymbol{\theta})|\hat{H}|\psi(\boldsymbol{\theta})\rangle.
\end{equation}
We can then minimize $\mathcal{E}(\boldsymbol{\theta})=\langle \psi(\boldsymbol{\theta})|\hat{H}|\psi(\boldsymbol{\theta})\rangle$ by viewing $\mathcal{E}$ as a black box cost function, as demonstrated in the previous work.\cite{QCMPS2023, QCMPS2024}

As mentioned earlier, the unitary evolution of the states is equivalent to tensor contraction, as both are essentially matrix multiplications. For a system with $N_b$ virtual qubits and one physical qubit, each block can be viewed as a linear isometry from $\mathbb{C}^D$ to $\mathbb{C}^D \otimes \mathbb{C}^2$:
\begin{equation}\label{eq:rc}
    [\mathscr{U}^i][\mathscr{U}^i]^\dagger = \mathrm{Id}_{\mathbb{C}^{D}}.
\end{equation}
Here $[\mathscr{U}^i]$ is a $D\times (2D)$ matrix, which is obtained by tracing out the physical qubit on the left side of the $(2D)\times (2D)$ unitary matrix $\mathscr{U}^i$ with state $\ket{0}$ and then reshaping it to the consistent shape, as shown in Figure \ref{fig:1}(c). Eq.(\ref{eq:rc}) signifies that QCMPS inherently satisfies the so-called \textit{right-orthogonal condition}, which is crucial for the classical DMRG algorithm to effectively perform variational optimization on MPS. In the quantum version, albeit direct application of the DMRG algorithm to optimize QCMPS is not feasible, the orthogonal condition plays a pivotal role in ensuring numerical stability during the optimization process of QCMPS.\cite{QCMPS2021, QCMPSchem2023}

The expressiveness of QCMPS depends not only by the bond dimension or the number of qubits but also on the number of layers $L$ of each $\mathscr{U}^i$. The total number of parameters $N_{\mathrm{params}}$ scales as $O(N_{\rm orb} \times L \times N_b)$, making it linear with respect to $N_{\rm orb}$ for fixed bond dimension and layer number.\cite{QCMPSchem2023} Generally, a larger $L$ value allows for more accurate approximation of the component tensors due to the increased degrees of freedom.  However, increasing $L$ can lead to optimization challenges, necessitating careful selection of circuit depth to balance expressiveness and optimizability as discussed by Larocca et al.\cite{barrenvqe2024}.

\subsection{\label{sec:qite}Circuit implementation for VarQITE optimization of QCMPS}
\begin{figure*}
    \centering
    \includegraphics[width=0.7\textwidth]{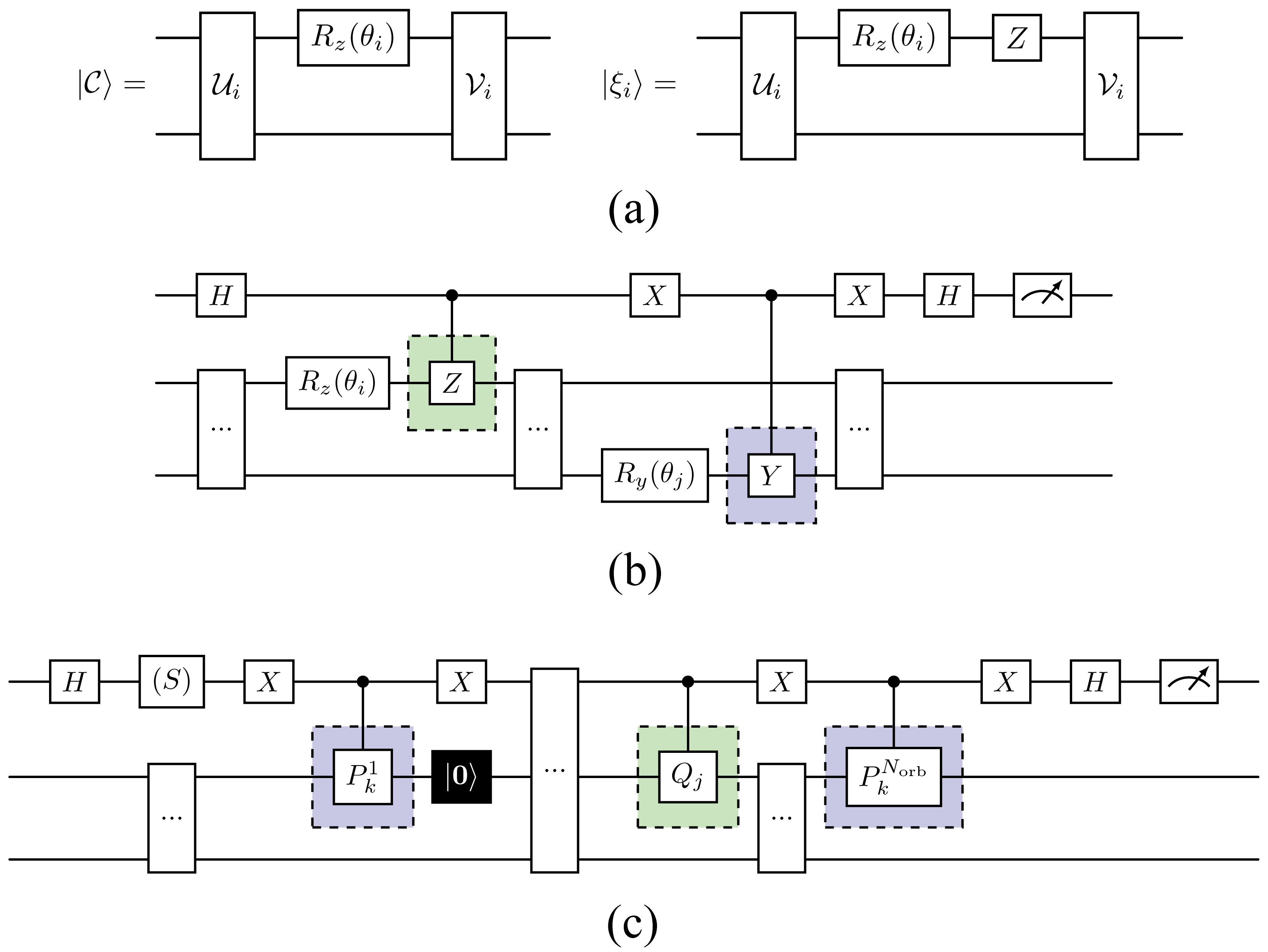}
    \caption{\label{fig:measurement_a_c}(\textbf{a}) The quantum circuits that represent the states $\ket{\mathcal{C}}$ and $\ket{\xi_i}$, here $q_i=z$ and $Q_i=Z$. (\textbf{b}) The Hadamard's test circuit to measure $\braket{\xi_j}{\xi_i}$. Here, the gate $Q_i=Z$ is controlled by the ancilla being $\ket{1}_{\rm anc}$ and the gate $Q_j=Y$ is controlled by the ancilla being $\ket{0}_{\rm anc}$. (\textbf{c}) The Hadamard's test circuit to measure $\mel{\xi_j}{\mathscr{P}_k}{\mathcal{C}}$, where the Pauli string $\mathscr{P}_k=\bigotimes_{\ell =1}^{N_{\mathrm{orb}}} P_k^\ell$ is controlled by $\ket{1}_{\mathrm{anc}}$ and $Q_j$ gate is controlled by $\ket{0}_{\rm anc}$. The $S$ gate is applied only when testing the imaginary part. } 
\end{figure*}

As mentioned in Section \ref{sec:mps}, the BFGS optimizer obtains the energy gradient using the finite difference method, while the inverse of the Hessian matrix is approximated through an iterative approach. These approximations can introduce errors, potentially leading to difficulties in optimization.\cite{tc_quantum2023} 
However, if VarQITE is applied to optimize QCMPS, then not only the energy gradient but also the Fisher information matrix (or the Fubini-Study metric on the ansatz space) are taken into consideration.\cite{varqite2019,qml2020} Specifically, the BFGS method updates the parameter using:
\begin{equation}
    \boldsymbol{\theta}' = \boldsymbol{\theta} + \Delta \tau \mathbf{B}^{-1}\mathbf{C}
\end{equation}
where $\mathbf{B}$ is the approximation of the Hessian matrix of the cost function $\mathcal{E}(\boldsymbol{\theta})$, whose inverse $\mathbf{B}^{-1}$ is obtained using a recursive algorithm. Thus, BFGS is a quasi-Newton method that approximates the first and second-order derivatives of the cost function based solely on the energy expectation value, without considering information about the ansatz itself. This information is included in the matrix $\mathbf{A}$, which is specifically used in VarQITE. Moreover, we emphasize again that the gradient vector $\mathbf{C}$ and $\mathbf{A}$ are analytically accessible from quantum circuit measurements and we do not to approximate them using finite difference method. The above discussions suggest that a VarQITE-based optimizer has the potential to outperform the BFGS method.

In the VarQITE algorithm, the vector of parameters is updated as follows at each iteration step:
\begin{equation}
    \boldsymbol{\theta}' = \boldsymbol{\theta} + \Delta\tau \mathbf{A}^{-1} \mathbf{C},
\end{equation}
where
\begin{equation}\label{eq:aij}
    \mathbf{A}_{i,j}  = \Re \braket{\pdv{\psi(\boldsymbol{\theta})}{\theta_i}}{\pdv{\psi(\boldsymbol{\theta})}{\theta_j}},
\end{equation}
\begin{equation}
    \mathbf{C}_{i}  = -\Re\mel{\pdv{\psi(\boldsymbol{\theta})}{\theta_i}}{ \hat{H}}{ \psi(\boldsymbol{\theta})}
\end{equation}
and $\Delta\tau$ is the step size or learning rate.

From the perspective of quantum machine learning,  $\mathbf{A}$ and $\mathbf{C}$ correspond to the quantum Fisher information matrix and quantum natural gradient respectively.\cite{qml2020,VQA2021} Since $\mathbf{A}_{i,j}$ and $\mathbf{C}_i$ can be directly measured from the quantum circuit, VarQITE makes the analytical metric and gradient accessible, thereby enabling efficient optimization of parameterized quantum circuits with respect to gate parameters.\cite{parameter2019} 

To measure the matrix $\mathbf{A}$ and the vector $\mathbf{C}$ on the quantum circuit, we follow the idea of ``parameter shift rule'' as showcased in the reference.\cite{parameter2019} For clarity, we consider the equivalent pure-state quantum circuit, as detailed in Appendix \ref{app:A}, instead of the circuit we actually use, which includes reset-to-zero channels.

Given a quantum circuit $\mathcal{C}$ with multiple parameters, we examine the $i,j$-entry $\mathbf{A}_{i,j}$ of the matrix $\mathbf{A}$. From Eq.(\ref{eq:aij}), we have
\begin{equation}
    \mathbf{A}_{i,j} = \Re \langle\partial_i\mathcal{C}|\partial_j\mathcal{C} \rangle.
\end{equation}
Thus, we need to evaluate $\Re\langle \partial_i \mathcal{C}|\partial_j\mathcal{C}\rangle$ on the quantum circuit. To make the measurement complexity more manageable to quantum devices, we design a special QCMPS circuit, where the types of the parametrized quantum gates are restricted to single-qubit rotation gates $R_q(\theta)$:
\begin{equation}
    R_q(\theta) = \exp(-\frac{\mathrm{i}\theta}{2}Q),
\end{equation}
where $Q\in \{X,Y,Z\}$. In this case, we define the unitary matrices $\mathcal{U}_i$ and $\mathcal{V}_i$ as shown in Figure \ref{fig:measurement_a_c}(a), then the $i$-th partial derivative of $\mathcal{C}$ can be expressed as\cite{derivation2021}

\begin{equation}
    \begin{aligned}
\ket{    \partial_i \mathcal{C}(\boldsymbol{\theta})}&= \partial_i[\mathcal{V}_iR_{q_i}(\theta_i)\mathcal{U}_i|{\overline{0}}\rangle]= \mathcal{V}_i[\partial_iR_{q_i}(\theta_i)]\mathcal{U}_i|{\overline{0}}\rangle\\
    &=-\frac{\mathrm{i}}{2}\mathcal{V}_iQ_i\exp(-\frac{\mathrm{i}\theta_i}{2} Q_i)\mathcal{U}_i|{\overline{0}}\rangle&\\&=-\frac{\mathrm{i}}{2}\mathcal{V}_iQ_iR_{q_i}(\theta_i)\mathcal{U}_i|{\overline{0}}\rangle,
    \end{aligned}
\end{equation}
where, we utilize the independence of $\mathcal{U}_i$ and $\mathcal{V}_i$ from $\theta_i$, as well as the linearity of partial derivatives. Then we can write
\begin{equation}
    \Re\langle\partial_i \mathcal{C}|\partial_j\mathcal{C}\rangle = \frac{1}{4}\Re\langle \xi_i|\xi_j\rangle,
\end{equation}
where $\ket{\xi_i}$ denotes the quantum circuit that coincides with $\mathcal{C}_i$, except for an additional gate $Q_i$ applied after the gate $R_{q_i}(\theta_i)$. For example, if $\theta_i$ is the parameter of an $R_z$ gate, then $\mathcal{C}(\boldsymbol{\theta})$ can be written as $\mathcal{C}(\boldsymbol{\theta}) = \mathcal{V}_i R_z(\theta_i) \mathcal{U}_i|{\overline{0}}\rangle$ with $\mathcal{U}_i$ and $\mathcal{V}_i$ independent of $\theta_i$. In this case, we have
\begin{equation}
    \ket{\xi_i} = \mathcal{V}_i Z R_z(\theta_i) \mathcal{U}_i|{\overline{0}}\rangle,
\end{equation} 
as graphically demonstrated in Figure \ref{fig:measurement_a_c}(a).

The diagonal elements of $\mathbf{A}$ are given by $\Re\langle \partial_i\mathcal{C}|\partial_i\mathcal{C}\rangle=\frac{1}{4}$. To compute $\Re\langle \xi_i|\xi_j\rangle$ for $i\ne j$ on the quantum device using Hadamard's test, what we need to prepare is the following state
\begin{equation}
    \ket{\psi_{ij}}=\frac{\ket{0}_{\mathrm{anc}}\otimes\ket{\xi_j}+\ket{1}_{\mathrm{anc}}\otimes \ket{\xi_i}}{\sqrt{2}}.
\end{equation}
A detailed explanation of the Hadamard's test process is provided in Appendix \ref{app:A}.

This preparation can be achieved by inserting a Controlled-$Q_k$ gate after the gate $R_{q_k}(\theta_k)$ for $k=i$ and $k=j$. Specifically, the C-$Q_i$ gate is controlled by the ancilla qubit being in state $\ket{1}{\rm anc}$, while the C-$Q_j$ gate is controlled by the ancilla qubit being in state $\ket{0}{\rm anc}$, as illustrated in Figure \ref{fig:measurement_a_c}(b).

It is worth noting that using only rotation gates for parametrization is beneficial, as it requires the introduction of only one additional gate to prepare the state $\ket{\xi_i}$ and just one extra measurement to compute the matrix element $\frac{1}{4}\Re\langle \xi_i|\xi_j\rangle$. As shown in Figure \ref{fig:1}(a), the decomposition of a general two-qubit ``entanglement block'' into single-parameter rotation gates is an effective approximation \cite{QCMPSkagome2024, decomp2004, decomp22004}, which facilitates the calculation of the partial derivatives of the circuit $\mathcal{C}$ at the same time.

For the gradient vector $\mathbf{C}$, 
\begin{equation}\label{eq:Cj}
    \mathbf{C}_j= \Re\langle\partial_j \mathcal{C}|\hat{H}\mathcal{C}\rangle=\Re\sum_{k} \left[\frac{\mathrm{i}}{2}c_k\langle\xi_j|\mathscr{P}_k |\mathcal{C}\rangle\right].
\end{equation}

And we need to prepare the state
\begin{equation}
    \ket{\varphi_j}=\frac{\ket{0}_{\mathrm{anc}}\otimes \ket{\mathscr{P}_k\mathcal{C}} +\ket{1}_{\mathrm{anc}}\otimes \ket{\xi_j}}{\sqrt{2}},
\end{equation}
using the quantum circuit shown in Figure \ref{fig:measurement_a_c}(c). For non-Hermitian Hamiltonians, the coefficients $c_k$ in Eq.(\ref{eq:Cj}) can have non-zero imaginary parts. Consequently, it is necessary to measure both the real and imaginary parts of $\mel{\xi_j}{\mathscr{P}_k}{\mathcal{C}}$.

At each iteration of VarQITE, the measurement complexity of $\mathbf{A}$ is $O(N_{\mathrm{params}}^2)$.\cite{tc_quantum2024} Nonetheless, it has been shown that the measurement cost can be reduced to linear by employing certain approximation strategies.\cite{approximateQFI2021,varqite_noqgt2024} Moreover, we want to note that there is a more hardware-efficient proposal of VarQITE approach that does not require the computation and inversion of the matrix $\mathbf{A}$.\cite{HEvarqite2021} Since the measurement of each element of $\mathbf{A}$ and $\mathbf{C}$ is independent, numerical simulations can be parallelized and sped up when multiple CPUs are available. The inverse of the matrix $\mathbf{A}$ needs to be evaluated at each step, and $\mathbf{A}$ may become ill-conditioned during the optimization process. Therefore, regularization is necessary to ensure the fidelity of $\mathbf{A}^{-1}\mathbf{C}$ and the numerical stability of the entire optimization process. To address this issue, we apply a diagonal correction to the matrix $\mathbf{A}$ at each iteration step,\cite{SCNQS2023, SCNQS2024}
\begin{equation}
    \mathbf{A} \mapsto \mathbf{A} + \delta\mathrm{Id}.
\end{equation}
Here, $\delta$ is a regularization parameter which is set to $10^{-5}$ unless otherwise specified.

\subsection{\label{sec:constrained}Transcorrelated Hamiltonians}

In quantum chemistry, adjustments to single-electron basis set functions, or spin orbitals, are often needed to improve accuracy. One example is the \textit{cusp condition correction} of Gaussian-type orbital (GTO) functions. Kato's work revealed that as the distance between electrons $|\mathbf{r}_i-\mathbf{r}_j|$ or between electrons and nuclei $|\mathbf{r}_i-\mathbf{R}_J|$ approaches zero, the first derivatives of wave functions become discontinuous and unbounded, known as Coulomb singularities. GTO basis sets are smooth, which is inconsistent with real wave function behavior at Coulomb singularities. Therefore, a large set of basis functions is typically required to accurately capture the electron-electron cusp condition.\cite{explicit2012}

To address this, explicit correlated methods are devised, incorporating interelectronic distances to construct efficient wave function ansatzes. For instance, employing a Jastrow factor on the wave function allows for the description of electronic correlation. This factor modifies the wave function as
\begin{equation}
  \ket{\Phi}\mapsto \ket{\Psi} = e^{\hat{J}(\{\mathbf{r}_i\})}\ket{\Phi},
\end{equation}
where
\begin{equation}
    \hat{J}(\{\mathbf{r}_i\}) = \sum_{i<j} \hat{u}(\mathbf{r}_i,\mathbf{r}_j),
\end{equation}
is Jastrow factor. There are various forms of function $\hat{u}$ satisfying the cusp condition for $\Psi(\{\mathbf{r}_i\})=(e ^{\hat{J}} \Phi)(\{\mathbf{r}_i\})$, and they typically include a set of optimizable variational parameters $\{\boldsymbol{\alpha}\}$. To determine the parameters that effectively correct the cusp condition of GTO basis set, we utilize a trial wave function $\ket{\Phi_T} = e^{\hat{J}(\{\boldsymbol{r}_i\};\{\boldsymbol{\alpha}\})}\ket{\Phi_R}$ and perform variational Monte Carlo (VMC) optimization to minimize the following objective function\cite{tc2010,tc2011,tc2012,tc_quantum2023,tc_quantum2024}
\begin{equation}
    \min_{\{\boldsymbol{\alpha}\}}\mathcal{E}_{\rm VMC}(\{\boldsymbol{\alpha}\}) =  \min_{\{\boldsymbol{\alpha}\}}\frac{\langle \Phi_T|\hat{H}|\Phi_T\rangle}{\langle \Phi_T | \Phi_T\rangle}.
\end{equation}

In practice, the reference wave function $\ket{\Phi_R}$ is chosen to be the Hartree-Fock wave function or its single reference Møller-Plesset 2 (MP2) wave function under a certain basis set.\cite{tc_quantum2024} The optimized $\hat{J}$ operator is then used to construct the Transcorrelated Hamiltonian. 

The basic idea of the  transcorrelated (TC) method is to incorperate the Jastrow factor into the system Hamiltonian, resulting in:
\begin{equation}
    \hat{H}_{\rm TC} := e^{-\hat{J}}\hat{H}e^{\hat{J}}.
\end{equation}
This is in general a similarity transformation, which means the TC Hamiltonian $\hat{H}_{\rm TC}$ is isospectral to the original Hamiltonian $\hat H$ as operators on the total Hilbert space. By taking advantage of the Jastrow factor incorperated in $\hat{H}_{\rm TC}$, we are enabled to obtain ground state energy closer to the CBS limit without the need for a large basis set. 

However, since $e^{\hat J({\{\mathbf{r}_i\}})}$ is not a unitary operator, $\hat{H}_{\rm TC}$ is non-Hermitian. Therefore, when we convert $\hat{H}_{\rm TC}$ to the qubit Hamiltonian via the Jordan-Wigner transformation, the coefficients of Pauli strings are complex numbers:
\begin{equation}
    \hat{H}_{\rm TC} =\sum_k c_k\mathscr{P}_k,\quad     c_k\in \mathbb{C},
\end{equation} 
and the expectation value of $\hat{H}_{\rm TC}$ operator, $\mathcal{E}_{\rm TC}(\boldsymbol{\theta})=\langle \psi(\boldsymbol{\theta}) |\hat{H}_{\rm TC}|\psi(\boldsymbol{\theta}) \rangle$, is generally a complex-valued function of the parameters $\boldsymbol{\theta}$. This makes the function-value-based optimizer inapplicable. Although it's possible to minimize the variance of the TC Hamiltonian
  instead of directly minimizing the complex-valued function $\mathcal{E}_{\rm TC}$ itself,\cite{tc1971variance}
 a more straightforward and effective approach might be to implement VarQITE to optimize $\mathcal{E}_{\rm TC}$. In fact, according to the ITE principle:
\begin{equation}
    \ket{\psi_{\rm GS}} = \lim_{\tau\to \infty}e^{-\hat H\tau}\ket{\psi(0)},
\end{equation} 
where $\ket{\psi_{\rm GS}}$ represents the ground-state eigenvector of $\hat{H}$. We denote the right ground-state eigenvector of $\hat{H}_{\rm TC}$ as $\ket{\phi_{\rm GS}^R}$ (the left and right eigenvector of $\hat{H}_{\rm TC}$ may differ due to the non-Hermiticity).\cite{tc_quantum2024} Notice that $\ket{\phi^R_{\rm GS}}$ is related to $\ket{\psi_{\rm GS}}$ via the similarity transformation operator $e^{\hat{J}}$, saying $\ket{\psi_{\rm GS}}=e^{\hat{J}}\ket{\phi_{\rm GS}^R}$, a direct calculation of $e^{\hat{J}}\ket{\phi_{\rm GS}^R}$ yields\cite{hermitian2020}
\begin{equation}
    \begin{aligned}
    e^{\hat{J}}\ket{\phi_{\rm GS}^R} &= \ket{\psi_{\rm GS}} = \lim_{\tau\to \infty}e^{-\hat H\tau}\ket{\psi(0)}\\
    &=\lim_{\tau\to \infty} \exp({-e^{\hat{J}}\hat{H}_{\rm TC} e^{-\hat{J}}}\tau)\ket{\psi(0)}\\
    &= \lim_{\tau\to\infty} \sum_{k=0}^\infty \frac{{e^{\hat{J}}(-\hat{H}_{\rm TC}\tau)^k e^{-\hat{J}}}}{k!}e^{\hat{J}}\ket{\phi(0)}\\
    &=e^{\hat{J}}\lim_{\tau\to\infty} e^{-\hat{H}_{\rm TC}\tau}\ket{\phi(0)}.
    \end{aligned}
\end{equation}
Eliminating $e^{\hat{J}}$ via multiplying $e^{-\hat{J}}$ on both sides gives $\ket{\phi_{\rm GS}^{R}}=\lim_{\tau\to\infty}e^{-\hat{H}_{\rm TC}\tau}\ket{\phi(0)}$. In other words, ITE also holds for non-Hermitian Hamiltonians, hence VarQITE is applicable for TC systems.

Apart from providing energy corrections approaching the CBS limit, the use of TC Hamiltonians offers additional advantages. Specifically, the ground state of $\hat{H}_{\rm TC}$ is shown to be more compact than that of the original Hamiltonian.\cite{explicit2012,explicit2017,similar_compact2019,compactsci2024} This is intuitively understandable, as the electron correlation due to the cusp condition is already incorporated into the TC Hamiltonian. Consequently, the CI expansion of the ground-state eigenvector of $\hat{H}_{\rm TC}$ does not need to encompass a large number of Slater determinants.\cite{tc_quantum2024}  This property is particularly crucial for the QCMPS ansatz. The reduced number of configurations needed to precisely describe the ground state allows us to use a lower bond dimension, which, in our QCMPS setting, further reduces the number of qubits needed for the simulation.

\section{\label{sec:Results}Results and Discussion}

In this section, we initiate our discussion by demonstrating the advantages of utilizing VarQITE to optimize QCMPS-based parametrized quantum circuit. Our analysis focuses on the $\rm H_2$, linear $\rm H_4$, and $\rm LiH$ molecules as our test cases. 

The regularization parameter $\delta$ is set to $10^{-5}$ in all the simulations. For the parameter-updating step $\Delta \tau$ in each VarQITE iteration of Hermitian Hamiltonians, we implement the adaptive-learning-rate strategy, where we select a step that minimizes the energy expectation $\langle\psi(\boldsymbol{\theta})|\hat{H}|\psi(\boldsymbol{\theta})\rangle$ within the set $\mathcal{S}=\{0.02, 0.05, 0.10, 0.20, \cdots, 0.70\}$. 
This strategy, as extensively discussed in the previous works,\cite{adaptlr2021,SCNQS2024} effectively capitalizes on the benefits of both large and small learning rates: the former enhances efficiency, while the latter ensures robustness in optimization.\cite{SCNQS2024}

To investigate the behavior of VarQITE-QCMPS simulation of non-Hermitian system, we also consider the transcorrelated Hamitonians of $\rm Be$ atom and $\rm LiH$ molecule. The TC qubit Hamiltonians are sourced from the GitHub repository provided by Dobrautz et al.\cite{github}

In this work, all quantum circuit simulations are executed using the Python package \texttt{Cirq}.\cite{cirq} VarQITE calculations terminate when the absolute difference between consecutive iteration expectation values falls below $10^{-7}$ or after 500 iterations. Qubit Hamiltonians are generated using \texttt{OpenFermion},\cite{of2020} and for LiH within the active space defined by natural molecular orbitals (NMOs), the Hamiltonian is computed using \texttt{PySCF}.\cite{PySCF18,PySCF20} Optimization employs the BFGS optimizer from \texttt{SciPy}.\cite{scipy2020}. For all the BFGS simulations, in practice, we adopt the following cost function proposed by Fan et al. to enforce the results of optimization to live in the sector of correct particle-number and zero total spin:\cite{QCMPSchem2023}
\begin{equation}
    \begin{aligned}
    \mathcal{F}(\boldsymbol{\theta}) &= \mel{\psi(\boldsymbol{\theta})}{\hat H}{\psi(\boldsymbol{\theta})} + \abs{\langle\psi(\boldsymbol{\theta})|\hat{S}^2 |\psi(\boldsymbol{\theta})\rangle}^2\\
    &+\abs{\mel{\psi(\boldsymbol{\theta})}{\hat{N}}{\psi(\boldsymbol{\theta})}- N_{\rm ele}}^2
    \end{aligned}
\end{equation}
where $\hat{S}$ and $\hat{N}$ are the total spin operator and the total particle number operator, respectively. Our Simulations assume an exact state density operator framework, neglecting qubit decoherence and gate infidelities. Benchmarking with DMRG calculations uses an in-house developed program, where two-site DMRG algorithm and random initialization with each tensor element sampled from the uniform distribution are used. 

For constructing NMOs of the LiH molecule, we adopt settings from prior work.\cite{tc_quantum2024} Initially, we consider the MP2 wave function under the cc-pVDZ basis set at a balanced bond length of $1.595$ Å. The 1-reduced density matrix (1-RDM), defined by $D_{m,n} :=\mel{\Psi_{\rm MP2}}{\hat{a}_m^\dagger \hat{a}_n}{\Psi_{\rm MP2}}$, is computed and diagonalized. Eigenvectors with significant occupation numbers form the basis for the new fermionic Hamiltonian, known as NMOs.\cite{nmo1955,nmo2021}

\subsection{\label{sec:hermite}Optimization of Hermitian Hamiltonians}

In this section, we examine the effectiveness of the VarQITE optimization of Hermitian systems, in comparison to the conventional BFGS optimizer. Due to the limitations of our both hardware and software, simulation of VarQITE algorithm is only applicable for systems of small sizes,\cite{tc_quantum2024} so in this work we only consider the QCMPS with at most $4$ qubits, which is equivalent to the MPS with virtual bond dimension $D=8$. Despite these limitations, we can still see the advantage of VarQITE-QCMPS in the optimization behavior and the preciseness of finding the ground state of the systems. For each system, we qualify the performance of the simulation using the energy error $\Delta E$ with respect to the FCI energy in the corresponding basis set, which can be computed by exactly diagonalizing the qubit Hamiltonians.

\begin{figure}[!h]
   
    \centering
   
    \includegraphics[height = 5.1cm]{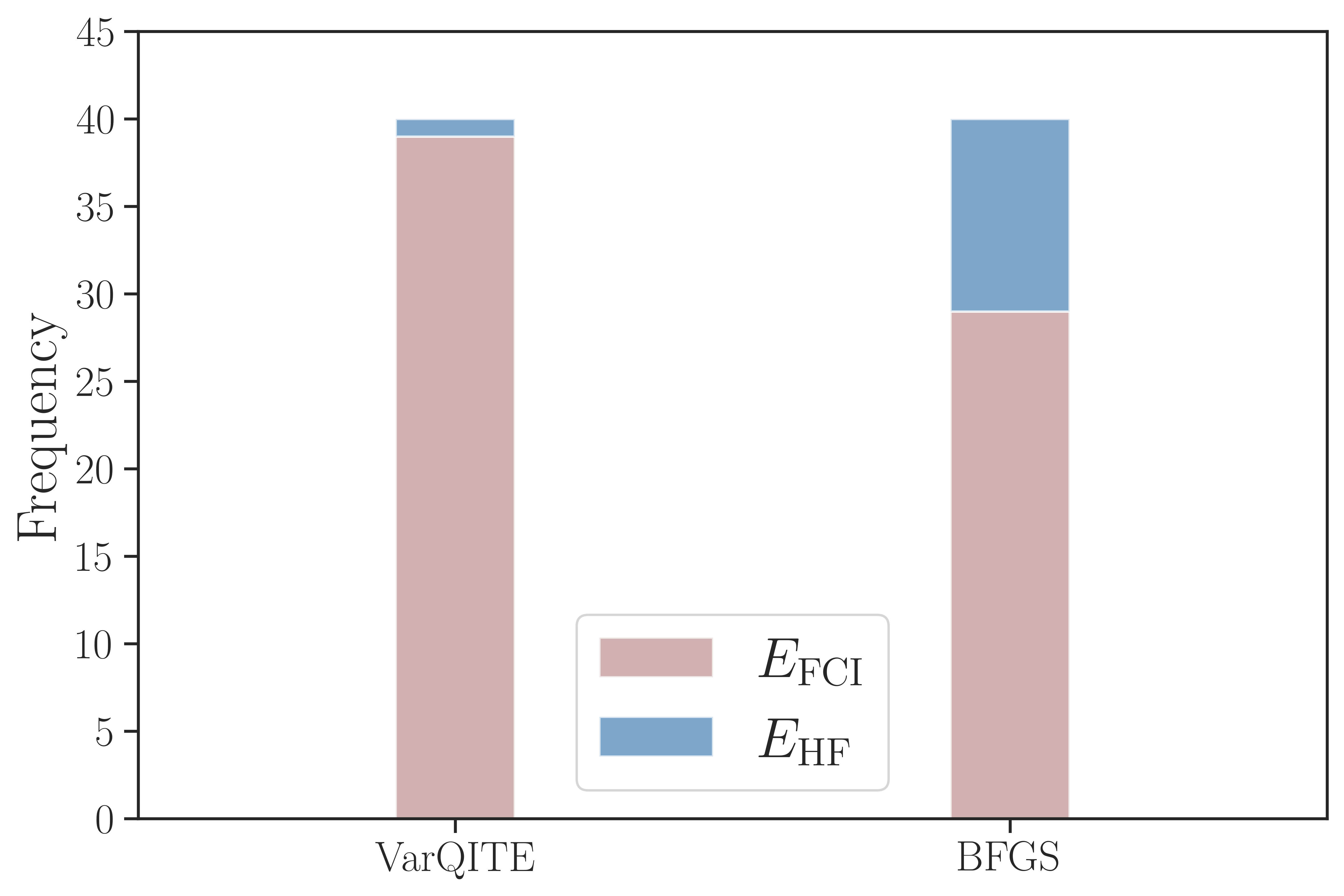}
    
    \includegraphics[height = 5.5cm]{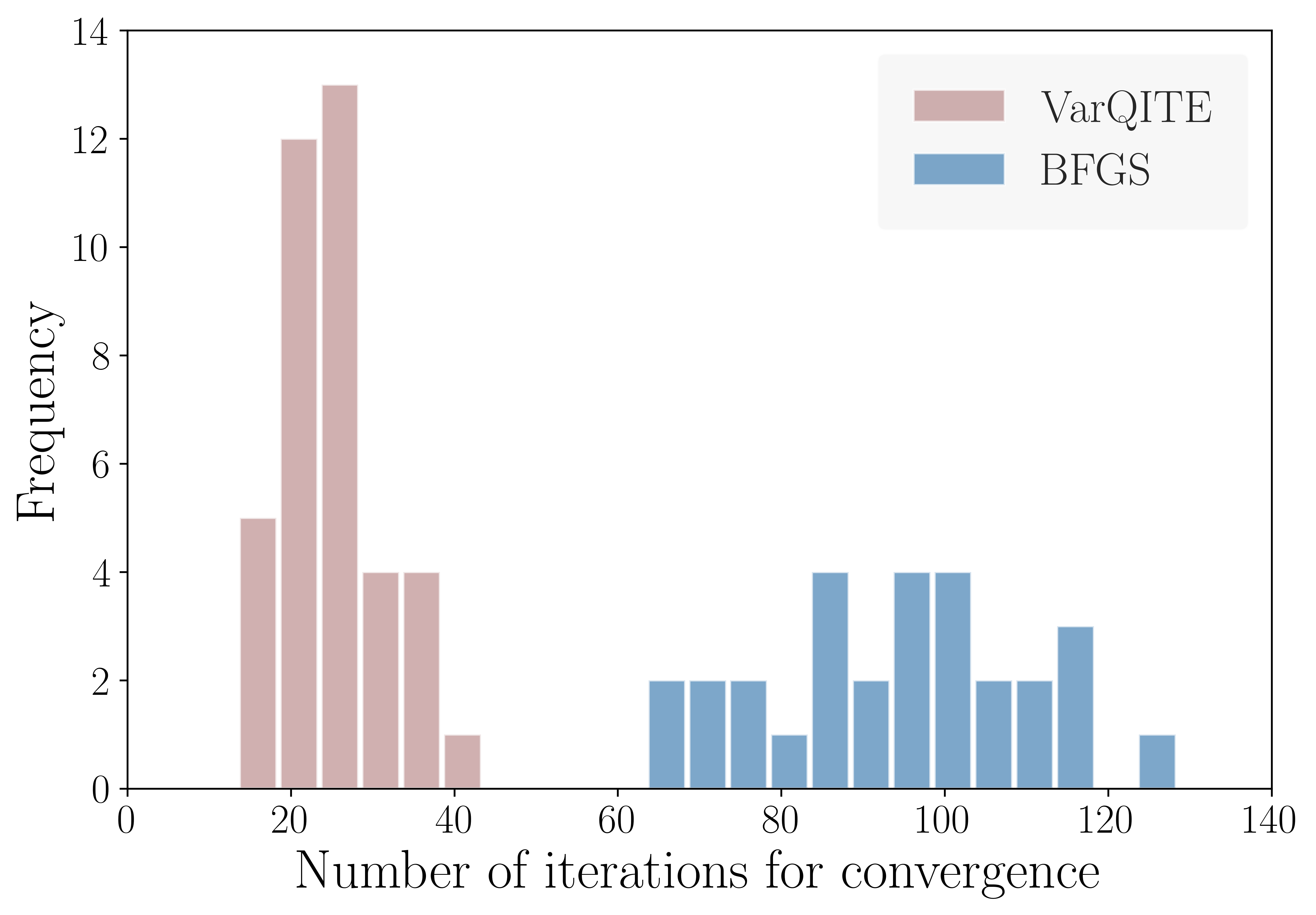}
    
    \caption{\label{fig:h2_result} Performance of VarQITE and BFGS optimizer in simulating the simplest $\rm H_2$ molecule, yielding either Hartree-Fock or FCI energy. The plot below depicts the distribution of the number of iterations required for convergence. Only the cases that converge to the FCI energy are considered.}
\end{figure}

For $\rm H_2$ molecule at a bond length of $0.7$ Å in the STO-3G basis,  a bond dimension of $D=2$ (equivalent to $2$ qubits) and a layer number of $L=1$ are sufficient to accurately prepare the ground state with an energy error relative to the FCI energy of less than $10^{-6}$. 

Even for this simplest system, we observe that BFGS optimization can encounter the issue of getting stuck at local minima, failing to reach the exact ground state energy. To rule out the influence of different random initial parameters, we employ 40 different random seeds to generate initial parameters for both VarQITE and BFGS optimizations. As shown in Figure \ref{fig:h2_result}, a significant proportion of BFGS cases converge to the Hartree-Fock energy, which is the most typical local minimum of the $\rm H_2$ molecule's energy expectation, $\mathcal{E}(\boldsymbol{\theta})$. In contrast, only one case of VarQITE optimization fails to achieve the FCI energy. Furthermore, among the initial parameters that do converge to the ground state, the number of steps required in VarQITE is significantly smaller than that in BFGS.

\begin{figure}[!h]
    \includegraphics[height = 6cm]{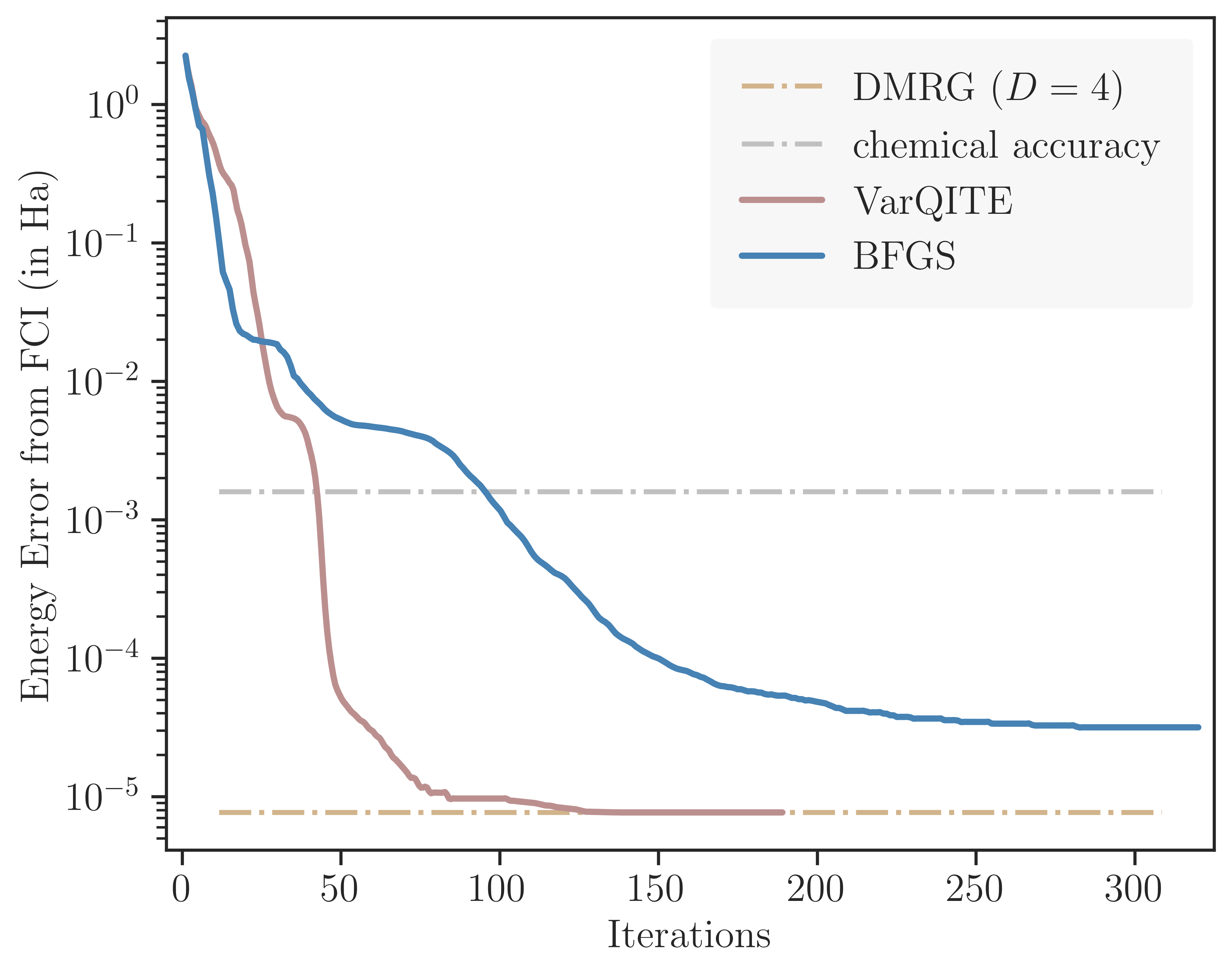}
    \includegraphics[height = 6cm]{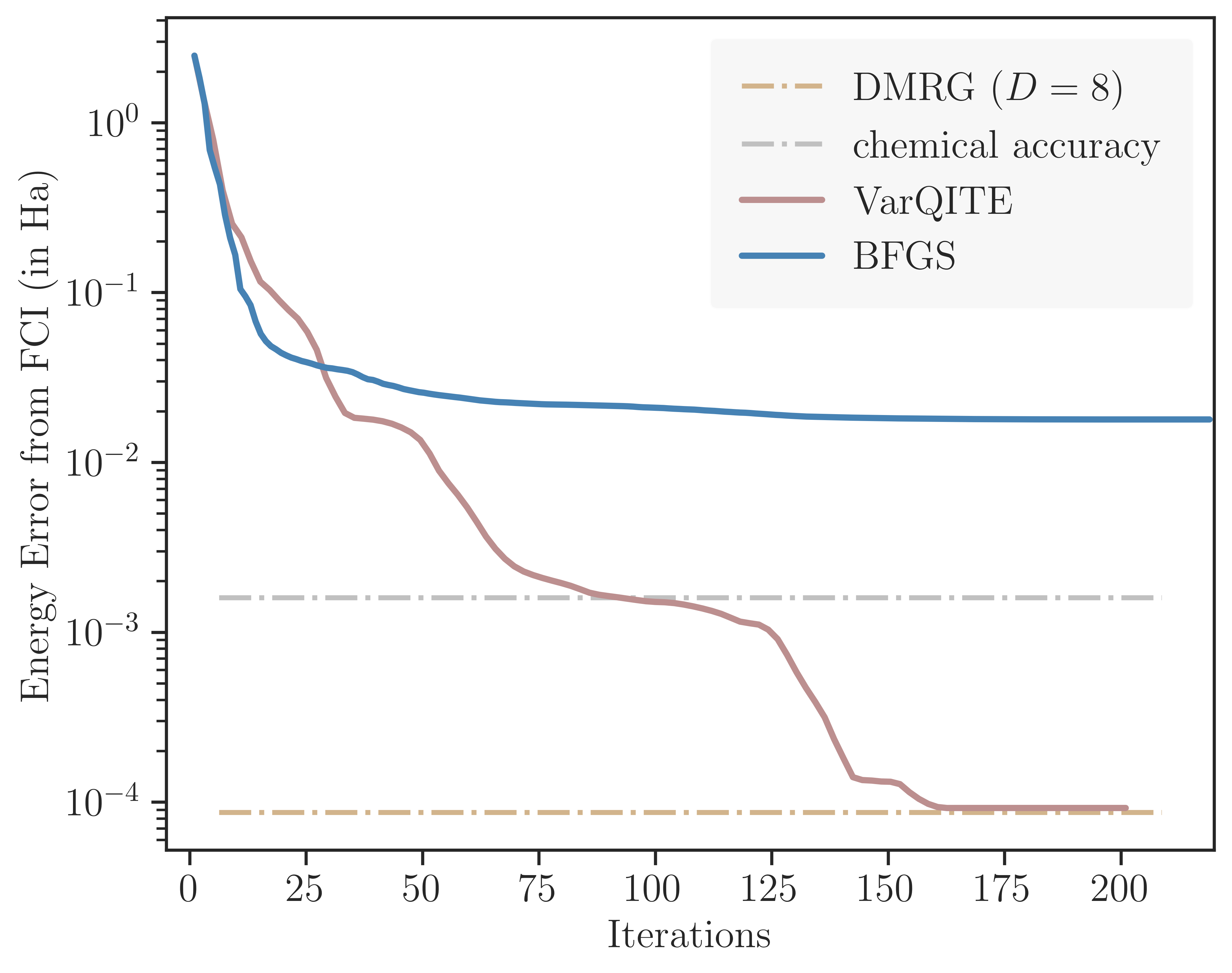}
    \caption{\label{fig:lost_function} Optimization performance of the standard BFGS iteration scheme and the VarQITE-QCMPS scheme on $\rm LiH$ at 1.595 Å ($D=4, L=2$, the upper panel) and $\rm H_4$ at $0.7$ Å ($D=8, L=2$, the lower panel). The energy error is relative to the FCI energy in both of the two panels. }
\end{figure}

The advantage of VarQITE is also evident in higher bond dimensions and larger layer number $L$, as well as in other molecular systems. As shown in Figure \ref{fig:lost_function}, we perform simulations on $\rm LiH$ ($D=4$, $L=2$) and $\rm H_4$ ($D=8$, $L=2$) systems. For the LiH molecule, we choose 4 NMOs to build the Hamiltonian, leading to 8-qubit Hamiltonian
after the Jordan-Wigner transformation. The qubit Hamiltonian for the $\rm H_4$ molecule is generated under the STO-3G basis at a bond length of 0.7 Å.

\begin{figure}[!h]
    \includegraphics[height = 6cm]{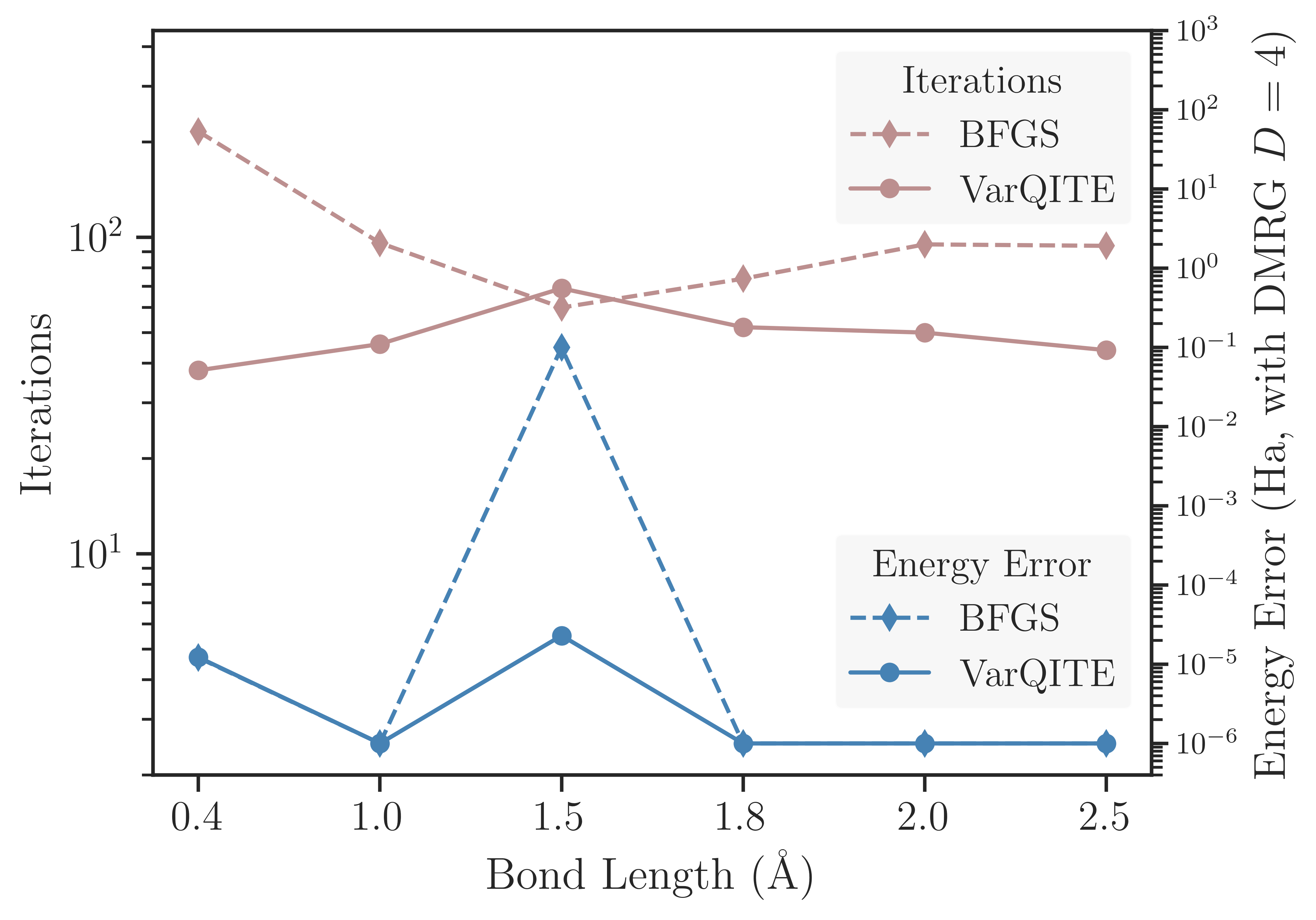}
     \includegraphics[height = 6cm]{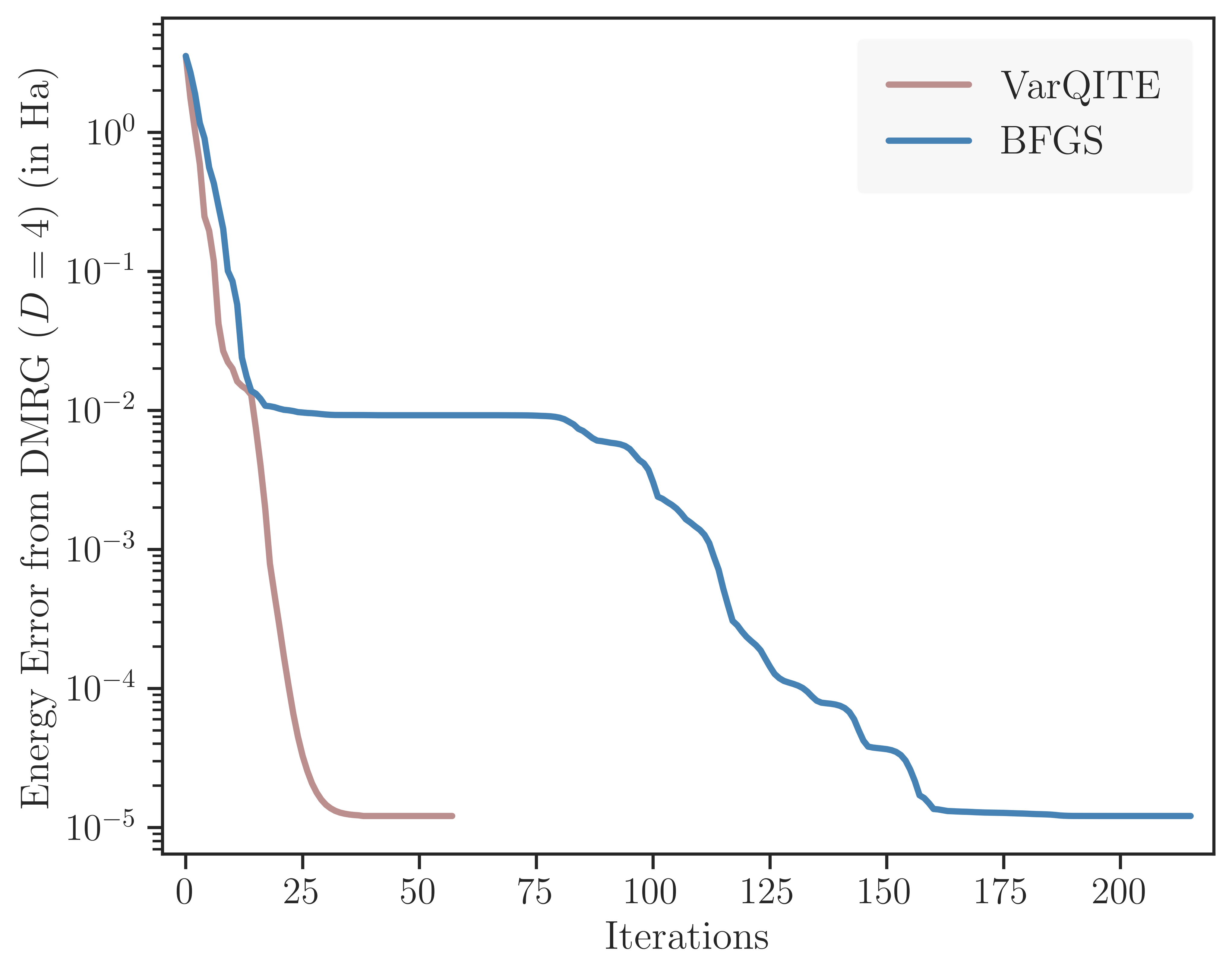}
    \caption{\label{fig:bdlength} Simulation of $\rm H_4$ molecule within the STO-3G basis set at different bond lengths, where the QCMPS ansatz with $D=4$ (3 qubits), $L=2$ is chosen. In the upper panel, the red line represents the number of iteration steps for the BFGS and VarQITE optimizers to conclude, while the blue one represents the error between the QCMPS results and the results obtained by DMRG calculation the equivalent classical MPS ($D=4$). In the lower panel, we showcase the optimization behavior at the 0.4 Å bond length.}
\end{figure}

For the LiH molecule, both BFGS and VarQITE yield energy results within chemical accuracy compared to the FCI result under the 4NMO basis. However, the VarQITE optimization requires significantly fewer steps and provides a more accurate result, aligning closely with the DMRG result. In the optimization of $\rm H_4$ with $D=8$ and $L=2$, the BFGS optimizer fails to achieve results within chemical accuracy compared to FCI, whereas VarQITE achieves this accuracy in fewer than 100 steps.

To explore the optimization performance at different bond length of linear $\rm H_4$, we generate the qubit Hamiltonians of $\rm H_4$ within the STO-3G basis at various bond lengths ranging from $0.4$ Å to $2.5$ Å and implement $D=4$ simulation to compare the optimization performance, as illustrated in Figure \ref{fig:bdlength}. We observe that BFGS optimizer struggles significantly in the optimization of $\rm H_4$ at the short bond length, where it got stuck in a local minimum for several iterations, ultimately requiring a considerable number of steps to converge. Additionally, BFGS exhibits significant errors for stretched bond lengths (1.5 Å), where stronger electron correlation exists. In contrast, VarQITE consistently achieves accuracy comparable to classical DMRG optimization and requires fewer overall convergence steps.
Our experiments demonstrate that VarQITE, which provides analytical metrics and gradients, indeed manifests favorable performance in terms of both efficiency and stability.

As with the simulations of $\rm H_2$ systems, to eliminate the impact of varying random initial parameters, we use 10 different random seeds to generate initial parameters for the 4-NMO $\rm LiH$, $\rm H_4$ at 0.7 {\AA}, and $\rm H_4$ at 0.4 {\AA} systems. The statistical results can be found in Appendix \ref{app:new}.
\begin{figure*}
    \centering
    \includegraphics[height = 7cm]{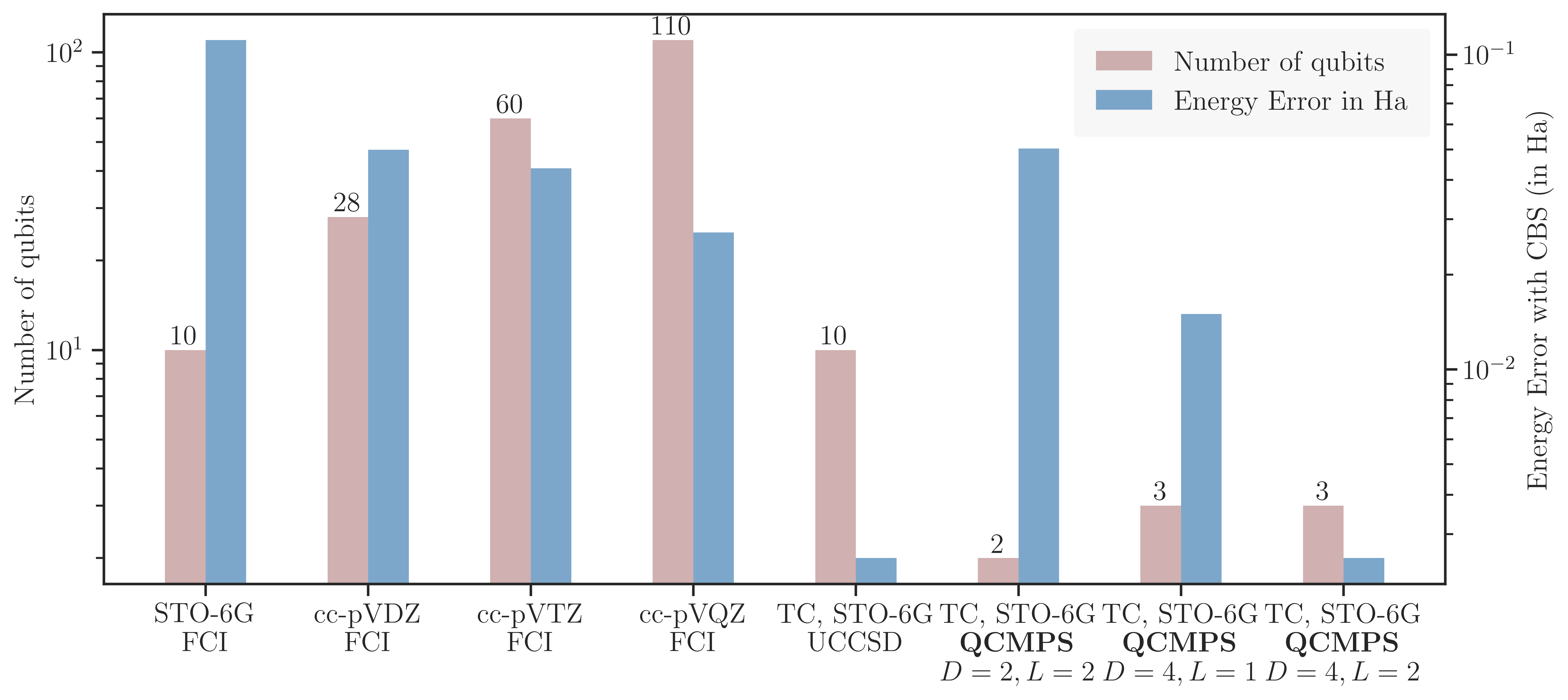}
    \caption{\label{fig:beatom} Comparison of non-TC FCI results and TC-VarQITE simulation results. The UCCSD\cite{tc_quantum2024} and CBS\cite{deepwf2019} results are extracted from the references for benchmarking. Except for QCMPS, all the required numbers of qubits is estimated based on the numbers of spin orbitals in the basis sets. }

\end{figure*}

\subsection{\label{sec:h2o_n2}Beryllium atom}

In this segment, we evaluate the effectiveness of our VarQITE-QCMPS algorithm for simulating the ground-state energy of the transcorrelated $\rm Be$ atom. The transcorrelated Hamiltonian of the $\rm Be$ atom is generated within the STO-6G basis and converted to a 10-qubit Hamiltonian. Our results indicate that the QCMPS ansatz can find a good approximation of the CBS energy of $\rm Be$ using only 3 qubits.

To qualify our algorithm, we showcase the FCI results for $\rm Be$ atom system without TC under rather large basis sets, such as the cc-pVDZ, cc-pVTZ and cc-pVQZ basis. As shown in Figure \ref{fig:beatom}, our algorithm indeed further reduces the number of qubits required for TC simulations. Specifically, a $D=2$, $L=2$ QCMPS with only $2$ qubits reaches energy comparable to the FCI energy within the cc-pVDZ basis set, which will require 28 qubits for an all-electron simulation if the conventional orbital-to-qubit transformation is used. And at the $D=4$ cases, where only $3$ qubits are necessary, the QCMPS result with $L=1$ already outperforms the FCI result in the cc-pVQZ basis set, which corresponds to $110$ qubits. Furthermore, the TC-VarQITE optimization under the Unitary Coupled Cluster with Singles and Doubles (UCCSD) ansatz requires only 10 qubits to achieve an energy difference of less than 2 mHa from the CBS limit.\cite{tc_quantum2024} However, through combination with QCMPS, this number can be further reduced to 3, as demonstrated in the case of $D=4, L=2$ in the Figure \ref{fig:beatom}.

The convergence of the complex cost function $\langle\psi(\boldsymbol{\theta})|\hat{H}_{\rm TC}|\psi(\boldsymbol{\theta})\rangle$ during the VarQITE iterations for the $D=4, L=2$ QCMPS ansatz is illustrated in Appendix \ref{app:C}. The real part of the cost function shows a monotonic decrease, ultimately converging to the minimum right eigenvalue of the TC Hamiltonian, which is obtainable through direct exact diagonalization. Regarding the imaginary part, oscillations between positive and negative values occur during the optimization process, yet its modulus gradually diminishes and eventually converges to zero. This indicates that the optimization process for the non-Hermitian TC Hamiltonian is both numerically stable and efficient.

\subsection{\label{sec:perform_block} Lithium hydride}

We conduct VarQITE-QCMPS calculations on transcorrelated LiH at its equilibrium bond length of $1.595$ Å. We employ the NMOs obtained from the MP2 wave function in the cc-pVDZ basis set, selecting either $3$ or $4$ orbitals with the most significant occupation numbers. As demonstrated in Table \ref{tab:energy}, the VarQITE-QCMPS method yields results with an accuracy within 
$10^{-6}$
  compared to the energies obtained directly from exact diagonalization (ED) of the TC Hamiltonian, using only $3$ qubits or a bond dimension of $D=4$, in both $3$ and $4$NMO-MP2 basis. This accuracy surpasses that of the FCI energy in the cc-pVDZ basis set, which requires up to $38$ qubits for conventional quantum simulation. The simulation can also be carried out using VarQITE-UCCSD ansatz, using $6$ and $8$ (or $4$ and $6$ if parity-encoding is implemented) qubits .\cite{tc_quantum2024} The comparison between the two ansatzes reveals that QCMPS still provides a noteworthy reduction in qubit requirements and improved accuracy at the same time.
  
\begin{table}[!h]
    \centering
    \caption{\label{tab:energy} Ground state energies of $\rm LiH$ using different methods. The TC-UCCSD results are extracted from the work of Dobrautz et al.\cite{tc_quantum2024} The number of required qubits in the first three rows is estimated based on the number of spin orbitals in the basis set, while in the two rows of TC-UCCSD, the numbers in brackets represent the requirements of qubits when parity encoding is used instead of Jordan-Wigner mapping. The bold faced items indicate the results obtained using TC-QCMPS method.}

    \begin{tabular}{cccc}
        \toprule
       \multicolumn{2}{c}{Method } & Number of Qubits&Energy \\
        \midrule
        \multirow{1}{*}{FCI}& cc-pVDZ & 38 & $-8.01473$\\
        \multirow{2}{*}{TC-UCCSD\cite{tc_quantum2024}}&3NMO-MP2&6(4)&$-8.02126$\\

        &4NMO-MP2&8(6)&$ -8.02307$\\
        \multirow{2}{*}{\textbf{TC-QCMPS}}&3NMO-MP2&\textbf{3}&$\mathbf{-8.02161}$\\
        &4NMO-MP2&\textbf{3}&$\mathbf{-8.02339}$\\
        \multirow{2}{*}{TC-ED}&3NMO-MP2&-&$-8.02161$\\
        &4NMO-MP2&-&$-8.02339$\\
        \bottomrule
    \end{tabular}
\end{table}

In previous work,\cite{tc2010,tc2011,tc2012,similar_compact2019,tc_quantum2023,tc_quantum2024,compactsci2024} the compactness of the ground state TC Hamiltonian has been thoroughly explored, revealing that the ground state of the TC Hamiltonian generally involves fewer determinants in its CI expansion compared to the non-TC case.
The specific manifestation is that the squared modulus of the coefficients for the Hartree-Fock state $|c_{\rm{HF}}|^2$ in the ground state is notably greater for the TC Hamiltonian in comparison to the non-TC counterpart.\cite{tc_quantum2024} In essence, this implies that the ground state exhibits weaker multi-configuration properties in the case of the TC Hamiltonian. For the conventional quantum algorithms based on orbital-to-qubit mapping, this property makes it possible to apply a shallower circuit ansatz and thus reduce the circuit complexity for simulation, making the algorithm more promising for applications on the NISQ devices. For example, the HEA has already been employed in experiments on today's actual superconducting quantum hardware for simulating transcorrelated LiH.\cite{tc_quantum2024}

These properties are particularly crucial for QCMPS as well. For MPS, the weak multi-configuration nature implies weak electronic correlation, allowing for the description of a quantum state with a low virtual bond dimension. In our algorithm, this directly corresponds to fewer qubits needed for simulating the TC Hamiltonian compared to the non-TC one, thanks to the compactness of the ground state.
As depicted in Table \ref{tab:tc_vs_notc}, for 3NMO-MP2, simulating the TC system requires only $D=4, L=1$ to achieve an accuracy of $10^{-6}$ compared to the ED result. In contrast, simulating the non-TC Hamiltonian necessitates an additional layer, effectively doubling the number of parameters, to reach the same level of precision relative to the corresponding FCI result. For 4NMO-MP2, simulating the TC Hamiltonian with $D=4, L=2$ is adequate to achieve an accuracy of $10^{-6}$, while the non-TC counterpart fails to attain this level of precision at the same bond dimension and number of layers.

\begin{table}[!h]
\centering
    \caption{\label{tab:tc_vs_notc} Performance of VarQITE-QCMPS method applying to TC and non-TC system. 
    All energy errors are relative to the exact diagonalization of their respective Hamiltonians. The bolded item in each line (if present) indicates the minimum QCMPS-VarQITE simulation resources required to achieve an energy accuracy of $10^{-3}$ mHa.}
\begin{tabular}{lcccc}
\toprule
\multirow{2}{*}{Energy Error in mHa} & \multicolumn{2}{c}{$D=2$} & \multicolumn{2}{c}{$D=4$} \\
\cmidrule(lr){2-3} \cmidrule(lr){4-5}
 & $ L=1 $ & $ L=2 $ & $ L=1 $ & $ L=2 $ \\
\midrule
{LiH (3NMO-MP2), non-TC} & 0.019 & 0.019 & 0.015 & \textbf{0.001} \\

LiH (3NMO-MP2), TC & 1.845 & 1.845 & \textbf{0.001} & 0.001 \\
{LiH (4NMO-MP2), non-TC} & 4.846 & 4.846 & 0.022 & 0.011 \\

LiH (4NMO-MP2), TC & 3.629 & 3.629 & 1.784 & \textbf{0.001} \\
\bottomrule
\end{tabular}
\end{table}

\section{\label{sec:conclusion}Conclusion}

This research introduces an algorithm that combines VarQITE and QCMPS, aimed at enhancing the optimization performance of the QCMPS ansatz and extending its applicability to non-Hermitian systems. This advancement addresses potential issues encountered during QCMPS optimization, significantly improving its efficiency, success rate, and stability. Traditionally, QCMPS optimization relied on an external optimizer such as BFGS, utilizing only the function values corresponding to the parameters during the optimization process. This approach may become not well-suited for QCMPS ansatz forms with complicated structures. In contrast, VarQITE leverages the analytical gradients, reducing the likelihood of getting trapped in local minima due to gradient errors. Additionally, the Fisher information matrix incorporates the structural information of the circuit itself, enhancing optimization efficiency. This improved strategy of optimization offers a potential method for applying QCMPS to various molecular systems. Furthermore, efforts to optimize QCMPS for non-Hermitian systems have demonstrated the potential to further reduce the resources required for quantum computing in TC systems. This advancement opens up possibilities for quantum simulations of TC calculations for a broader range of molecules and non-Hermitian problems in both physics and chemistry.

Despite these advancements, optimizing QCMPS remains a highly nonlinear and non-convex problem and currently lacks the robust optimization methods available as DMRG for classical MPS. Specifically, the power of classial DMRG algorithm stems from the efficient truncation based on singular value decomposition (SVD) to a large extent, and this mileage is not straightforwardly available in QCMPS, leading to the difficulties in optimization compared to the classical DMRG algorithm. Additionally, increasing the number of orbitals ($N_{\rm orb}$), which leads to greater circuit depth, and enlarging the MPS bond dimension, which necessitates a higher number of qubits, may present further challenges for optimization that warrant additional investigation. Due to computational resource limitations, we have not yet explored larger molecular systems or systems with stronger correlations. Future research should focus on further modifying the structure of each component tensor encoder to better adapt to VarQITE optimization. Additionally, translating classical MPS-based algorithms for solving problems beyond ground state energy into variational optimization problems for QCMPS presents an intriguing avenue for exploration. We hope that this research will improve the optimization behavior of QCMPS and further expand its applications in quantum physics and chemistry.
\begin{acknowledgments}

This work was financially
supported by the National Natural Science Foundation
of China (Grant Nos. 223B1011, 22222605 and 22076095). The Tsinghua Xuetang Talents Program and High-Performance Computing Center of Tsinghua University were acknowledged for
providing computational resources. H.-E. Li and H.-S. Hu thank Dr. Dingshun Lv (ByteDance Research), Dr. Chang-Su Cao (ByteDance Research), Prof. Zhenyu Li (University of Science and Technology of China), Dr. Yi Fan (University of Science and Technology of China), Prof. Xiao Yuan (Peking University) and Dr. Werner Dobrautz (Chalmers University of Technology) for helpful discussions.
\end{acknowledgments}
\section*{Author Declarations}
\subsection*{Conflict of Interest}

The authors have no conflicts to disclose.
\subsection*{Author Contributions}

\textbf{Hao-En Li:} Conceptualization (lead); Data Curation (lead); Formal Analysis (lead); Investigation (lead); Methodology (lead); Software (equal); Validation (equal); Visualization (equal); Writing – original draft (lead); Writing – review \& editing (equal)
\textbf{Xiang Li:} Conceptualization (equal); Formal Analysis (equal); Investigation (equal); Methodology (equal); Software (equal); Validation (equal); Writing – review \& editing (equal)
\textbf{Jia-Cheng Huang:} Conceptualization (equal); Formal Analysis (equal); Investigation (equal); Methodology (equal); Validation (equal); Visualization (equal); Writing – review \& editing (equal)
\textbf{Guang-Ze Zhang:} Conceptualization (equal); Formal Analysis (equal); Investigation (equal); Methodology (equal); Validation (equal); Visualization (equal); Writing – review \& editing (equal)
\textbf{Zhu-Ping Shen:} Validation (equal); Visualization (equal); Writing – review \& editing (supporting)
\textbf{Chen Zhao:} Visualization (equal); Writing – review \& editing (supporting)
\textbf{Jun Li:} Funding Acquisition (supporting); Project Administration (supporting); Supervision (supporting); Writing – review \& editing (supporting)
\textbf{Han-Shi Hu:} Funding Acquisition (equal); Project Administration (equal); Resources (equal); Supervision (equal); Writing – review \& editing (equal)
\section*{Data Availability Statement}

The data that support the findings of this study are
available from the corresponding author upon reasonable
request.

\appendix

\section{Details in measurement of QCMPS}\label{app:A}

We discuss the method to measure the expectation value of QCMPS ansatz with respect to the qubit Hamiltonian. In the QCMPS setting, we need to reset on a qubit after applying each $\mathscr{U}^i$ in order to reuse the physical index. The reset-to-zero channel is equivalent to evolving quantum circuit density matrix into the form of Kraus representation:\cite{kraus1975,kraus2003}

\begin{equation}\label{eq:trs}
    \rho \mapsto M_0\rho M_0^\dagger + M_1\rho M_1^\dagger,
\end{equation}

\begin{figure}[!h]
    \centering
    \includegraphics[scale = 1.2]{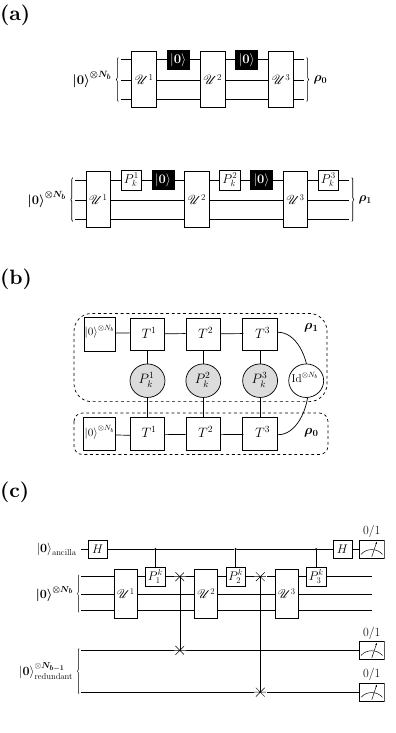}
    \caption{\label{fig:appendA}(\textbf{a}) Quantum circuits with output states $\rho_0$ and $\rho_1$. (\textbf{b}) Graphical representation of $\mel{\psi}{\mathscr{P}_k}{\psi}$ using classical MPS. (\textbf{c}) The equivalent pure-state quantum circuit with SWAP gates and redundant qubits. }
\end{figure}
where
\begin{equation}
    M_0=\mathrm{Id}_{\rm anc}\otimes \dyad{0}{0}\otimes \mathrm{Id}^{\otimes N_b-1},\quad M_1=\mathrm{Id}_{\rm anc}\otimes \dyad{0}{1}\otimes \mathrm{Id}^{\otimes N_b-1}.
\end{equation}
The subscript ``$\mathrm{anc}$'' represents that the operator is applied to the ancilla subsystem.

According to the correspondence between quantum circuits and classical MPS, the measurement of $\mel{\psi}{\mathscr{P}_k}{\psi}$ entails evaluating the ``inner product'' between the states $\rho_0$ and $\rho_1$, illustrated in Figure \ref{fig:appendA}(a), which can be graphically represented using classical MPS as depicted in Figure \ref{fig:appendA}(b). In fact, if we interpret the measurement operation—corresponds to taking trace on the last tensor $T^3$ in Figure \ref{fig:appendA} (b)—as a contraction with the operator 
$\mathrm{Id}^{\otimes {N_b}}$, then the wave function encoded by QCMPS can also be viewed as an MPS with $N_b$  additional physical indices on the last tensor (denoted as $\ket{\psi'(\boldsymbol{\theta})}$). Consequently, the optimization problem we are actually dealing with becomes:
\begin{equation}\label{eq:eqproblem}
    \boldsymbol{\theta}^\ast = \mathrm{argmin}_{\boldsymbol{\theta}}\mathcal{E}'(\boldsymbol{\theta})=\mathrm{argmin}_{\boldsymbol{\theta}}\mel{\psi'(\boldsymbol{\theta})}{\hat{H}\otimes \mathrm{Id}^{\otimes N_b}}{\psi'(\boldsymbol{\theta})}
\end{equation}
because
\begin{equation}
    \sum_{k}c_k \mathscr{P}_k\otimes \mathrm{Id}^{\otimes N_b}=\hat{H}\otimes\mathrm{Id}^{\otimes N_b}.
\end{equation}
The eigenvalues of $\hat{H}\otimes \mathrm{Id}^{\otimes N_b}$ are each eigenvalue of $\hat{H}$ repeated $2^{N_b}$ times, therefore the problem in Eq.(\ref{eq:eqproblem}) also finds the ground-state energy of $\hat{H}$.

We notice that after the transformation Eq.(\ref{eq:trs}), the state on the quantum circuit is no longer a pure state since $\mathrm{Tr}\rho^2 <1$. However, for the purpose of making the derivation easier to understand, it's helpful to contemplate an equivalent pure-state quantum circuit where SWAP gates and $N_r=N_b-1$ ``redundant qubits'' are included to replace the reset-to-zero channels. This is illustrated in Figure\ref{fig:appendA}(c), where we establish the state

\begin{equation}\label{eq:state}
    \small \frac{\ket{0}_{\mathrm{anc}}\left(\sum_{\mathbf{f}\in \{0,1\}^{N_r}}\ket{\psi_\mathbf{f}}\ket{\mathbf{f}}_{\mathrm{red}}\right)+\ket{1}_{\mathrm{anc}}\left(\sum_{\mathbf{f}\in \{0,1\}^{N_r}}|{\widetilde{\psi}_\mathbf{f}}\rangle\ket{\mathbf{f}}_{\mathrm{red}}\right)}{\sqrt{2}},
\end{equation}
where $\ket{\psi_\mathbf{0}}$ and $|\widetilde{\psi}_{\mathbf{0}}\rangle$ are exactly the equivalent ``subsystem state'' to the aforementioned output states $\rho_0$ and $\rho_1$. Once the state in Eq.(\ref{eq:state}) is prepared, applying a final Hadamard gate on the ancilla qubits yields
\begin{equation}
    \begin{aligned}
       &\frac{1}{2}\ket{0}_{\mathrm{anc}}\sum_{\mathbf{f}\in \{0,1\}^{N_r}}\left(\ket{\psi_\mathbf{f}}+|\widetilde{\psi}_{\mathbf{f}}\rangle\right)\ket{\mathbf{f}}_{\mathrm{red}} \\     
       +&\frac{1}{2}\ket{1}_{\mathrm{anc}}\sum_{\mathbf{f}\in \{0,1\}^{N_r}}\left(\ket{\psi_\mathbf{f}}-|\widetilde{\psi}_{\mathbf{f}}\rangle\right)\ket{\mathbf{f}}_{\mathrm{red}},
    \end{aligned}
\end{equation}
and measuring the ancilla and redundant qubits yields the joint probabilities,
\begin{equation}
    \small
    \begin{aligned}
   &P(\mathrm{anc}=0,\mathrm{red}=\mathbf{0})= \frac{1}{4}\left(\norm{\ket{\psi_{\mathbf{0}}}}_2^2+\lVert|\widetilde{\psi}_{\mathbf{0}}\rangle\rVert_2^2+2\Re\langle \psi_{\mathbf{0}}|\widetilde{\psi}_\mathbf{0}\rangle\right),\\
    &P(\mathrm{anc}=1,\mathrm{red}=\mathbf{0})= \frac{1}{4}\left(\norm{\ket{\psi_{\mathbf{0}}}}_2^2+\lVert|\widetilde{\psi}_{\mathbf{0}}\rangle\rVert_2^2-2\Re\langle \psi_{\mathbf{0}}|\widetilde{\psi}_\mathbf{0}\rangle\right),\\
    &\Rightarrow \Re\langle \psi_{\mathbf{0}}|\widetilde{\psi}_\mathbf{0}\rangle=P(\mathrm{anc}=0,\mathrm{red}=\mathbf{0})-P(\mathrm{anc}=1,\mathrm{red}=\mathbf{0}).
    \end{aligned}
\end{equation}

This process is consistent with the results obtained from the standard Hadamard's test on the original QCMPS with the reset-to-zero channels.

For the TC Hamiltonian, the expectation value of the wave function with respect to Pauli strings is generally not real, requiring separate calculations for the real and imaginary parts. To obtain the imaginary part, we simply need to follow the initial Hadamard gate with an $S = \exp(-\mathrm{i}\pi Z/2)$ gate. Consequently, the state prepared by Eq.(\ref{eq:state}) will be,
\begin{equation}
    \small \frac{\ket{0}_{\mathrm{anc}}\left(\sum_{\mathbf{f}\in \{0,1\}^{N_b}}\ket{\psi_\mathbf{f}}\ket{\mathbf{f}}_{\mathrm{red}}\right)+\ket{1}_{\mathrm{anc}}\left(\sum_{\mathbf{f}\in \{0,1\}^{N_b}}\mathrm{i}|{\widetilde{\psi}_\mathbf{f}}\rangle\ket{\mathbf{f}}_{\mathrm{red}}\right)}{\sqrt{2}},
\end{equation}
allowing us to compute the imaginary part as:
\begin{equation}
    \mathrm{Im}\langle \psi_{\mathbf{0}}|\widetilde{\psi}_\mathbf{0}\rangle = P(\mathrm{anc}=0,\mathrm{red}=\mathbf{0})-P(\mathrm{anc}=1,\mathrm{red}=\mathbf{0}).
\end{equation}
\section{Variational quantum imaginary time evolution algorithm}\label{app:B}

The VarQITE has already been sufficiently developed by the previous works\cite{varqite2019, varqite2023, varqite_theory2019}. Based on the principle of imaginary time evolution (ITE) in quantum mechanics, VarQITE offers an effective approach for converging PQCs toward accurate approximations of the ground states in physical and chemical systems. In the ITE formalism, we consider the imaginary-time Schrödinger's Equation
\begin{equation}\label{eq:itse}
    \pdv{\ket{\psi(\tau)}}{\tau} = -\hat{H}\ket{\psi(\tau)}.
\end{equation}
Given the initial state $\ket{\psi_0}$, we can express the explicit solution of Eq.(\ref{eq:itse}) as
\begin{equation}
    \ket{\psi(\tau)} = \frac{e^{-\hat{H}\tau}\ket{\psi_0}}{\sqrt{\mel{\psi_0}{e^{-2\hat{H}\tau}}{\psi_0}}}.
\end{equation}
Note that $e^{-\hat{H}\tau}$ is generally not unitary, thus the normalization factor $1/\sqrt{\mel{\psi_0}{e^{-2\hat{H}\tau}}{\psi_0}}$ must be included in the expression of $\ket{\psi(\tau)}$. If the initial state $\ket{\psi_0}$ has a nontrivial overlap with the true ground state of the Hamiltonian $\hat{H}$, then $\ket{\psi(\tau)}$ will converge to the ground state as $\tau \to\infty$. However, due to the non-unitarity of the state-propagation operator $e^{-\hat{H}\tau}$, directly implementing this evolution on quantum devices may pose challenges. Fortunately, we can encode $\ket{\psi(\tau)}$ using a PQC controlled by set of parameters $\boldsymbol{\theta}(\tau)$.\cite{varqite2019} In this way, we adopt a variational approach to approximate the ITE process within an \textit{ansatz submanifold} of the total Hilbert space. This submanifold is generated by the PQC denoted as $\ket{\phi(\boldsymbol{\theta}(\tau))}$. Drawing from MacLachlan's variational principle,\cite{mac1964} this algorithm finds the optimal approximation of the exact ITE state in the ansatz submanifold by minimizing the distance:
\begin{equation}\label{eq:mac}
    \delta \norm{\left(\pdv{\tau}+ \hat{H}-\langle{\hat{H}}\rangle_{\boldsymbol{\theta}(\tau)}\right)\ket{\phi(\boldsymbol{\theta}(\tau))}}=0,
\end{equation}
where $\norm{\rho} := \Tr[\sqrt{\rho\rho^\dagger}]$ denotes the trace-norm of a state and $\langle{\hat{H}}\rangle_{\boldsymbol{\theta}(\tau)}:=\mel{\phi(\boldsymbol{\theta}(\tau))}{\hat{H}}{\phi(\boldsymbol{\theta}(\tau))}$. Eq.(\ref{eq:mac}) leads the following system of ordinary differential equations:
\begin{equation}
    \sum_{j} \mathbf{A}_{i,j} (\tau) \dot{\theta}_j(\tau)= \mathbf{C}_i(\tau),
\end{equation}
where the coefficients $A_{i,j}$ and $C_i$, dependent of the imaginary time $\tau$, are given by
\begin{equation}\label{eq:a}
    \mathbf{A}_{i,j} (\tau)= 
    \Re \braket{\pdv{\phi(\boldsymbol{\theta}(\tau))}{\theta_i}}{\pdv{\phi(\boldsymbol{\theta}(\tau))}{\theta_j}},
\end{equation}
and
\begin{equation}\label{eq:c}
    \mathbf{C}_{i} (\tau)=  -\Re\mel{\pdv{\phi(\boldsymbol{\theta}(\tau))}{\theta_i}}{ \hat{H}}{ \phi(\boldsymbol{\theta}(\tau))}.
\end{equation}

The detailed derivation of Eq.(\ref{eq:a}) and Eq.(\ref{eq:c}) can be found in the reference.\cite{varqite_theory2019}

Inspired by the numerical ODE solvers, such as forward Euler method and Runge-Kutta method,\cite{varqite_theory2023} we select a step $\Delta \tau$ and then updates the parameters $\boldsymbol{\theta}_n$ at the $n$-th iteration according to 
\begin{equation}
    \boldsymbol{\theta}_{n+1} = \boldsymbol{\theta}_n + \Delta \tau \mathbf{A}^{-1} \mathbf{C}.
\end{equation}

Here, $\mathbf{A}$ and $\mathbf{C}$ represent the matrix $\mathbf{A}=(\mathbf{A}_{i,j})$ and the vector $\mathbf{C} = (\mathbf{C}_{i})$ at $\boldsymbol{\theta}_n$ respectively.

\section{Further statistical results for the optimization performance of VarQITE and BFGS methods}\label{app:new}

\begin{figure}[htbp]
    \centering
    \subfigure[ $\rm LiH$ at 1.595 {\AA}, $D = 4$, $L=2$]{
        \centering
        \includegraphics[width = 0.49\textwidth]{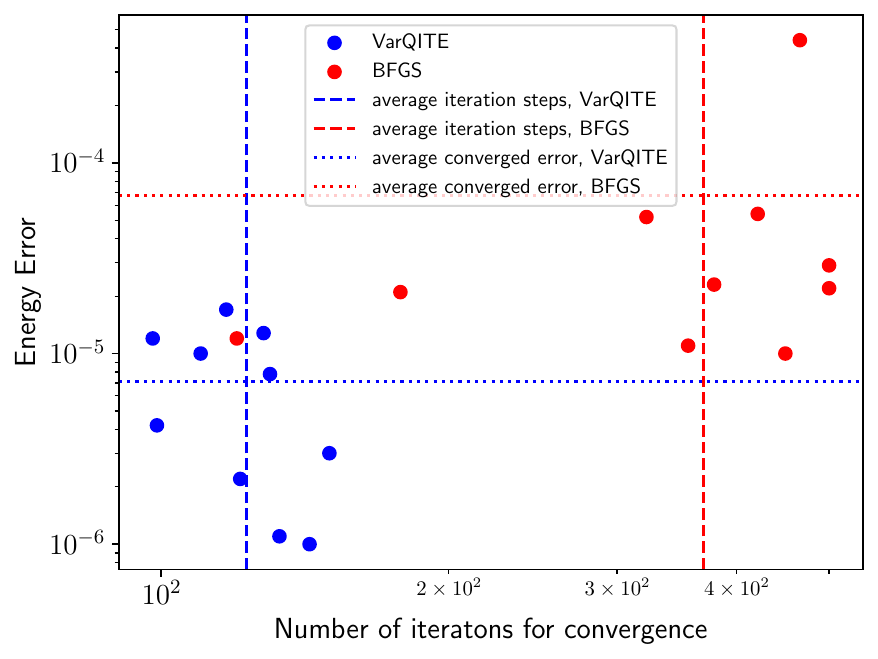}
    }
    \newline
   \subfigure[ $\rm H_4$ at 0.7 {\AA}, $D = 8$, $L=2$]{
        \centering
        \includegraphics[width = 0.49\textwidth]{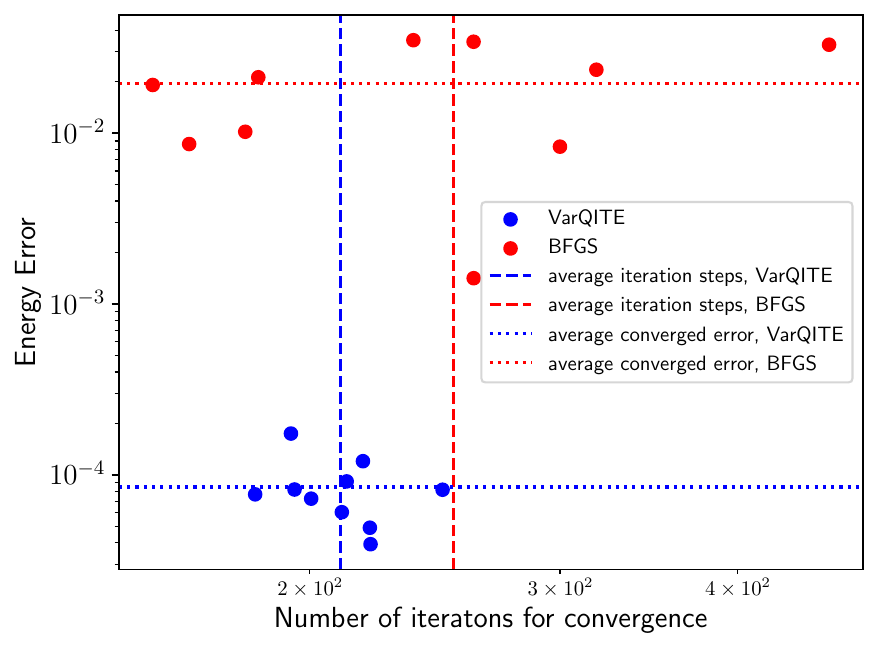}
   }
   \newline
   \subfigure[ $\rm H_4$ at 0.4 {\AA}, $D = 4$, $L=2$]{
        \centering
        \includegraphics[width = 0.49\textwidth]{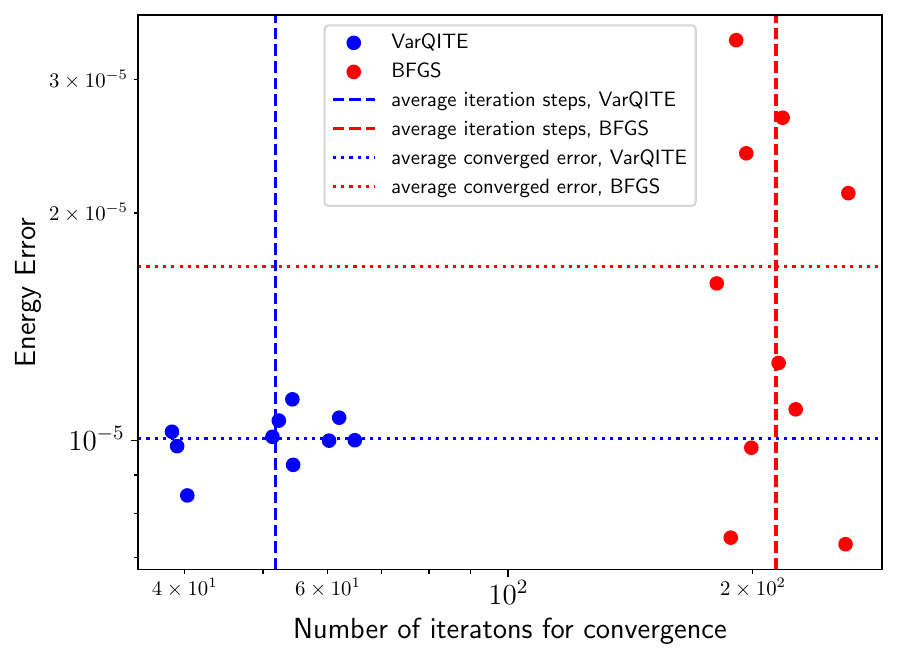}
   }
  
    \caption{  \label{fig:stat}Statistical results for the optimization performance of VarQITE and BFGS methods}
\end{figure}

We conducted simulations utilizing 10 distinct sets of initial parameters for the 4-NMO $\rm LiH$, $\rm H_4$ at 0.7 {\AA}, and $\rm H_4$ at 0.4 {\AA} systems, as depicted in Figures \ref{fig:lost_function} and \ref{fig:bdlength} of the manuscript. Our findings indicate that, across all three scenarios, the VarQITE method consistently surpasses the BFGS algorithm in terms of both energy precision and the number of iterations required for convergence, as shown in Figure \ref{fig:stat}.

\section{Convergence of the complex cost function in the numerical simulation of Be atom system}\label{app:C}

\begin{figure}[!h]
    \centering
    \includegraphics[height = 5.5cm]{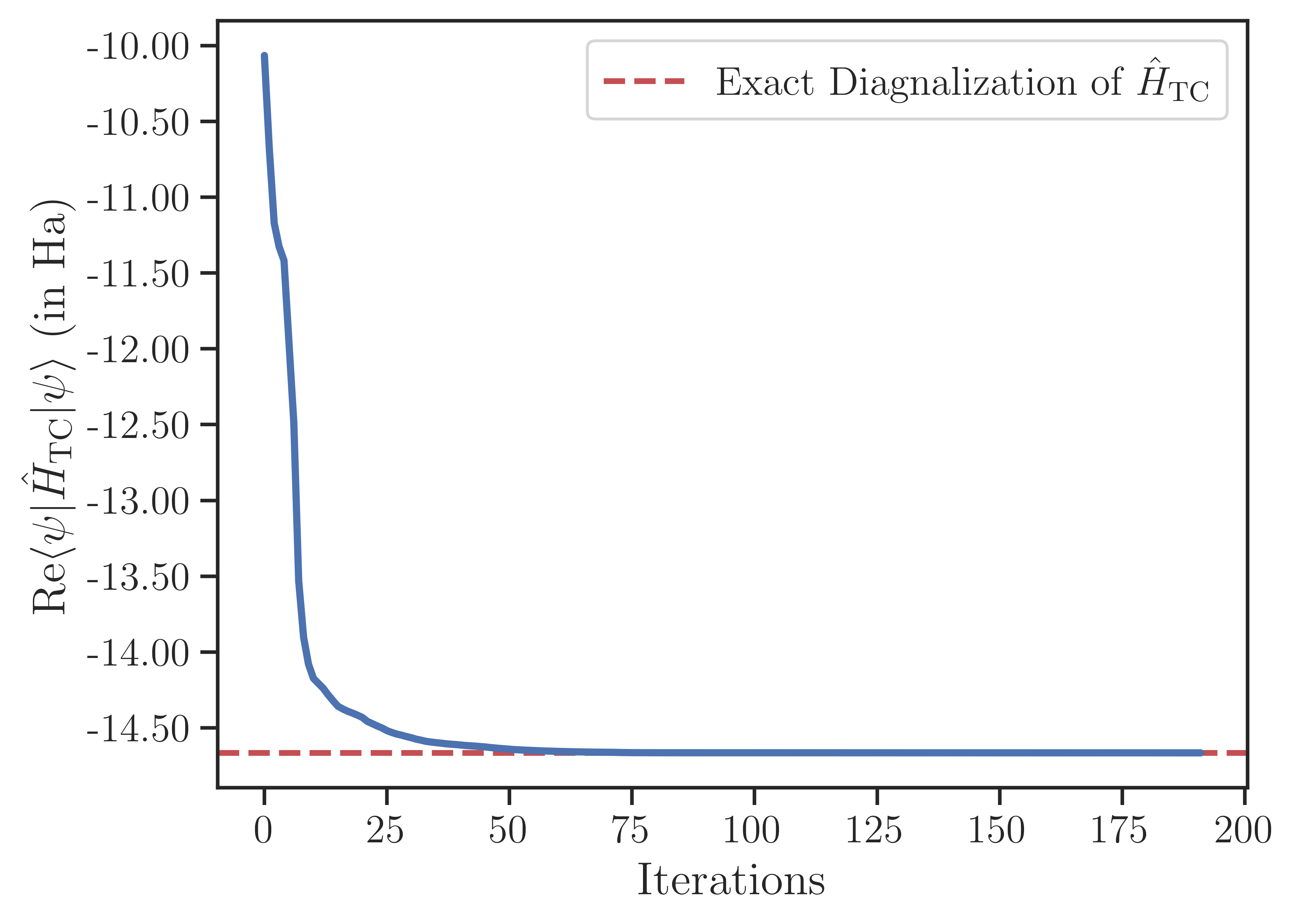}

    \includegraphics[height = 5.5cm]{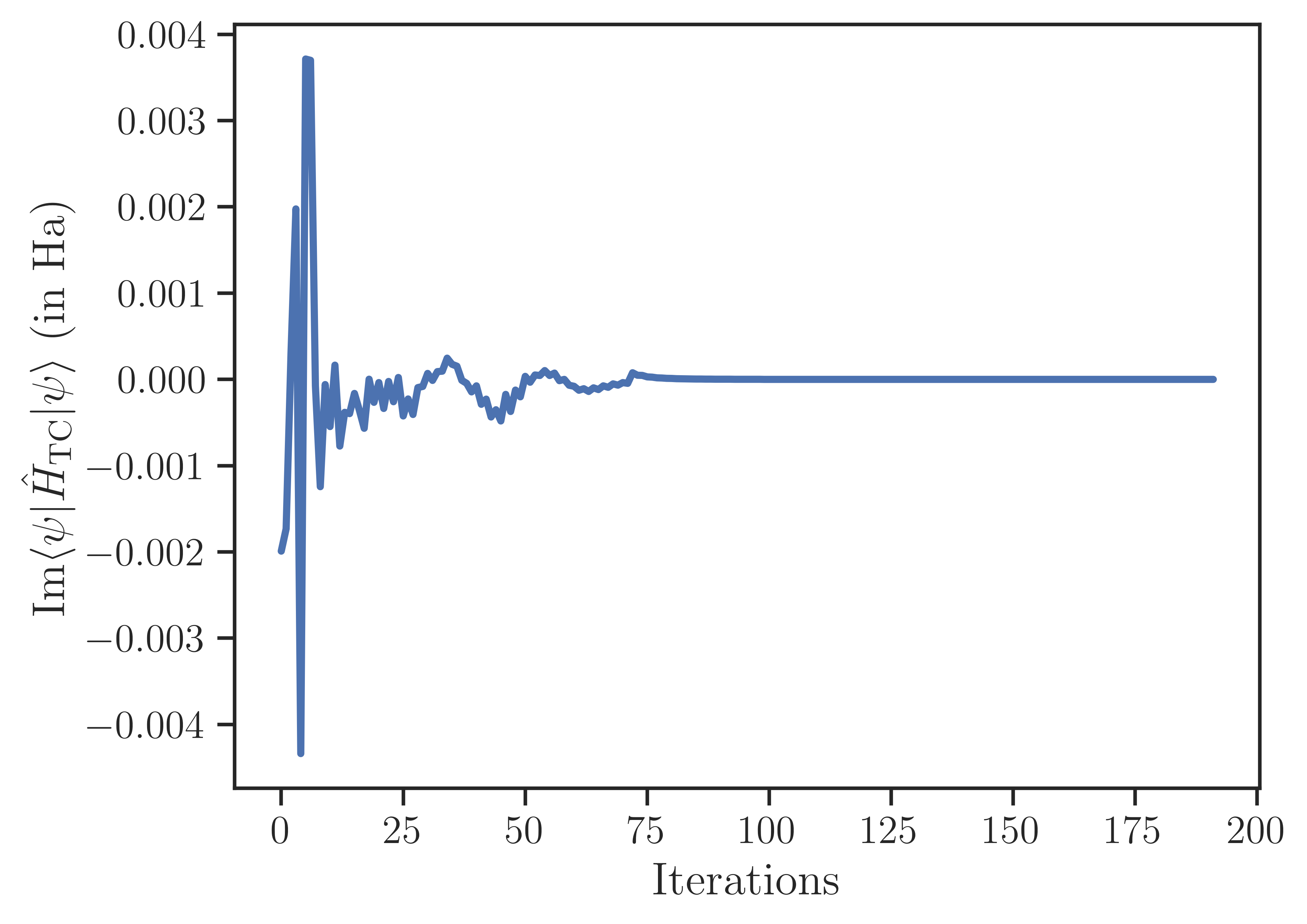}
    \caption{\label{fig:re_im}Evolution of the real (upper panel) and imaginary part (lower panel) of $\mel{\psi(\boldsymbol{\theta})}{\hat{H}_{\rm TC}}{\psi(\boldsymbol{\theta})}$ in the VarQITE optimization of the QCMPS ansatz, with $D=4$ ($3$ qubits) and $L=2$.}
\end{figure}

We employ VarQITE simulation with a fixed step size of $\delta=0.05$ for the beryllium atom system. The convergence of the cost function with iteration steps is shown in Figure \ref{fig:re_im}.
\nocite{*}
\section*{References}
\bibliography{aipsamp}

\providecommand{\noopsort}[1]{}\providecommand{\singleletter}[1]{#1}%
\begin{thebibliography}{82}%
\makeatletter
\providecommand \@ifxundefined [1]{%
 \@ifx{#1\undefined}
}%
\providecommand \@ifnum [1]{%
 \ifnum #1\expandafter \@firstoftwo
 \else \expandafter \@secondoftwo
 \fi
}%
\providecommand \@ifx [1]{%
 \ifx #1\expandafter \@firstoftwo
 \else \expandafter \@secondoftwo
 \fi
}%
\providecommand \natexlab [1]{#1}%
\providecommand \enquote  [1]{``#1''}%
\providecommand \bibnamefont  [1]{#1}%
\providecommand \bibfnamefont [1]{#1}%
\providecommand \citenamefont [1]{#1}%
\providecommand \href@noop [0]{\@secondoftwo}%
\providecommand \href [0]{\begingroup \@sanitize@url \@href}%
\providecommand \@href[1]{\@@startlink{#1}\@@href}%
\providecommand \@@href[1]{\endgroup#1\@@endlink}%
\providecommand \@sanitize@url [0]{\catcode `\\12\catcode `\$12\catcode `\&12\catcode `\#12\catcode `\^12\catcode `\_12\catcode `\%12\relax}%
\providecommand \@@startlink[1]{}%
\providecommand \@@endlink[0]{}%
\providecommand \url  [0]{\begingroup\@sanitize@url \@url }%
\providecommand \@url [1]{\endgroup\@href {#1}{\urlprefix }}%
\providecommand \urlprefix  [0]{URL }%
\providecommand \Eprint [0]{\href }%
\providecommand \doibase [0]{http://dx.doi.org/}%
\providecommand \selectlanguage [0]{\@gobble}%
\providecommand \bibinfo  [0]{\@secondoftwo}%
\providecommand \bibfield  [0]{\@secondoftwo}%
\providecommand \translation [1]{[#1]}%
\providecommand \BibitemOpen [0]{}%
\providecommand \bibitemStop [0]{}%
\providecommand \bibitemNoStop [0]{.\EOS\space}%
\providecommand \EOS [0]{\spacefactor3000\relax}%
\providecommand \BibitemShut  [1]{\csname bibitem#1\endcsname}%
\let\auto@bib@innerbib\@empty
\bibitem [{\citenamefont {Eriksen}(2020)}]{fullci2020}%
  \BibitemOpen
  \bibfield  {author} {\bibinfo {author} {\bibfnamefont {J.~J.}\ \bibnamefont {Eriksen}},\ }\href {https://doi/abs/10.1021/acs.jpclett.0c03225} {\bibfield  {journal} {\bibinfo  {journal} {J. Phys. Chem. Lett.}\ }\textbf {\bibinfo {volume} {12}},\ \bibinfo {pages} {418--432} (\bibinfo {year} {2020})}\BibitemShut {NoStop}%
\bibitem [{\citenamefont {Gao}\ \emph {et~al.}(2024)\citenamefont {Gao}, \citenamefont {Imamura}, \citenamefont {Kasagi},\ and\ \citenamefont {Yoshida}}]{fullci2024}%
  \BibitemOpen
  \bibfield  {author} {\bibinfo {author} {\bibfnamefont {H.}~\bibnamefont {Gao}}, \bibinfo {author} {\bibfnamefont {S.}~\bibnamefont {Imamura}}, \bibinfo {author} {\bibfnamefont {A.}~\bibnamefont {Kasagi}}, \ and\ \bibinfo {author} {\bibfnamefont {E.}~\bibnamefont {Yoshida}},\ }\href {https://doi.org/10.1021/acs.jctc.3c01190} {\bibfield  {journal} {\bibinfo  {journal} {J. Chem. Theory Comput.}\ }\textbf {\bibinfo {volume} {20}},\ \bibinfo {pages} {1185--1192} (\bibinfo {year} {2024})}\BibitemShut {NoStop}%
\bibitem [{\citenamefont {Kong}, \citenamefont {Bischoff},\ and\ \citenamefont {Valeev}(2012)}]{explicit2012}%
  \BibitemOpen
  \bibfield  {author} {\bibinfo {author} {\bibfnamefont {L.}~\bibnamefont {Kong}}, \bibinfo {author} {\bibfnamefont {F.~A.}\ \bibnamefont {Bischoff}}, \ and\ \bibinfo {author} {\bibfnamefont {E.~F.}\ \bibnamefont {Valeev}},\ }\href {https://doi.org/10.1021/cr200204r} {\bibfield  {journal} {\bibinfo  {journal} {Chem. Rev.}\ }\textbf {\bibinfo {volume} {112}},\ \bibinfo {pages} {75--107} (\bibinfo {year} {2012})}\BibitemShut {NoStop}%
\bibitem [{\citenamefont {Gr{\"u}neis}\ \emph {et~al.}(2017)\citenamefont {Gr{\"u}neis}, \citenamefont {Hirata}, \citenamefont {Ohnishi},\ and\ \citenamefont {Ten-No}}]{explicit2017}%
  \BibitemOpen
  \bibfield  {author} {\bibinfo {author} {\bibfnamefont {A.}~\bibnamefont {Gr{\"u}neis}}, \bibinfo {author} {\bibfnamefont {S.}~\bibnamefont {Hirata}}, \bibinfo {author} {\bibfnamefont {Y.-y.}\ \bibnamefont {Ohnishi}}, \ and\ \bibinfo {author} {\bibfnamefont {S.}~\bibnamefont {Ten-No}},\ }\href {https://doi.org/10.1063/1.4976974} {\bibfield  {journal} {\bibinfo  {journal} {J. Chem. Phys.}\ }\textbf {\bibinfo {volume} {146}},\ \bibinfo {pages} {080901} (\bibinfo {year} {2017})}\BibitemShut {NoStop}%
\bibitem [{\citenamefont {Luo}(2010)}]{tc2010}%
  \BibitemOpen
  \bibfield  {author} {\bibinfo {author} {\bibfnamefont {H.}~\bibnamefont {Luo}},\ }\href {https://doi.org/10.1063/1.3505037} {\bibfield  {journal} {\bibinfo  {journal} {J. Chem. Phys.}\ }\textbf {\bibinfo {volume} {133}},\ \bibinfo {pages} {154109} (\bibinfo {year} {2010})}\BibitemShut {NoStop}%
\bibitem [{\citenamefont {Luo}(2011)}]{tc2011}%
  \BibitemOpen
  \bibfield  {author} {\bibinfo {author} {\bibfnamefont {H.}~\bibnamefont {Luo}},\ }\href {https://doi.org/10.1063/1.3607990} {\bibfield  {journal} {\bibinfo  {journal} {J. Chem. Phys.}\ }\textbf {\bibinfo {volume} {135}},\ \bibinfo {pages} {024109} (\bibinfo {year} {2011})}\BibitemShut {NoStop}%
\bibitem [{\citenamefont {Yanai}\ and\ \citenamefont {Shiozaki}(2012)}]{tc2012}%
  \BibitemOpen
  \bibfield  {author} {\bibinfo {author} {\bibfnamefont {T.}~\bibnamefont {Yanai}}\ and\ \bibinfo {author} {\bibfnamefont {T.}~\bibnamefont {Shiozaki}},\ }\href {https://doi.org/10.1063/1.3688225} {\bibfield  {journal} {\bibinfo  {journal} {J. Chem. Phys.}\ }\textbf {\bibinfo {volume} {136}},\ \bibinfo {pages} {084107} (\bibinfo {year} {2012})}\BibitemShut {NoStop}%
\bibitem [{\citenamefont {White}(1992)}]{DMRG1992}%
  \BibitemOpen
  \bibfield  {author} {\bibinfo {author} {\bibfnamefont {S.~R.}\ \bibnamefont {White}},\ }\href {\doibase 10.1103/PhysRevLett.69.2863} {\bibfield  {journal} {\bibinfo  {journal} {Phys. Rev. Lett.}\ }\textbf {\bibinfo {volume} {69}},\ \bibinfo {pages} {2863--2866} (\bibinfo {year} {1992})}\BibitemShut {NoStop}%
\bibitem [{\citenamefont {Schollwöck}(2011)}]{DMRG2011}%
  \BibitemOpen
  \bibfield  {author} {\bibinfo {author} {\bibfnamefont {U.}~\bibnamefont {Schollwöck}},\ }\href {https://www.sciencedirect.com/science/article/pii/S0003491610001752} {\bibfield  {journal} {\bibinfo  {journal} {Ann. Phys.}\ }\textbf {\bibinfo {volume} {326}},\ \bibinfo {pages} {96--192} (\bibinfo {year} {2011})}\BibitemShut {NoStop}%
\bibitem [{\citenamefont {Baiardi}\ and\ \citenamefont {Reiher}(2020)}]{DMRG2020}%
  \BibitemOpen
  \bibfield  {author} {\bibinfo {author} {\bibfnamefont {A.}~\bibnamefont {Baiardi}}\ and\ \bibinfo {author} {\bibfnamefont {M.}~\bibnamefont {Reiher}},\ }\href {https://doi.org/10.1063/1.5129672} {\bibfield  {journal} {\bibinfo  {journal} {J. Chem. Phys.}\ }\textbf {\bibinfo {volume} {152}},\ \bibinfo {pages} {040903} (\bibinfo {year} {2020})}\BibitemShut {NoStop}%
\bibitem [{\citenamefont {Li}, \citenamefont {Ren},\ and\ \citenamefont {Shuai}(2020)}]{TDDMRG2020}%
  \BibitemOpen
  \bibfield  {author} {\bibinfo {author} {\bibfnamefont {W.}~\bibnamefont {Li}}, \bibinfo {author} {\bibfnamefont {J.}~\bibnamefont {Ren}}, \ and\ \bibinfo {author} {\bibfnamefont {Z.}~\bibnamefont {Shuai}},\ }\href {https://doi.org/10.1021/acs.jpclett.0c01072} {\bibfield  {journal} {\bibinfo  {journal} {J. Phys. Chem. Lett.}\ }\textbf {\bibinfo {volume} {11}},\ \bibinfo {pages} {4930--4936} (\bibinfo {year} {2020})}\BibitemShut {NoStop}%
\bibitem [{\citenamefont {Wang}, \citenamefont {Ren},\ and\ \citenamefont {Shuai}(2021)}]{TDDMRG2021}%
  \BibitemOpen
  \bibfield  {author} {\bibinfo {author} {\bibfnamefont {Y.}~\bibnamefont {Wang}}, \bibinfo {author} {\bibfnamefont {J.}~\bibnamefont {Ren}}, \ and\ \bibinfo {author} {\bibfnamefont {Z.}~\bibnamefont {Shuai}},\ }\href {https://doi.org/10.1063/5.0052804} {\bibfield  {journal} {\bibinfo  {journal} {J. Chem. Phys.}\ }\textbf {\bibinfo {volume} {154}},\ \bibinfo {pages} {214109} (\bibinfo {year} {2021})}\BibitemShut {NoStop}%
\bibitem [{\citenamefont {Nakatani}\ and\ \citenamefont {Chan}(2013)}]{tensor2013}%
  \BibitemOpen
  \bibfield  {author} {\bibinfo {author} {\bibfnamefont {N.}~\bibnamefont {Nakatani}}\ and\ \bibinfo {author} {\bibfnamefont {G.~K.-L.}\ \bibnamefont {Chan}},\ }\href {https://doi.org/10.1063/1.4798639} {\bibfield  {journal} {\bibinfo  {journal} {J. Chem. Phys.}\ }\textbf {\bibinfo {volume} {138}},\ \bibinfo {pages} {134113} (\bibinfo {year} {2013})}\BibitemShut {NoStop}%
\bibitem [{\citenamefont {Murg}\ \emph {et~al.}(2015)\citenamefont {Murg}, \citenamefont {Verstraete}, \citenamefont {Schneider}, \citenamefont {Nagy},\ and\ \citenamefont {Legeza}}]{tensor2015}%
  \BibitemOpen
  \bibfield  {author} {\bibinfo {author} {\bibfnamefont {V.}~\bibnamefont {Murg}}, \bibinfo {author} {\bibfnamefont {F.}~\bibnamefont {Verstraete}}, \bibinfo {author} {\bibfnamefont {R.}~\bibnamefont {Schneider}}, \bibinfo {author} {\bibfnamefont {P.~R.}\ \bibnamefont {Nagy}}, \ and\ \bibinfo {author} {\bibfnamefont {O.}~\bibnamefont {Legeza}},\ }\href {https://doi.org/10.1021/ct501187j} {\bibfield  {journal} {\bibinfo  {journal} {J. Chem. Theory Comput.}\ }\textbf {\bibinfo {volume} {11}},\ \bibinfo {pages} {1027--1036} (\bibinfo {year} {2015})}\BibitemShut {NoStop}%
\bibitem [{\citenamefont {Li}, \citenamefont {O'Rourke},\ and\ \citenamefont {Chan}(2019)}]{tensor2019}%
  \BibitemOpen
  \bibfield  {author} {\bibinfo {author} {\bibfnamefont {Z.}~\bibnamefont {Li}}, \bibinfo {author} {\bibfnamefont {M.~J.}\ \bibnamefont {O'Rourke}}, \ and\ \bibinfo {author} {\bibfnamefont {G.~K.-L.}\ \bibnamefont {Chan}},\ }\href {\doibase 10.1103/PhysRevB.100.155121} {\bibfield  {journal} {\bibinfo  {journal} {Phys. Rev. B}\ }\textbf {\bibinfo {volume} {100}},\ \bibinfo {pages} {155121} (\bibinfo {year} {2019})}\BibitemShut {NoStop}%
\bibitem [{\citenamefont {Li}(2021)}]{tensor2021}%
  \BibitemOpen
  \bibfield  {author} {\bibinfo {author} {\bibfnamefont {Z.}~\bibnamefont {Li}},\ }\href {\doibase 10.1088/2516-1075/abe192} {\bibfield  {journal} {\bibinfo  {journal} {Electronic Structure}\ }\textbf {\bibinfo {volume} {3}},\ \bibinfo {pages} {014001} (\bibinfo {year} {2021})}\BibitemShut {NoStop}%
\bibitem [{\citenamefont {Carleo}\ and\ \citenamefont {Troyer}(2017)}]{NQS2017}%
  \BibitemOpen
  \bibfield  {author} {\bibinfo {author} {\bibfnamefont {G.}~\bibnamefont {Carleo}}\ and\ \bibinfo {author} {\bibfnamefont {M.}~\bibnamefont {Troyer}},\ }\href {\doibase 10.1126/science.aag2302} {\bibfield  {journal} {\bibinfo  {journal} {Science}\ }\textbf {\bibinfo {volume} {355}},\ \bibinfo {pages} {602--606} (\bibinfo {year} {2017})}\BibitemShut {NoStop}%
\bibitem [{\citenamefont {Choo}, \citenamefont {Mezzacapo},\ and\ \citenamefont {Carleo}(2020)}]{NQS2020}%
  \BibitemOpen
  \bibfield  {author} {\bibinfo {author} {\bibfnamefont {K.}~\bibnamefont {Choo}}, \bibinfo {author} {\bibfnamefont {A.}~\bibnamefont {Mezzacapo}}, \ and\ \bibinfo {author} {\bibfnamefont {G.}~\bibnamefont {Carleo}},\ }\href {https://doi.org/10.1038/s41467-020-15724-9} {\bibfield  {journal} {\bibinfo  {journal} {Nat. Commun.}\ }\textbf {\bibinfo {volume} {11}},\ \bibinfo {pages} {2368} (\bibinfo {year} {2020})}\BibitemShut {NoStop}%
\bibitem [{\citenamefont {Hermann}\ \emph {et~al.}(2023)\citenamefont {Hermann}, \citenamefont {Spencer}, \citenamefont {Choo}, \citenamefont {Mezzacapo}, \citenamefont {Foulkes}, \citenamefont {Pfau}, \citenamefont {Carleo},\ and\ \citenamefont {No{\'e}}}]{NQS2023}%
  \BibitemOpen
  \bibfield  {author} {\bibinfo {author} {\bibfnamefont {J.}~\bibnamefont {Hermann}}, \bibinfo {author} {\bibfnamefont {J.}~\bibnamefont {Spencer}}, \bibinfo {author} {\bibfnamefont {K.}~\bibnamefont {Choo}}, \bibinfo {author} {\bibfnamefont {A.}~\bibnamefont {Mezzacapo}}, \bibinfo {author} {\bibfnamefont {W.~M.~C.}\ \bibnamefont {Foulkes}}, \bibinfo {author} {\bibfnamefont {D.}~\bibnamefont {Pfau}}, \bibinfo {author} {\bibfnamefont {G.}~\bibnamefont {Carleo}}, \ and\ \bibinfo {author} {\bibfnamefont {F.}~\bibnamefont {No{\'e}}},\ }\href {https://doi.org/10.1038/s41570-023-00516-8} {\bibfield  {journal} {\bibinfo  {journal} {Nat. Rev. Phys.}\ }\textbf {\bibinfo {volume} {7}},\ \bibinfo {pages} {692--709} (\bibinfo {year} {2023})}\BibitemShut {NoStop}%
\bibitem [{\citenamefont {Li}\ \emph {et~al.}(2023)\citenamefont {Li}, \citenamefont {Huang}, \citenamefont {Zhang}, \citenamefont {Li}, \citenamefont {Cao}, \citenamefont {Lv},\ and\ \citenamefont {Hu}}]{SCNQS2023}%
  \BibitemOpen
  \bibfield  {author} {\bibinfo {author} {\bibfnamefont {X.}~\bibnamefont {Li}}, \bibinfo {author} {\bibfnamefont {J.-C.}\ \bibnamefont {Huang}}, \bibinfo {author} {\bibfnamefont {G.-Z.}\ \bibnamefont {Zhang}}, \bibinfo {author} {\bibfnamefont {H.-E.}\ \bibnamefont {Li}}, \bibinfo {author} {\bibfnamefont {C.-S.}\ \bibnamefont {Cao}}, \bibinfo {author} {\bibfnamefont {D.}~\bibnamefont {Lv}}, \ and\ \bibinfo {author} {\bibfnamefont {H.-S.}\ \bibnamefont {Hu}},\ }\href {https://doi.org/10.1021/acs.jctc.3c00831} {\bibfield  {journal} {\bibinfo  {journal} {J. Chem. Theory Comput.}\ }\textbf {\bibinfo {volume} {19}},\ \bibinfo {pages} {8156--8165} (\bibinfo {year} {2023})}\BibitemShut {NoStop}%
\bibitem [{\citenamefont {Hartmann}\ and\ \citenamefont {Carleo}(2019)}]{DYNNQS2019}%
  \BibitemOpen
  \bibfield  {author} {\bibinfo {author} {\bibfnamefont {M.~J.}\ \bibnamefont {Hartmann}}\ and\ \bibinfo {author} {\bibfnamefont {G.}~\bibnamefont {Carleo}},\ }\href {\doibase 10.1103/PhysRevLett.122.250502} {\bibfield  {journal} {\bibinfo  {journal} {Phys. Rev. Lett.}\ }\textbf {\bibinfo {volume} {122}},\ \bibinfo {pages} {250502} (\bibinfo {year} {2019})}\BibitemShut {NoStop}%
\bibitem [{\citenamefont {McArdle}\ \emph {et~al.}(2020)\citenamefont {McArdle}, \citenamefont {Endo}, \citenamefont {Aspuru-Guzik}, \citenamefont {Benjamin},\ and\ \citenamefont {Yuan}}]{QuantumRev2020}%
  \BibitemOpen
  \bibfield  {author} {\bibinfo {author} {\bibfnamefont {S.}~\bibnamefont {McArdle}}, \bibinfo {author} {\bibfnamefont {S.}~\bibnamefont {Endo}}, \bibinfo {author} {\bibfnamefont {A.}~\bibnamefont {Aspuru-Guzik}}, \bibinfo {author} {\bibfnamefont {S.~C.}\ \bibnamefont {Benjamin}}, \ and\ \bibinfo {author} {\bibfnamefont {X.}~\bibnamefont {Yuan}},\ }\href {\doibase 10.1103/RevModPhys.92.015003} {\bibfield  {journal} {\bibinfo  {journal} {Rev. Mod. Phys.}\ }\textbf {\bibinfo {volume} {92}},\ \bibinfo {pages} {015003} (\bibinfo {year} {2020})}\BibitemShut {NoStop}%
\bibitem [{\citenamefont {Peruzzo}\ \emph {et~al.}(2014)\citenamefont {Peruzzo}, \citenamefont {McClean}, \citenamefont {Shadbolt}, \citenamefont {Yung}, \citenamefont {Zhou}, \citenamefont {Love}, \citenamefont {Aspuru-Guzik},\ and\ \citenamefont {O’brien}}]{vqe2014}%
  \BibitemOpen
  \bibfield  {author} {\bibinfo {author} {\bibfnamefont {A.}~\bibnamefont {Peruzzo}}, \bibinfo {author} {\bibfnamefont {J.}~\bibnamefont {McClean}}, \bibinfo {author} {\bibfnamefont {P.}~\bibnamefont {Shadbolt}}, \bibinfo {author} {\bibfnamefont {M.-H.}\ \bibnamefont {Yung}}, \bibinfo {author} {\bibfnamefont {X.-Q.}\ \bibnamefont {Zhou}}, \bibinfo {author} {\bibfnamefont {P.~J.}\ \bibnamefont {Love}}, \bibinfo {author} {\bibfnamefont {A.}~\bibnamefont {Aspuru-Guzik}}, \ and\ \bibinfo {author} {\bibfnamefont {J.~L.}\ \bibnamefont {O’brien}},\ }\href {https://doi.org/10.1038/ncomms5213} {\bibfield  {journal} {\bibinfo  {journal} {Nat. commun.}\ }\textbf {\bibinfo {volume} {5}},\ \bibinfo {pages} {4213} (\bibinfo {year} {2014})}\BibitemShut {NoStop}%
\bibitem [{\citenamefont {Kandala}\ \emph {et~al.}(2017)\citenamefont {Kandala}, \citenamefont {Mezzacapo}, \citenamefont {Temme}, \citenamefont {Takita}, \citenamefont {Brink}, \citenamefont {Chow},\ and\ \citenamefont {Gambetta}}]{HEA2017}%
  \BibitemOpen
  \bibfield  {author} {\bibinfo {author} {\bibfnamefont {A.}~\bibnamefont {Kandala}}, \bibinfo {author} {\bibfnamefont {A.}~\bibnamefont {Mezzacapo}}, \bibinfo {author} {\bibfnamefont {K.}~\bibnamefont {Temme}}, \bibinfo {author} {\bibfnamefont {M.}~\bibnamefont {Takita}}, \bibinfo {author} {\bibfnamefont {M.}~\bibnamefont {Brink}}, \bibinfo {author} {\bibfnamefont {J.~M.}\ \bibnamefont {Chow}}, \ and\ \bibinfo {author} {\bibfnamefont {J.~M.}\ \bibnamefont {Gambetta}},\ }\href {https://doi.org/10.1038/nature23879} {\bibfield  {journal} {\bibinfo  {journal} {Nature}\ }\textbf {\bibinfo {volume} {549}},\ \bibinfo {pages} {242--246} (\bibinfo {year} {2017})}\BibitemShut {NoStop}%
\bibitem [{\citenamefont {Cerezo}\ \emph {et~al.}(2021)\citenamefont {Cerezo}, \citenamefont {Arrasmith}, \citenamefont {Babbush}, \citenamefont {Benjamin}, \citenamefont {Endo}, \citenamefont {Fujii}, \citenamefont {McClean}, \citenamefont {Mitarai}, \citenamefont {Yuan}, \citenamefont {Cincio} \emph {et~al.}}]{VQA2021}%
  \BibitemOpen
  \bibfield  {author} {\bibinfo {author} {\bibfnamefont {M.}~\bibnamefont {Cerezo}}, \bibinfo {author} {\bibfnamefont {A.}~\bibnamefont {Arrasmith}}, \bibinfo {author} {\bibfnamefont {R.}~\bibnamefont {Babbush}}, \bibinfo {author} {\bibfnamefont {S.~C.}\ \bibnamefont {Benjamin}}, \bibinfo {author} {\bibfnamefont {S.}~\bibnamefont {Endo}}, \bibinfo {author} {\bibfnamefont {K.}~\bibnamefont {Fujii}}, \bibinfo {author} {\bibfnamefont {J.~R.}\ \bibnamefont {McClean}}, \bibinfo {author} {\bibfnamefont {K.}~\bibnamefont {Mitarai}}, \bibinfo {author} {\bibfnamefont {X.}~\bibnamefont {Yuan}}, \bibinfo {author} {\bibfnamefont {L.}~\bibnamefont {Cincio}},  \emph {et~al.},\ }\href {https://doi.org/10.1038/s42254-021-00348-9} {\bibfield  {journal} {\bibinfo  {journal} {Nat. Rev. Phys.}\ }\textbf {\bibinfo {volume} {3}},\ \bibinfo {pages} {625--644} (\bibinfo {year} {2021})}\BibitemShut {NoStop}%
\bibitem [{\citenamefont {Farhi}, \citenamefont {Goldstone},\ and\ \citenamefont {Gutmann}(2014)}]{QAOA}%
  \BibitemOpen
  \bibfield  {author} {\bibinfo {author} {\bibfnamefont {E.}~\bibnamefont {Farhi}}, \bibinfo {author} {\bibfnamefont {J.}~\bibnamefont {Goldstone}}, \ and\ \bibinfo {author} {\bibfnamefont {S.}~\bibnamefont {Gutmann}},\ }\href {https://doi.org/10.48550/arXiv.1411.4028} {\bibfield  {journal} {\bibinfo  {journal} {arXiv:1411.4028}\ } (\bibinfo {year} {2014})}\BibitemShut {NoStop}%
\bibitem [{\citenamefont {Stair}, \citenamefont {Huang},\ and\ \citenamefont {Evangelista}(2020)}]{mrsqk2020}%
  \BibitemOpen
  \bibfield  {author} {\bibinfo {author} {\bibfnamefont {N.~H.}\ \bibnamefont {Stair}}, \bibinfo {author} {\bibfnamefont {R.}~\bibnamefont {Huang}}, \ and\ \bibinfo {author} {\bibfnamefont {F.~A.}\ \bibnamefont {Evangelista}},\ }\href {\doibase 10.1021/acs.jctc.9b01125} {\bibfield  {journal} {\bibinfo  {journal} {J. Chem. Theory Comput.}\ }\textbf {\bibinfo {volume} {16}},\ \bibinfo {pages} {2236--2245} (\bibinfo {year} {2020})}\BibitemShut {NoStop}%
\bibitem [{\citenamefont {Lubasch}\ \emph {et~al.}(2020)\citenamefont {Lubasch}, \citenamefont {Joo}, \citenamefont {Moinier}, \citenamefont {Kiffner},\ and\ \citenamefont {Jaksch}}]{nonlinear2020}%
  \BibitemOpen
  \bibfield  {author} {\bibinfo {author} {\bibfnamefont {M.}~\bibnamefont {Lubasch}}, \bibinfo {author} {\bibfnamefont {J.}~\bibnamefont {Joo}}, \bibinfo {author} {\bibfnamefont {P.}~\bibnamefont {Moinier}}, \bibinfo {author} {\bibfnamefont {M.}~\bibnamefont {Kiffner}}, \ and\ \bibinfo {author} {\bibfnamefont {D.}~\bibnamefont {Jaksch}},\ }\href {\doibase 10.1103/PhysRevA.101.010301} {\bibfield  {journal} {\bibinfo  {journal} {Phys. Rev. A}\ }\textbf {\bibinfo {volume} {101}},\ \bibinfo {pages} {010301} (\bibinfo {year} {2020})}\BibitemShut {NoStop}%
\bibitem [{\citenamefont {Shen}\ \emph {et~al.}(2017)\citenamefont {Shen}, \citenamefont {Zhang}, \citenamefont {Zhang}, \citenamefont {Zhang}, \citenamefont {Yung},\ and\ \citenamefont {Kim}}]{ucc2017}%
  \BibitemOpen
  \bibfield  {author} {\bibinfo {author} {\bibfnamefont {Y.}~\bibnamefont {Shen}}, \bibinfo {author} {\bibfnamefont {X.}~\bibnamefont {Zhang}}, \bibinfo {author} {\bibfnamefont {S.}~\bibnamefont {Zhang}}, \bibinfo {author} {\bibfnamefont {J.-N.}\ \bibnamefont {Zhang}}, \bibinfo {author} {\bibfnamefont {M.-H.}\ \bibnamefont {Yung}}, \ and\ \bibinfo {author} {\bibfnamefont {K.}~\bibnamefont {Kim}},\ }\href {\doibase 10.1103/PhysRevA.95.020501} {\bibfield  {journal} {\bibinfo  {journal} {Phys. Rev. A}\ }\textbf {\bibinfo {volume} {95}},\ \bibinfo {pages} {020501} (\bibinfo {year} {2017})}\BibitemShut {NoStop}%
\bibitem [{\citenamefont {Romero}\ \emph {et~al.}(2018)\citenamefont {Romero}, \citenamefont {Babbush}, \citenamefont {McClean}, \citenamefont {Hempel}, \citenamefont {Love},\ and\ \citenamefont {Aspuru-Guzik}}]{ucc2018}%
  \BibitemOpen
  \bibfield  {author} {\bibinfo {author} {\bibfnamefont {J.}~\bibnamefont {Romero}}, \bibinfo {author} {\bibfnamefont {R.}~\bibnamefont {Babbush}}, \bibinfo {author} {\bibfnamefont {J.~R.}\ \bibnamefont {McClean}}, \bibinfo {author} {\bibfnamefont {C.}~\bibnamefont {Hempel}}, \bibinfo {author} {\bibfnamefont {P.~J.}\ \bibnamefont {Love}}, \ and\ \bibinfo {author} {\bibfnamefont {A.}~\bibnamefont {Aspuru-Guzik}},\ }\href {https://iopscience.iop.org/article/10.1088/2058-9565/aad3e4/meta} {\bibfield  {journal} {\bibinfo  {journal} {Quantum Sci. Technol.}\ }\textbf {\bibinfo {volume} {4}},\ \bibinfo {pages} {014008} (\bibinfo {year} {2018})}\BibitemShut {NoStop}%
\bibitem [{\citenamefont {Anand}\ \emph {et~al.}(2022)\citenamefont {Anand}, \citenamefont {Schleich}, \citenamefont {Alperin-Lea}, \citenamefont {Jensen}, \citenamefont {Sim}, \citenamefont {D{\'\i}az-Tinoco}, \citenamefont {Kottmann}, \citenamefont {Degroote}, \citenamefont {Izmaylov},\ and\ \citenamefont {Aspuru-Guzik}}]{ucc2022}%
  \BibitemOpen
  \bibfield  {author} {\bibinfo {author} {\bibfnamefont {A.}~\bibnamefont {Anand}}, \bibinfo {author} {\bibfnamefont {P.}~\bibnamefont {Schleich}}, \bibinfo {author} {\bibfnamefont {S.}~\bibnamefont {Alperin-Lea}}, \bibinfo {author} {\bibfnamefont {P.~W.}\ \bibnamefont {Jensen}}, \bibinfo {author} {\bibfnamefont {S.}~\bibnamefont {Sim}}, \bibinfo {author} {\bibfnamefont {M.}~\bibnamefont {D{\'\i}az-Tinoco}}, \bibinfo {author} {\bibfnamefont {J.~S.}\ \bibnamefont {Kottmann}}, \bibinfo {author} {\bibfnamefont {M.}~\bibnamefont {Degroote}}, \bibinfo {author} {\bibfnamefont {A.~F.}\ \bibnamefont {Izmaylov}}, \ and\ \bibinfo {author} {\bibfnamefont {A.}~\bibnamefont {Aspuru-Guzik}},\ }\href {https://doi.org/10.1039/D1CS00932J} {\bibfield  {journal} {\bibinfo  {journal} {Chem. Soc. Rev.}\ }\textbf {\bibinfo {volume} {51}},\ \bibinfo {pages} {1659--1684} (\bibinfo {year} {2022})}\BibitemShut {NoStop}%
\bibitem [{\citenamefont {Cao}\ \emph {et~al.}(2022)\citenamefont {Cao}, \citenamefont {Hu}, \citenamefont {Zhang}, \citenamefont {Xu}, \citenamefont {Chen}, \citenamefont {Yu}, \citenamefont {Li}, \citenamefont {Hu}, \citenamefont {Lv},\ and\ \citenamefont {Yung}}]{symUCCSD2022}%
  \BibitemOpen
  \bibfield  {author} {\bibinfo {author} {\bibfnamefont {C.}~\bibnamefont {Cao}}, \bibinfo {author} {\bibfnamefont {J.}~\bibnamefont {Hu}}, \bibinfo {author} {\bibfnamefont {W.}~\bibnamefont {Zhang}}, \bibinfo {author} {\bibfnamefont {X.}~\bibnamefont {Xu}}, \bibinfo {author} {\bibfnamefont {D.}~\bibnamefont {Chen}}, \bibinfo {author} {\bibfnamefont {F.}~\bibnamefont {Yu}}, \bibinfo {author} {\bibfnamefont {J.}~\bibnamefont {Li}}, \bibinfo {author} {\bibfnamefont {H.-S.}\ \bibnamefont {Hu}}, \bibinfo {author} {\bibfnamefont {D.}~\bibnamefont {Lv}}, \ and\ \bibinfo {author} {\bibfnamefont {M.-H.}\ \bibnamefont {Yung}},\ }\href {\doibase 10.1103/PhysRevA.105.062452} {\bibfield  {journal} {\bibinfo  {journal} {Phys. Rev. A}\ }\textbf {\bibinfo {volume} {105}},\ \bibinfo {pages} {062452} (\bibinfo {year} {2022})}\BibitemShut {NoStop}%
\bibitem [{\citenamefont {Grimsley}\ \emph {et~al.}(2019)\citenamefont {Grimsley}, \citenamefont {Economou}, \citenamefont {Barnes},\ and\ \citenamefont {Mayhall}}]{ADAPT}%
  \BibitemOpen
  \bibfield  {author} {\bibinfo {author} {\bibfnamefont {H.~R.}\ \bibnamefont {Grimsley}}, \bibinfo {author} {\bibfnamefont {S.~E.}\ \bibnamefont {Economou}}, \bibinfo {author} {\bibfnamefont {E.}~\bibnamefont {Barnes}}, \ and\ \bibinfo {author} {\bibfnamefont {N.~J.}\ \bibnamefont {Mayhall}},\ }\href {https://doi.org/10.1038/s41467-019-10988-2} {\bibfield  {journal} {\bibinfo  {journal} {Nat. Commun.}\ }\textbf {\bibinfo {volume} {10}},\ \bibinfo {pages} {3007} (\bibinfo {year} {2019})}\BibitemShut {NoStop}%
\bibitem [{\citenamefont {Tang}\ \emph {et~al.}(2021{\natexlab{a}})\citenamefont {Tang}, \citenamefont {Shkolnikov}, \citenamefont {Barron}, \citenamefont {Grimsley}, \citenamefont {Mayhall}, \citenamefont {Barnes},\ and\ \citenamefont {Economou}}]{qbitADAPT}%
  \BibitemOpen
  \bibfield  {author} {\bibinfo {author} {\bibfnamefont {H.~L.}\ \bibnamefont {Tang}}, \bibinfo {author} {\bibfnamefont {V.}~\bibnamefont {Shkolnikov}}, \bibinfo {author} {\bibfnamefont {G.~S.}\ \bibnamefont {Barron}}, \bibinfo {author} {\bibfnamefont {H.~R.}\ \bibnamefont {Grimsley}}, \bibinfo {author} {\bibfnamefont {N.~J.}\ \bibnamefont {Mayhall}}, \bibinfo {author} {\bibfnamefont {E.}~\bibnamefont {Barnes}}, \ and\ \bibinfo {author} {\bibfnamefont {S.~E.}\ \bibnamefont {Economou}},\ }\href {\doibase 10.1103/PRXQuantum.2.020310} {\bibfield  {journal} {\bibinfo  {journal} {PRX Quantum}\ }\textbf {\bibinfo {volume} {2}},\ \bibinfo {pages} {020310} (\bibinfo {year} {2021}{\natexlab{a}})}\BibitemShut {NoStop}%
\bibitem [{\citenamefont {Motta}\ \emph {et~al.}(2023)\citenamefont {Motta}, \citenamefont {Sung}, \citenamefont {Whaley}, \citenamefont {Head-Gordon},\ and\ \citenamefont {Shee}}]{lucj2023}%
  \BibitemOpen
  \bibfield  {author} {\bibinfo {author} {\bibfnamefont {M.}~\bibnamefont {Motta}}, \bibinfo {author} {\bibfnamefont {K.~J.}\ \bibnamefont {Sung}}, \bibinfo {author} {\bibfnamefont {K.~B.}\ \bibnamefont {Whaley}}, \bibinfo {author} {\bibfnamefont {M.}~\bibnamefont {Head-Gordon}}, \ and\ \bibinfo {author} {\bibfnamefont {J.}~\bibnamefont {Shee}},\ }\href {https://doi.org/10.1039/D3SC02516K} {\bibfield  {journal} {\bibinfo  {journal} {Chem. Sci.}\ }\textbf {\bibinfo {volume} {14}},\ \bibinfo {pages} {11213--11227} (\bibinfo {year} {2023})}\BibitemShut {NoStop}%
\bibitem [{\citenamefont {Xiao}\ \emph {et~al.}(2024)\citenamefont {Xiao}, \citenamefont {Zhao}, \citenamefont {Ren}, \citenamefont {Fang},\ and\ \citenamefont {Li}}]{PCHEA2024}%
  \BibitemOpen
  \bibfield  {author} {\bibinfo {author} {\bibfnamefont {X.}~\bibnamefont {Xiao}}, \bibinfo {author} {\bibfnamefont {H.}~\bibnamefont {Zhao}}, \bibinfo {author} {\bibfnamefont {J.}~\bibnamefont {Ren}}, \bibinfo {author} {\bibfnamefont {W.-H.}\ \bibnamefont {Fang}}, \ and\ \bibinfo {author} {\bibfnamefont {Z.}~\bibnamefont {Li}},\ }\href {\doibase 10.1021/acs.jctc.3c00966} {\bibfield  {journal} {\bibinfo  {journal} {J. Chem. Theory Comput.}\ }\textbf {\bibinfo {volume} {20}},\ \bibinfo {pages} {1912--1922} (\bibinfo {year} {2024})}\BibitemShut {NoStop}%
\bibitem [{\citenamefont {Liu}\ \emph {et~al.}(2019)\citenamefont {Liu}, \citenamefont {Zhang}, \citenamefont {Wan},\ and\ \citenamefont {Wang}}]{QCMPS2019}%
  \BibitemOpen
  \bibfield  {author} {\bibinfo {author} {\bibfnamefont {J.-G.}\ \bibnamefont {Liu}}, \bibinfo {author} {\bibfnamefont {Y.-H.}\ \bibnamefont {Zhang}}, \bibinfo {author} {\bibfnamefont {Y.}~\bibnamefont {Wan}}, \ and\ \bibinfo {author} {\bibfnamefont {L.}~\bibnamefont {Wang}},\ }\href {\doibase 10.1103/PhysRevResearch.1.023025} {\bibfield  {journal} {\bibinfo  {journal} {Phys. Rev. Res.}\ }\textbf {\bibinfo {volume} {1}},\ \bibinfo {pages} {023025} (\bibinfo {year} {2019})}\BibitemShut {NoStop}%
\bibitem [{\citenamefont {Ran}(2020)}]{QCMPSSU22020}%
  \BibitemOpen
  \bibfield  {author} {\bibinfo {author} {\bibfnamefont {S.-J.}\ \bibnamefont {Ran}},\ }\href {\doibase 10.1103/PhysRevA.101.032310} {\bibfield  {journal} {\bibinfo  {journal} {Phys. Rev. A}\ }\textbf {\bibinfo {volume} {101}},\ \bibinfo {pages} {032310} (\bibinfo {year} {2020})}\BibitemShut {NoStop}%
\bibitem [{\citenamefont {Foss-Feig}\ \emph {et~al.}(2021)\citenamefont {Foss-Feig}, \citenamefont {Hayes}, \citenamefont {Dreiling}, \citenamefont {Figgatt}, \citenamefont {Gaebler}, \citenamefont {Moses}, \citenamefont {Pino},\ and\ \citenamefont {Potter}}]{QCMPS2021}%
  \BibitemOpen
  \bibfield  {author} {\bibinfo {author} {\bibfnamefont {M.}~\bibnamefont {Foss-Feig}}, \bibinfo {author} {\bibfnamefont {D.}~\bibnamefont {Hayes}}, \bibinfo {author} {\bibfnamefont {J.~M.}\ \bibnamefont {Dreiling}}, \bibinfo {author} {\bibfnamefont {C.}~\bibnamefont {Figgatt}}, \bibinfo {author} {\bibfnamefont {J.~P.}\ \bibnamefont {Gaebler}}, \bibinfo {author} {\bibfnamefont {S.~A.}\ \bibnamefont {Moses}}, \bibinfo {author} {\bibfnamefont {J.~M.}\ \bibnamefont {Pino}}, \ and\ \bibinfo {author} {\bibfnamefont {A.~C.}\ \bibnamefont {Potter}},\ }\href {\doibase 10.1103/PhysRevResearch.3.033002} {\bibfield  {journal} {\bibinfo  {journal} {Phys. Rev. Res.}\ }\textbf {\bibinfo {volume} {3}},\ \bibinfo {pages} {033002} (\bibinfo {year} {2021})}\BibitemShut {NoStop}%
\bibitem [{\citenamefont {Lee}\ \emph {et~al.}(2021)\citenamefont {Lee}, \citenamefont {Patil}, \citenamefont {Zhang},\ and\ \citenamefont {Hsieh}}]{QRBM2021}%
  \BibitemOpen
  \bibfield  {author} {\bibinfo {author} {\bibfnamefont {C.~K.}\ \bibnamefont {Lee}}, \bibinfo {author} {\bibfnamefont {P.}~\bibnamefont {Patil}}, \bibinfo {author} {\bibfnamefont {S.}~\bibnamefont {Zhang}}, \ and\ \bibinfo {author} {\bibfnamefont {C.~Y.}\ \bibnamefont {Hsieh}},\ }\href {\doibase 10.1103/PhysRevResearch.3.023095} {\bibfield  {journal} {\bibinfo  {journal} {Phys. Rev. Res.}\ }\textbf {\bibinfo {volume} {3}},\ \bibinfo {pages} {023095} (\bibinfo {year} {2021})}\BibitemShut {NoStop}%
\bibitem [{\citenamefont {Haghshenas}\ \emph {et~al.}(2022)\citenamefont {Haghshenas}, \citenamefont {Gray}, \citenamefont {Potter},\ and\ \citenamefont {Chan}}]{QCMPS2022}%
  \BibitemOpen
  \bibfield  {author} {\bibinfo {author} {\bibfnamefont {R.}~\bibnamefont {Haghshenas}}, \bibinfo {author} {\bibfnamefont {J.}~\bibnamefont {Gray}}, \bibinfo {author} {\bibfnamefont {A.~C.}\ \bibnamefont {Potter}}, \ and\ \bibinfo {author} {\bibfnamefont {G.~K.-L.}\ \bibnamefont {Chan}},\ }\href {\doibase 10.1103/PhysRevX.12.011047} {\bibfield  {journal} {\bibinfo  {journal} {Phys. Rev. X}\ }\textbf {\bibinfo {volume} {12}},\ \bibinfo {pages} {011047} (\bibinfo {year} {2022})}\BibitemShut {NoStop}%
\bibitem [{\citenamefont {Anand}\ \emph {et~al.}(2023)\citenamefont {Anand}, \citenamefont {Hauschild}, \citenamefont {Zhang}, \citenamefont {Potter},\ and\ \citenamefont {Zaletel}}]{QCMPS2023}%
  \BibitemOpen
  \bibfield  {author} {\bibinfo {author} {\bibfnamefont {S.}~\bibnamefont {Anand}}, \bibinfo {author} {\bibfnamefont {J.}~\bibnamefont {Hauschild}}, \bibinfo {author} {\bibfnamefont {Y.}~\bibnamefont {Zhang}}, \bibinfo {author} {\bibfnamefont {A.~C.}\ \bibnamefont {Potter}}, \ and\ \bibinfo {author} {\bibfnamefont {M.~P.}\ \bibnamefont {Zaletel}},\ }\href {\doibase 10.1103/PRXQuantum.4.030334} {\bibfield  {journal} {\bibinfo  {journal} {PRX Quantum}\ }\textbf {\bibinfo {volume} {4}},\ \bibinfo {pages} {030334} (\bibinfo {year} {2023})}\BibitemShut {NoStop}%
\bibitem [{\citenamefont {Fan}\ \emph {et~al.}(2023)\citenamefont {Fan}, \citenamefont {Liu}, \citenamefont {Li},\ and\ \citenamefont {Yang}}]{QCMPSchem2023}%
  \BibitemOpen
  \bibfield  {author} {\bibinfo {author} {\bibfnamefont {Y.}~\bibnamefont {Fan}}, \bibinfo {author} {\bibfnamefont {J.}~\bibnamefont {Liu}}, \bibinfo {author} {\bibfnamefont {Z.}~\bibnamefont {Li}}, \ and\ \bibinfo {author} {\bibfnamefont {J.}~\bibnamefont {Yang}},\ }\href {\doibase 10.1021/acs.jctc.3c00068} {\bibfield  {journal} {\bibinfo  {journal} {J. Chem. Theory. Comput.}\ }\textbf {\bibinfo {volume} {19}},\ \bibinfo {pages} {5407--5417} (\bibinfo {year} {2023})}\BibitemShut {NoStop}%
\bibitem [{\citenamefont {Malz}\ \emph {et~al.}(2024)\citenamefont {Malz}, \citenamefont {Styliaris}, \citenamefont {Wei},\ and\ \citenamefont {Cirac}}]{QCMPS2024}%
  \BibitemOpen
  \bibfield  {author} {\bibinfo {author} {\bibfnamefont {D.}~\bibnamefont {Malz}}, \bibinfo {author} {\bibfnamefont {G.}~\bibnamefont {Styliaris}}, \bibinfo {author} {\bibfnamefont {Z.-Y.}\ \bibnamefont {Wei}}, \ and\ \bibinfo {author} {\bibfnamefont {J.~I.}\ \bibnamefont {Cirac}},\ }\href {\doibase 10.1103/PhysRevLett.132.040404} {\bibfield  {journal} {\bibinfo  {journal} {Phys. Rev. Lett.}\ }\textbf {\bibinfo {volume} {132}},\ \bibinfo {pages} {040404} (\bibinfo {year} {2024})}\BibitemShut {NoStop}%
\bibitem [{\citenamefont {Cervero~Mart{\'{i}}n}, \citenamefont {Plekhanov},\ and\ \citenamefont {Lubasch}(2023)}]{barren2023}%
  \BibitemOpen
  \bibfield  {author} {\bibinfo {author} {\bibfnamefont {E.}~\bibnamefont {Cervero~Mart{\'{i}}n}}, \bibinfo {author} {\bibfnamefont {K.}~\bibnamefont {Plekhanov}}, \ and\ \bibinfo {author} {\bibfnamefont {M.}~\bibnamefont {Lubasch}},\ }\bibfield  {title} {\enquote {\bibinfo {title} {Barren plateaus in quantum tensor network optimization},}\ }\href {\doibase 10.22331/q-2023-04-13-974} {\bibfield  {journal} {\bibinfo  {journal} {Quantum}\ }\textbf {\bibinfo {volume} {7}},\ \bibinfo {pages} {974} (\bibinfo {year} {2023})}\BibitemShut {NoStop}%
\bibitem [{\citenamefont {Sokolov}\ \emph {et~al.}(2023)\citenamefont {Sokolov}, \citenamefont {Dobrautz}, \citenamefont {Luo}, \citenamefont {Alavi},\ and\ \citenamefont {Tavernelli}}]{tc_quantum2023}%
  \BibitemOpen
  \bibfield  {author} {\bibinfo {author} {\bibfnamefont {I.~O.}\ \bibnamefont {Sokolov}}, \bibinfo {author} {\bibfnamefont {W.}~\bibnamefont {Dobrautz}}, \bibinfo {author} {\bibfnamefont {H.}~\bibnamefont {Luo}}, \bibinfo {author} {\bibfnamefont {A.}~\bibnamefont {Alavi}}, \ and\ \bibinfo {author} {\bibfnamefont {I.}~\bibnamefont {Tavernelli}},\ }\href {\doibase 10.1103/PhysRevResearch.5.023174} {\bibfield  {journal} {\bibinfo  {journal} {Phys. Rev. Res.}\ }\textbf {\bibinfo {volume} {5}},\ \bibinfo {pages} {023174} (\bibinfo {year} {2023})}\BibitemShut {NoStop}%
\bibitem [{\citenamefont {Larocca}\ \emph {et~al.}(2024)\citenamefont {Larocca}, \citenamefont {Thanasilp}, \citenamefont {Wang}, \citenamefont {Sharma}, \citenamefont {Biamonte}, \citenamefont {Coles}, \citenamefont {Cincio}, \citenamefont {McClean}, \citenamefont {Holmes},\ and\ \citenamefont {Cerezo}}]{barrenvqe2024}%
  \BibitemOpen
  \bibfield  {author} {\bibinfo {author} {\bibfnamefont {M.}~\bibnamefont {Larocca}}, \bibinfo {author} {\bibfnamefont {S.}~\bibnamefont {Thanasilp}}, \bibinfo {author} {\bibfnamefont {S.}~\bibnamefont {Wang}}, \bibinfo {author} {\bibfnamefont {K.}~\bibnamefont {Sharma}}, \bibinfo {author} {\bibfnamefont {J.}~\bibnamefont {Biamonte}}, \bibinfo {author} {\bibfnamefont {P.~J.}\ \bibnamefont {Coles}}, \bibinfo {author} {\bibfnamefont {L.}~\bibnamefont {Cincio}}, \bibinfo {author} {\bibfnamefont {J.~R.}\ \bibnamefont {McClean}}, \bibinfo {author} {\bibfnamefont {Z.}~\bibnamefont {Holmes}}, \ and\ \bibinfo {author} {\bibfnamefont {M.}~\bibnamefont {Cerezo}},\ }\href {https://doi.org/10.48550/arXiv.2405.00781} {\bibfield  {journal} {\bibinfo  {journal} {arXiv:2405.00781}\ } (\bibinfo {year} {2024})}\BibitemShut {NoStop}%
\bibitem [{\citenamefont {Dobrautz}\ \emph {et~al.}(2024)\citenamefont {Dobrautz}, \citenamefont {Sokolov}, \citenamefont {Liao}, \citenamefont {Ríos}, \citenamefont {Rahm}, \citenamefont {Alavi},\ and\ \citenamefont {Tavernelli}}]{tc_quantum2024}%
  \BibitemOpen
  \bibfield  {author} {\bibinfo {author} {\bibfnamefont {W.}~\bibnamefont {Dobrautz}}, \bibinfo {author} {\bibfnamefont {I.~O.}\ \bibnamefont {Sokolov}}, \bibinfo {author} {\bibfnamefont {K.}~\bibnamefont {Liao}}, \bibinfo {author} {\bibfnamefont {P.~L.}\ \bibnamefont {Ríos}}, \bibinfo {author} {\bibfnamefont {M.}~\bibnamefont {Rahm}}, \bibinfo {author} {\bibfnamefont {A.}~\bibnamefont {Alavi}}, \ and\ \bibinfo {author} {\bibfnamefont {I.}~\bibnamefont {Tavernelli}},\ }\href {\doibase 10.1021/acs.jctc.4c00070} {\bibfield  {journal} {\bibinfo  {journal} {J. Chem. Theory Comput.}\ }\textbf {\bibinfo {volume} {20}},\ \bibinfo {pages} {4146--4160} (\bibinfo {year} {2024})}\BibitemShut {NoStop}%
\bibitem [{\citenamefont {Magnusson}\ \emph {et~al.}(2024)\citenamefont {Magnusson}, \citenamefont {Fitzpatrick}, \citenamefont {Knecht}, \citenamefont {Rahm},\ and\ \citenamefont {Dobrautz}}]{adapt_tc2024}%
  \BibitemOpen
  \bibfield  {author} {\bibinfo {author} {\bibfnamefont {E.}~\bibnamefont {Magnusson}}, \bibinfo {author} {\bibfnamefont {A.}~\bibnamefont {Fitzpatrick}}, \bibinfo {author} {\bibfnamefont {S.}~\bibnamefont {Knecht}}, \bibinfo {author} {\bibfnamefont {M.}~\bibnamefont {Rahm}}, \ and\ \bibinfo {author} {\bibfnamefont {W.}~\bibnamefont {Dobrautz}},\ }\href {\doibase 10.1039/D4FD00039K} {\bibfield  {journal} {\bibinfo  {journal} {Faraday Discuss.}\ ,\ \bibinfo {pages} {--}} (\bibinfo {year} {2024})}\BibitemShut {NoStop}%
\bibitem [{\citenamefont {McArdle}\ \emph {et~al.}(2019)\citenamefont {McArdle}, \citenamefont {Jones}, \citenamefont {Endo}, \citenamefont {Li}, \citenamefont {Benjamin},\ and\ \citenamefont {Yuan}}]{varqite2019}%
  \BibitemOpen
  \bibfield  {author} {\bibinfo {author} {\bibfnamefont {S.}~\bibnamefont {McArdle}}, \bibinfo {author} {\bibfnamefont {T.}~\bibnamefont {Jones}}, \bibinfo {author} {\bibfnamefont {S.}~\bibnamefont {Endo}}, \bibinfo {author} {\bibfnamefont {Y.}~\bibnamefont {Li}}, \bibinfo {author} {\bibfnamefont {S.~C.}\ \bibnamefont {Benjamin}}, \ and\ \bibinfo {author} {\bibfnamefont {X.}~\bibnamefont {Yuan}},\ }\href {https://doi.org/10.1038/s41534-019-0187-2} {\bibfield  {journal} {\bibinfo  {journal} {npj Quantum Inf.}\ }\textbf {\bibinfo {volume} {5}},\ \bibinfo {pages} {75} (\bibinfo {year} {2019})}\BibitemShut {NoStop}%
\bibitem [{\citenamefont {Koch}\ \emph {et~al.}(2023)\citenamefont {Koch}, \citenamefont {Schaudt}, \citenamefont {Mogk}, \citenamefont {Mrziglod}, \citenamefont {Berg},\ and\ \citenamefont {Beck}}]{varqite2023}%
  \BibitemOpen
  \bibfield  {author} {\bibinfo {author} {\bibfnamefont {M.}~\bibnamefont {Koch}}, \bibinfo {author} {\bibfnamefont {O.}~\bibnamefont {Schaudt}}, \bibinfo {author} {\bibfnamefont {G.}~\bibnamefont {Mogk}}, \bibinfo {author} {\bibfnamefont {T.}~\bibnamefont {Mrziglod}}, \bibinfo {author} {\bibfnamefont {H.}~\bibnamefont {Berg}}, \ and\ \bibinfo {author} {\bibfnamefont {M.~E.}\ \bibnamefont {Beck}},\ }\href {\doibase 10.1021/acsomega.3c01060} {\bibfield  {journal} {\bibinfo  {journal} {ACS Omega}\ }\textbf {\bibinfo {volume} {8}},\ \bibinfo {pages} {22596--22602} (\bibinfo {year} {2023})}\BibitemShut {NoStop}%
\bibitem [{\citenamefont {Schuld}\ \emph {et~al.}(2019)\citenamefont {Schuld}, \citenamefont {Bergholm}, \citenamefont {Gogolin}, \citenamefont {Izaac},\ and\ \citenamefont {Killoran}}]{parameter2019}%
  \BibitemOpen
  \bibfield  {author} {\bibinfo {author} {\bibfnamefont {M.}~\bibnamefont {Schuld}}, \bibinfo {author} {\bibfnamefont {V.}~\bibnamefont {Bergholm}}, \bibinfo {author} {\bibfnamefont {C.}~\bibnamefont {Gogolin}}, \bibinfo {author} {\bibfnamefont {J.}~\bibnamefont {Izaac}}, \ and\ \bibinfo {author} {\bibfnamefont {N.}~\bibnamefont {Killoran}},\ }\href {\doibase 10.1103/PhysRevA.99.032331} {\bibfield  {journal} {\bibinfo  {journal} {Phys. Rev. A}\ }\textbf {\bibinfo {volume} {99}},\ \bibinfo {pages} {032331} (\bibinfo {year} {2019})}\BibitemShut {NoStop}%
\bibitem [{\citenamefont {McArdle}\ and\ \citenamefont {Tew}(2020)}]{hermitian2020}%
  \BibitemOpen
  \bibfield  {author} {\bibinfo {author} {\bibfnamefont {S.}~\bibnamefont {McArdle}}\ and\ \bibinfo {author} {\bibfnamefont {D.~P.}\ \bibnamefont {Tew}},\ }\href {https://doi.org/10.48550/arXiv.2006.11181} {\bibfield  {journal} {\bibinfo  {journal} {arXiv:2006.11181}\ } (\bibinfo {year} {2020})}\BibitemShut {NoStop}%
\bibitem [{\citenamefont {Javanmard}\ \emph {et~al.}(2024)\citenamefont {Javanmard}, \citenamefont {Liaubaite}, \citenamefont {Osborne}, \citenamefont {Xu},\ and\ \citenamefont {Yung}}]{QCMPSkagome2024}%
  \BibitemOpen
  \bibfield  {author} {\bibinfo {author} {\bibfnamefont {Y.}~\bibnamefont {Javanmard}}, \bibinfo {author} {\bibfnamefont {U.}~\bibnamefont {Liaubaite}}, \bibinfo {author} {\bibfnamefont {T.~J.}\ \bibnamefont {Osborne}}, \bibinfo {author} {\bibfnamefont {X.}~\bibnamefont {Xu}}, \ and\ \bibinfo {author} {\bibfnamefont {M.-H.}\ \bibnamefont {Yung}},\ }\href {https://doi.org/10.48550/arXiv.2401.02355} {\bibfield  {journal} {\bibinfo  {journal} {arXiv:2401.02355}\ } (\bibinfo {year} {2024})}\BibitemShut {NoStop}%
\bibitem [{\citenamefont {Jordan}\ and\ \citenamefont {Wigner}(1928)}]{JWT1928}%
  \BibitemOpen
  \bibfield  {author} {\bibinfo {author} {\bibfnamefont {P.}~\bibnamefont {Jordan}}\ and\ \bibinfo {author} {\bibfnamefont {E.}~\bibnamefont {Wigner}},\ }\href {https://doi.org/10.1007/BF01331938} {\bibfield  {journal} {\bibinfo  {journal} {Z. Phys.}\ }\textbf {\bibinfo {volume} {47}},\ \bibinfo {pages} {631--651} (\bibinfo {year} {1928})}\BibitemShut {NoStop}%
\bibitem [{\citenamefont {Stokes}\ \emph {et~al.}(2020)\citenamefont {Stokes}, \citenamefont {Izaac}, \citenamefont {Killoran},\ and\ \citenamefont {Carleo}}]{qml2020}%
  \BibitemOpen
  \bibfield  {author} {\bibinfo {author} {\bibfnamefont {J.}~\bibnamefont {Stokes}}, \bibinfo {author} {\bibfnamefont {J.}~\bibnamefont {Izaac}}, \bibinfo {author} {\bibfnamefont {N.}~\bibnamefont {Killoran}}, \ and\ \bibinfo {author} {\bibfnamefont {G.}~\bibnamefont {Carleo}},\ }\href {\doibase 10.22331/q-2020-05-25-269} {\bibfield  {journal} {\bibinfo  {journal} {{Quantum}}\ }\textbf {\bibinfo {volume} {4}},\ \bibinfo {pages} {269} (\bibinfo {year} {2020})}\BibitemShut {NoStop}%
\bibitem [{\citenamefont {Funcke}\ \emph {et~al.}(2021)\citenamefont {Funcke}, \citenamefont {Hartung}, \citenamefont {Jansen}, \citenamefont {K{\"{u}}hn},\ and\ \citenamefont {Stornati}}]{derivation2021}%
  \BibitemOpen
  \bibfield  {author} {\bibinfo {author} {\bibfnamefont {L.}~\bibnamefont {Funcke}}, \bibinfo {author} {\bibfnamefont {T.}~\bibnamefont {Hartung}}, \bibinfo {author} {\bibfnamefont {K.}~\bibnamefont {Jansen}}, \bibinfo {author} {\bibfnamefont {S.}~\bibnamefont {K{\"{u}}hn}}, \ and\ \bibinfo {author} {\bibfnamefont {P.}~\bibnamefont {Stornati}},\ }\href {\doibase 10.22331/q-2021-03-29-422} {\bibfield  {journal} {\bibinfo  {journal} {{Quantum}}\ }\textbf {\bibinfo {volume} {5}},\ \bibinfo {pages} {422} (\bibinfo {year} {2021})}\BibitemShut {NoStop}%
\bibitem [{\citenamefont {Shende}, \citenamefont {Markov},\ and\ \citenamefont {Bullock}(2004)}]{decomp2004}%
  \BibitemOpen
  \bibfield  {author} {\bibinfo {author} {\bibfnamefont {V.~V.}\ \bibnamefont {Shende}}, \bibinfo {author} {\bibfnamefont {I.~L.}\ \bibnamefont {Markov}}, \ and\ \bibinfo {author} {\bibfnamefont {S.~S.}\ \bibnamefont {Bullock}},\ }\href {\doibase 10.1103/PhysRevA.69.062321} {\bibfield  {journal} {\bibinfo  {journal} {Phys. Rev. A}\ }\textbf {\bibinfo {volume} {69}},\ \bibinfo {pages} {062321} (\bibinfo {year} {2004})}\BibitemShut {NoStop}%
\bibitem [{\citenamefont {Shende}, \citenamefont {Bullock},\ and\ \citenamefont {Markov}(2004)}]{decomp22004}%
  \BibitemOpen
  \bibfield  {author} {\bibinfo {author} {\bibfnamefont {V.~V.}\ \bibnamefont {Shende}}, \bibinfo {author} {\bibfnamefont {S.~S.}\ \bibnamefont {Bullock}}, \ and\ \bibinfo {author} {\bibfnamefont {I.~L.}\ \bibnamefont {Markov}},\ }\href {\doibase 10.1103/PhysRevA.70.012310} {\bibfield  {journal} {\bibinfo  {journal} {Phys. Rev. A}\ }\textbf {\bibinfo {volume} {70}},\ \bibinfo {pages} {012310} (\bibinfo {year} {2004})}\BibitemShut {NoStop}%
\bibitem [{\citenamefont {Gacon}\ \emph {et~al.}(2021)\citenamefont {Gacon}, \citenamefont {Zoufal}, \citenamefont {Carleo},\ and\ \citenamefont {Woerner}}]{approximateQFI2021}%
  \BibitemOpen
  \bibfield  {author} {\bibinfo {author} {\bibfnamefont {J.}~\bibnamefont {Gacon}}, \bibinfo {author} {\bibfnamefont {C.}~\bibnamefont {Zoufal}}, \bibinfo {author} {\bibfnamefont {G.}~\bibnamefont {Carleo}}, \ and\ \bibinfo {author} {\bibfnamefont {S.}~\bibnamefont {Woerner}},\ }\href {\doibase 10.22331/q-2021-10-20-567} {\bibfield  {journal} {\bibinfo  {journal} {{Quantum}}\ }\textbf {\bibinfo {volume} {5}},\ \bibinfo {pages} {567} (\bibinfo {year} {2021})}\BibitemShut {NoStop}%
\bibitem [{\citenamefont {Gacon}\ \emph {et~al.}(2024)\citenamefont {Gacon}, \citenamefont {Nys}, \citenamefont {Rossi}, \citenamefont {Woerner},\ and\ \citenamefont {Carleo}}]{varqite_noqgt2024}%
  \BibitemOpen
  \bibfield  {author} {\bibinfo {author} {\bibfnamefont {J.}~\bibnamefont {Gacon}}, \bibinfo {author} {\bibfnamefont {J.}~\bibnamefont {Nys}}, \bibinfo {author} {\bibfnamefont {R.}~\bibnamefont {Rossi}}, \bibinfo {author} {\bibfnamefont {S.}~\bibnamefont {Woerner}}, \ and\ \bibinfo {author} {\bibfnamefont {G.}~\bibnamefont {Carleo}},\ }\href {\doibase 10.1103/PhysRevResearch.6.013143} {\bibfield  {journal} {\bibinfo  {journal} {Phys. Rev. Res.}\ }\textbf {\bibinfo {volume} {6}},\ \bibinfo {pages} {013143} (\bibinfo {year} {2024})}\BibitemShut {NoStop}%
\bibitem [{\citenamefont {Benedetti}, \citenamefont {Fiorentini},\ and\ \citenamefont {Lubasch}(2021)}]{HEvarqite2021}%
  \BibitemOpen
  \bibfield  {author} {\bibinfo {author} {\bibfnamefont {M.}~\bibnamefont {Benedetti}}, \bibinfo {author} {\bibfnamefont {M.}~\bibnamefont {Fiorentini}}, \ and\ \bibinfo {author} {\bibfnamefont {M.}~\bibnamefont {Lubasch}},\ }\bibfield  {title} {\enquote {\bibinfo {title} {Hardware-efficient variational quantum algorithms for time evolution},}\ }\href {\doibase 10.1103/PhysRevResearch.3.033083} {\bibfield  {journal} {\bibinfo  {journal} {Phys. Rev. Res.}\ }\textbf {\bibinfo {volume} {3}},\ \bibinfo {pages} {033083} (\bibinfo {year} {2021})}\BibitemShut {NoStop}%
\bibitem [{\citenamefont {Li}\ \emph {et~al.}(2024)\citenamefont {Li}, \citenamefont {Huang}, \citenamefont {Zhang}, \citenamefont {Li}, \citenamefont {Shen}, \citenamefont {Zhao}, \citenamefont {Li},\ and\ \citenamefont {Hu}}]{SCNQS2024}%
  \BibitemOpen
  \bibfield  {author} {\bibinfo {author} {\bibfnamefont {X.}~\bibnamefont {Li}}, \bibinfo {author} {\bibfnamefont {J.-C.}\ \bibnamefont {Huang}}, \bibinfo {author} {\bibfnamefont {G.-Z.}\ \bibnamefont {Zhang}}, \bibinfo {author} {\bibfnamefont {H.-E.}\ \bibnamefont {Li}}, \bibinfo {author} {\bibfnamefont {Z.-P.}\ \bibnamefont {Shen}}, \bibinfo {author} {\bibfnamefont {C.}~\bibnamefont {Zhao}}, \bibinfo {author} {\bibfnamefont {J.}~\bibnamefont {Li}}, \ and\ \bibinfo {author} {\bibfnamefont {H.-S.}\ \bibnamefont {Hu}},\ }\href {\doibase 10.1063/5.0214150} {\bibfield  {journal} {\bibinfo  {journal} {J. Chem. Phys.}\ }\textbf {\bibinfo {volume} {160}},\ \bibinfo {pages} {234102} (\bibinfo {year} {2024})}\BibitemShut {NoStop}%
\bibitem [{\citenamefont {Handy}(1971)}]{tc1971variance}%
  \BibitemOpen
  \bibfield  {author} {\bibinfo {author} {\bibfnamefont {N.}~\bibnamefont {Handy}},\ }\href {\doibase 10.1080/00268977100101961} {\bibfield  {journal} {\bibinfo  {journal} {Mol. Phys.}\ }\textbf {\bibinfo {volume} {21}},\ \bibinfo {pages} {817--828} (\bibinfo {year} {1971})}\BibitemShut {NoStop}%
\bibitem [{\citenamefont {Dobrautz}, \citenamefont {Luo},\ and\ \citenamefont {Alavi}(2019)}]{similar_compact2019}%
  \BibitemOpen
  \bibfield  {author} {\bibinfo {author} {\bibfnamefont {W.}~\bibnamefont {Dobrautz}}, \bibinfo {author} {\bibfnamefont {H.}~\bibnamefont {Luo}}, \ and\ \bibinfo {author} {\bibfnamefont {A.}~\bibnamefont {Alavi}},\ }\href {\doibase 10.1103/PhysRevB.99.075119} {\bibfield  {journal} {\bibinfo  {journal} {Phys. Rev. B}\ }\textbf {\bibinfo {volume} {99}},\ \bibinfo {pages} {075119} (\bibinfo {year} {2019})}\BibitemShut {NoStop}%
\bibitem [{\citenamefont {Ammar}\ \emph {et~al.}(2024)\citenamefont {Ammar}, \citenamefont {Scemama}, \citenamefont {Loos},\ and\ \citenamefont {Giner}}]{compactsci2024}%
  \BibitemOpen
  \bibfield  {author} {\bibinfo {author} {\bibfnamefont {A.}~\bibnamefont {Ammar}}, \bibinfo {author} {\bibfnamefont {A.}~\bibnamefont {Scemama}}, \bibinfo {author} {\bibfnamefont {P.-F.}\ \bibnamefont {Loos}}, \ and\ \bibinfo {author} {\bibfnamefont {E.}~\bibnamefont {Giner}},\ }\href {https://doi.org/10.48550/2405.02640} {\bibfield  {journal} {\bibinfo  {journal} {arXiv:2405.02640}\ } (\bibinfo {year} {2024})}\BibitemShut {NoStop}%
\bibitem [{\citenamefont {Adams}\ \emph {et~al.}(2021)\citenamefont {Adams}, \citenamefont {Carleo}, \citenamefont {Lovato},\ and\ \citenamefont {Rocco}}]{adaptlr2021}%
  \BibitemOpen
  \bibfield  {author} {\bibinfo {author} {\bibfnamefont {C.}~\bibnamefont {Adams}}, \bibinfo {author} {\bibfnamefont {G.}~\bibnamefont {Carleo}}, \bibinfo {author} {\bibfnamefont {A.}~\bibnamefont {Lovato}}, \ and\ \bibinfo {author} {\bibfnamefont {N.}~\bibnamefont {Rocco}},\ }\href {\doibase 10.1103/PhysRevLett.127.022502} {\bibfield  {journal} {\bibinfo  {journal} {Phys. Rev. Lett.}\ }\textbf {\bibinfo {volume} {127}},\ \bibinfo {pages} {022502} (\bibinfo {year} {2021})}\BibitemShut {NoStop}%
\bibitem [{\citenamefont {Dobrautz}(2023)}]{github}%
  \BibitemOpen
  \bibfield  {author} {\bibinfo {author} {\bibfnamefont {W.}~\bibnamefont {Dobrautz}},\ }\href {https://github.com/dobrautz/tc-varqite-hamiltonians} {\enquote {\bibinfo {title} {tc-varqite-hamiltonians},}\ } (\bibinfo {year} {2023})\BibitemShut {NoStop}%
\bibitem [{\citenamefont {Developers}(2023)}]{cirq}%
  \BibitemOpen
  \bibfield  {author} {\bibinfo {author} {\bibfnamefont {C.}~\bibnamefont {Developers}},\ }\href {\doibase 10.5281/zenodo.10247207} {\enquote {\bibinfo {title} {Cirq},}\ } (\bibinfo {year} {2023})\BibitemShut {NoStop}%
\bibitem [{\citenamefont {McClean}\ \emph {et~al.}(2020)\citenamefont {McClean}, \citenamefont {Rubin}, \citenamefont {Sung}, \citenamefont {Kivlichan}, \citenamefont {Bonet-Monroig}, \citenamefont {Cao}, \citenamefont {Dai}, \citenamefont {Fried}, \citenamefont {Gidney}, \citenamefont {Gimby}, \citenamefont {Gokhale}, \citenamefont {Häner}, \citenamefont {Hardikar}, \citenamefont {Havlíček}, \citenamefont {Higgott}, \citenamefont {Huang}, \citenamefont {Izaac}, \citenamefont {Jiang}, \citenamefont {Liu}, \citenamefont {McArdle}, \citenamefont {Neeley}, \citenamefont {O’Brien}, \citenamefont {O’Gorman}, \citenamefont {Ozfidan}, \citenamefont {Radin}, \citenamefont {Romero}, \citenamefont {Sawaya}, \citenamefont {Senjean}, \citenamefont {Setia}, \citenamefont {Sim}, \citenamefont {Steiger}, \citenamefont {Steudtner}, \citenamefont {Sun}, \citenamefont {Sun}, \citenamefont {Wang}, \citenamefont {Zhang},\ and\ \citenamefont {Babbush}}]{of2020}%
  \BibitemOpen
  \bibfield  {author} {\bibinfo {author} {\bibfnamefont {J.~R.}\ \bibnamefont {McClean}}, \bibinfo {author} {\bibfnamefont {N.~C.}\ \bibnamefont {Rubin}}, \bibinfo {author} {\bibfnamefont {K.~J.}\ \bibnamefont {Sung}}, \bibinfo {author} {\bibfnamefont {I.~D.}\ \bibnamefont {Kivlichan}}, \bibinfo {author} {\bibfnamefont {X.}~\bibnamefont {Bonet-Monroig}}, \bibinfo {author} {\bibfnamefont {Y.}~\bibnamefont {Cao}}, \bibinfo {author} {\bibfnamefont {C.}~\bibnamefont {Dai}}, \bibinfo {author} {\bibfnamefont {E.~S.}\ \bibnamefont {Fried}}, \bibinfo {author} {\bibfnamefont {C.}~\bibnamefont {Gidney}}, \bibinfo {author} {\bibfnamefont {B.}~\bibnamefont {Gimby}}, \bibinfo {author} {\bibfnamefont {P.}~\bibnamefont {Gokhale}}, \bibinfo {author} {\bibfnamefont {T.}~\bibnamefont {Häner}}, \bibinfo {author} {\bibfnamefont {T.}~\bibnamefont {Hardikar}}, \bibinfo {author} {\bibfnamefont {V.}~\bibnamefont {Havlíček}}, \bibinfo {author} {\bibfnamefont {O.}~\bibnamefont {Higgott}}, \bibinfo {author} {\bibfnamefont {C.}~\bibnamefont {Huang}}, \bibinfo {author} {\bibfnamefont {J.}~\bibnamefont {Izaac}}, \bibinfo {author} {\bibfnamefont {Z.}~\bibnamefont {Jiang}}, \bibinfo {author} {\bibfnamefont {X.}~\bibnamefont {Liu}}, \bibinfo {author} {\bibfnamefont {S.}~\bibnamefont {McArdle}}, \bibinfo {author} {\bibfnamefont {M.}~\bibnamefont {Neeley}}, \bibinfo {author} {\bibfnamefont {T.}~\bibnamefont {O’Brien}}, \bibinfo {author} {\bibfnamefont {B.}~\bibnamefont {O’Gorman}}, \bibinfo {author} {\bibfnamefont {I.}~\bibnamefont {Ozfidan}}, \bibinfo {author} {\bibfnamefont {M.~D.}\ \bibnamefont {Radin}}, \bibinfo {author} {\bibfnamefont {J.}~\bibnamefont {Romero}}, \bibinfo {author} {\bibfnamefont {N.~P.~D.}\ \bibnamefont {Sawaya}}, \bibinfo {author} {\bibfnamefont {B.}~\bibnamefont {Senjean}}, \bibinfo {author} {\bibfnamefont {K.}~\bibnamefont {Setia}}, \bibinfo {author} {\bibfnamefont {S.}~\bibnamefont {Sim}}, \bibinfo {author} {\bibfnamefont {D.~S.}\ \bibnamefont {Steiger}}, \bibinfo {author} {\bibfnamefont {M.}~\bibnamefont {Steudtner}}, \bibinfo {author} {\bibfnamefont {Q.}~\bibnamefont {Sun}}, \bibinfo {author} {\bibfnamefont {W.}~\bibnamefont {Sun}}, \bibinfo {author} {\bibfnamefont {D.}~\bibnamefont {Wang}}, \bibinfo {author} {\bibfnamefont {F.}~\bibnamefont {Zhang}}, \ and\ \bibinfo {author} {\bibfnamefont {R.}~\bibnamefont {Babbush}},\ }\href {\doibase 10.1088/2058-9565/ab8ebc} {\bibfield  {journal} {\bibinfo  {journal} {Quantum Sci. Technol.}\ }\textbf {\bibinfo {volume} {5}},\ \bibinfo {pages} {034014} (\bibinfo {year} {2020})}\BibitemShut {NoStop}%
\bibitem [{\citenamefont {Sun}\ \emph {et~al.}(2018)\citenamefont {Sun}, \citenamefont {Berkelbach}, \citenamefont {Blunt}, \citenamefont {Booth}, \citenamefont {Guo}, \citenamefont {Li}, \citenamefont {Liu}, \citenamefont {McClain}, \citenamefont {Sayfutyarova}, \citenamefont {Sharma}, \citenamefont {Wouters},\ and\ \citenamefont {Chan}}]{PySCF18}%
  \BibitemOpen
  \bibfield  {author} {\bibinfo {author} {\bibfnamefont {Q.}~\bibnamefont {Sun}}, \bibinfo {author} {\bibfnamefont {T.~C.}\ \bibnamefont {Berkelbach}}, \bibinfo {author} {\bibfnamefont {N.~S.}\ \bibnamefont {Blunt}}, \bibinfo {author} {\bibfnamefont {G.~H.}\ \bibnamefont {Booth}}, \bibinfo {author} {\bibfnamefont {S.}~\bibnamefont {Guo}}, \bibinfo {author} {\bibfnamefont {Z.}~\bibnamefont {Li}}, \bibinfo {author} {\bibfnamefont {J.}~\bibnamefont {Liu}}, \bibinfo {author} {\bibfnamefont {J.~D.}\ \bibnamefont {McClain}}, \bibinfo {author} {\bibfnamefont {E.~R.}\ \bibnamefont {Sayfutyarova}}, \bibinfo {author} {\bibfnamefont {S.}~\bibnamefont {Sharma}}, \bibinfo {author} {\bibfnamefont {S.}~\bibnamefont {Wouters}}, \ and\ \bibinfo {author} {\bibfnamefont {G.~K.-L.}\ \bibnamefont {Chan}},\ }\href {\doibase https://doi.org/10.1002/wcms.1340} {\bibfield  {journal} {\bibinfo  {journal} {WIREs Comput. Mol. Sci.}\ }\textbf {\bibinfo {volume} {8}},\ \bibinfo {pages} {e1340} (\bibinfo {year} {2018})}\BibitemShut {NoStop}%
\bibitem [{\citenamefont {Sun}\ \emph {et~al.}(2020)\citenamefont {Sun}, \citenamefont {Zhang}, \citenamefont {Banerjee}, \citenamefont {Bao}, \citenamefont {Barbry}, \citenamefont {Blunt}, \citenamefont {Bogdanov}, \citenamefont {Booth}, \citenamefont {Chen}, \citenamefont {Cui}, \citenamefont {Eriksen}, \citenamefont {Gao}, \citenamefont {Guo}, \citenamefont {Hermann}, \citenamefont {Hermes}, \citenamefont {Koh}, \citenamefont {Koval}, \citenamefont {Lehtola}, \citenamefont {Li}, \citenamefont {Liu}, \citenamefont {Mardirossian}, \citenamefont {McClain}, \citenamefont {Motta}, \citenamefont {Mussard}, \citenamefont {Pham}, \citenamefont {Pulkin}, \citenamefont {Purwanto}, \citenamefont {Robinson}, \citenamefont {Ronca}, \citenamefont {Sayfutyarova}, \citenamefont {Scheurer}, \citenamefont {Schurkus}, \citenamefont {Smith}, \citenamefont {Sun}, \citenamefont {Sun}, \citenamefont {Upadhyay}, \citenamefont {Wagner}, \citenamefont {Wang}, \citenamefont {White}, \citenamefont {Whitfield}, \citenamefont {Williamson}, \citenamefont {Wouters}, \citenamefont {Yang}, \citenamefont {Yu}, \citenamefont {Zhu}, \citenamefont {Berkelbach}, \citenamefont {Sharma}, \citenamefont {Sokolov},\ and\ \citenamefont {Chan}}]{PySCF20}%
  \BibitemOpen
  \bibfield  {author} {\bibinfo {author} {\bibfnamefont {Q.}~\bibnamefont {Sun}}, \bibinfo {author} {\bibfnamefont {X.}~\bibnamefont {Zhang}}, \bibinfo {author} {\bibfnamefont {S.}~\bibnamefont {Banerjee}}, \bibinfo {author} {\bibfnamefont {P.}~\bibnamefont {Bao}}, \bibinfo {author} {\bibfnamefont {M.}~\bibnamefont {Barbry}}, \bibinfo {author} {\bibfnamefont {N.~S.}\ \bibnamefont {Blunt}}, \bibinfo {author} {\bibfnamefont {N.~A.}\ \bibnamefont {Bogdanov}}, \bibinfo {author} {\bibfnamefont {G.~H.}\ \bibnamefont {Booth}}, \bibinfo {author} {\bibfnamefont {J.}~\bibnamefont {Chen}}, \bibinfo {author} {\bibfnamefont {Z.-H.}\ \bibnamefont {Cui}}, \bibinfo {author} {\bibfnamefont {J.~J.}\ \bibnamefont {Eriksen}}, \bibinfo {author} {\bibfnamefont {Y.}~\bibnamefont {Gao}}, \bibinfo {author} {\bibfnamefont {S.}~\bibnamefont {Guo}}, \bibinfo {author} {\bibfnamefont {J.}~\bibnamefont {Hermann}}, \bibinfo {author} {\bibfnamefont {M.~R.}\ \bibnamefont {Hermes}}, \bibinfo {author} {\bibfnamefont {K.}~\bibnamefont {Koh}}, \bibinfo {author} {\bibfnamefont {P.}~\bibnamefont {Koval}}, \bibinfo {author} {\bibfnamefont {S.}~\bibnamefont {Lehtola}}, \bibinfo {author} {\bibfnamefont {Z.}~\bibnamefont {Li}}, \bibinfo {author} {\bibfnamefont {J.}~\bibnamefont {Liu}}, \bibinfo {author} {\bibfnamefont {N.}~\bibnamefont {Mardirossian}}, \bibinfo {author} {\bibfnamefont {J.~D.}\ \bibnamefont {McClain}}, \bibinfo {author} {\bibfnamefont {M.}~\bibnamefont {Motta}}, \bibinfo {author} {\bibfnamefont {B.}~\bibnamefont {Mussard}}, \bibinfo {author} {\bibfnamefont {H.~Q.}\ \bibnamefont {Pham}}, \bibinfo {author} {\bibfnamefont {A.}~\bibnamefont {Pulkin}}, \bibinfo {author} {\bibfnamefont {W.}~\bibnamefont {Purwanto}}, \bibinfo {author} {\bibfnamefont {P.~J.}\ \bibnamefont {Robinson}}, \bibinfo {author} {\bibfnamefont {E.}~\bibnamefont {Ronca}}, \bibinfo {author} {\bibfnamefont {E.~R.}\ \bibnamefont {Sayfutyarova}}, \bibinfo {author} {\bibfnamefont {M.}~\bibnamefont {Scheurer}}, \bibinfo {author} {\bibfnamefont {H.~F.}\ \bibnamefont {Schurkus}}, \bibinfo {author} {\bibfnamefont {J.~E.~T.}\ \bibnamefont {Smith}}, \bibinfo {author} {\bibfnamefont {C.}~\bibnamefont {Sun}}, \bibinfo {author} {\bibfnamefont {S.-N.}\ \bibnamefont {Sun}}, \bibinfo {author} {\bibfnamefont {S.}~\bibnamefont {Upadhyay}}, \bibinfo {author} {\bibfnamefont {L.~K.}\ \bibnamefont {Wagner}}, \bibinfo {author} {\bibfnamefont {X.}~\bibnamefont {Wang}}, \bibinfo {author} {\bibfnamefont {A.}~\bibnamefont {White}}, \bibinfo {author} {\bibfnamefont {J.~D.}\ \bibnamefont {Whitfield}}, \bibinfo {author} {\bibfnamefont {M.~J.}\ \bibnamefont {Williamson}}, \bibinfo {author} {\bibfnamefont {S.}~\bibnamefont {Wouters}}, \bibinfo {author} {\bibfnamefont {J.}~\bibnamefont {Yang}}, \bibinfo {author} {\bibfnamefont {J.~M.}\ \bibnamefont {Yu}}, \bibinfo {author} {\bibfnamefont {T.}~\bibnamefont {Zhu}}, \bibinfo {author} {\bibfnamefont {T.~C.}\ \bibnamefont {Berkelbach}}, \bibinfo {author} {\bibfnamefont {S.}~\bibnamefont {Sharma}}, \bibinfo {author} {\bibfnamefont {A.~Y.}\ \bibnamefont {Sokolov}}, \ and\ \bibinfo {author} {\bibfnamefont {G.~K.-L.}\ \bibnamefont {Chan}},\ }\href {https://doi.org/10.1063/5.0006074} {\bibfield  {journal} {\bibinfo  {journal} {J. Chem. Phys.}\ }\textbf {\bibinfo {volume} {153}},\ \bibinfo {pages} {024109} (\bibinfo {year} {2020})}\BibitemShut {NoStop}%
\bibitem [{\citenamefont {Virtanen}\ \emph {et~al.}(2020)\citenamefont {Virtanen}, \citenamefont {Gommers}, \citenamefont {Oliphant}, \citenamefont {Haberland}, \citenamefont {Reddy}, \citenamefont {Cournapeau}, \citenamefont {Burovski}, \citenamefont {Peterson}, \citenamefont {Weckesser}, \citenamefont {Bright} \emph {et~al.}}]{scipy2020}%
  \BibitemOpen
  \bibfield  {author} {\bibinfo {author} {\bibfnamefont {P.}~\bibnamefont {Virtanen}}, \bibinfo {author} {\bibfnamefont {R.}~\bibnamefont {Gommers}}, \bibinfo {author} {\bibfnamefont {T.~E.}\ \bibnamefont {Oliphant}}, \bibinfo {author} {\bibfnamefont {M.}~\bibnamefont {Haberland}}, \bibinfo {author} {\bibfnamefont {T.}~\bibnamefont {Reddy}}, \bibinfo {author} {\bibfnamefont {D.}~\bibnamefont {Cournapeau}}, \bibinfo {author} {\bibfnamefont {E.}~\bibnamefont {Burovski}}, \bibinfo {author} {\bibfnamefont {P.}~\bibnamefont {Peterson}}, \bibinfo {author} {\bibfnamefont {W.}~\bibnamefont {Weckesser}}, \bibinfo {author} {\bibfnamefont {J.}~\bibnamefont {Bright}},  \emph {et~al.},\ }\href {https://doi.org/10.1038/s41592-019-0686-2} {\bibfield  {journal} {\bibinfo  {journal} {Nat. Methods}\ }\textbf {\bibinfo {volume} {17}},\ \bibinfo {pages} {261--272} (\bibinfo {year} {2020})}\BibitemShut {NoStop}%
\bibitem [{\citenamefont {L\"owdin}(1955)}]{nmo1955}%
  \BibitemOpen
  \bibfield  {author} {\bibinfo {author} {\bibfnamefont {P.-O.}\ \bibnamefont {L\"owdin}},\ }\href {\doibase 10.1103/PhysRev.97.1474} {\bibfield  {journal} {\bibinfo  {journal} {Phys. Rev.}\ }\textbf {\bibinfo {volume} {97}},\ \bibinfo {pages} {1474--1489} (\bibinfo {year} {1955})}\BibitemShut {NoStop}%
\bibitem [{\citenamefont {Verma}\ \emph {et~al.}(2021)\citenamefont {Verma}, \citenamefont {Huntington}, \citenamefont {Coons}, \citenamefont {Kawashima}, \citenamefont {Yamazaki},\ and\ \citenamefont {Zaribafiyan}}]{nmo2021}%
  \BibitemOpen
  \bibfield  {author} {\bibinfo {author} {\bibfnamefont {P.}~\bibnamefont {Verma}}, \bibinfo {author} {\bibfnamefont {L.}~\bibnamefont {Huntington}}, \bibinfo {author} {\bibfnamefont {M.~P.}\ \bibnamefont {Coons}}, \bibinfo {author} {\bibfnamefont {Y.}~\bibnamefont {Kawashima}}, \bibinfo {author} {\bibfnamefont {T.}~\bibnamefont {Yamazaki}}, \ and\ \bibinfo {author} {\bibfnamefont {A.}~\bibnamefont {Zaribafiyan}},\ }\href {\doibase 10.1063/5.0054647} {\bibfield  {journal} {\bibinfo  {journal} {J. Chem. Phys.}\ }\textbf {\bibinfo {volume} {155}},\ \bibinfo {pages} {034110} (\bibinfo {year} {2021})}\BibitemShut {NoStop}%
\bibitem [{\citenamefont {Han}, \citenamefont {Zhang},\ and\ \citenamefont {E}(2019)}]{deepwf2019}%
  \BibitemOpen
  \bibfield  {author} {\bibinfo {author} {\bibfnamefont {J.}~\bibnamefont {Han}}, \bibinfo {author} {\bibfnamefont {L.}~\bibnamefont {Zhang}}, \ and\ \bibinfo {author} {\bibfnamefont {W.}~\bibnamefont {E}},\ }\href {\doibase https://doi.org/10.1016/j.jcp.2019.108929} {\bibfield  {journal} {\bibinfo  {journal} {J. Comput. Phys.}\ }\textbf {\bibinfo {volume} {399}},\ \bibinfo {pages} {108929} (\bibinfo {year} {2019})}\BibitemShut {NoStop}%
\bibitem [{\citenamefont {Choi}(1975)}]{kraus1975}%
  \BibitemOpen
  \bibfield  {author} {\bibinfo {author} {\bibfnamefont {M.-D.}\ \bibnamefont {Choi}},\ }\href {\doibase https://doi.org/10.1016/0024-3795(75)90075-0} {\bibfield  {journal} {\bibinfo  {journal} {Linear Algebra Appl.}\ }\textbf {\bibinfo {volume} {10}},\ \bibinfo {pages} {285--290} (\bibinfo {year} {1975})}\BibitemShut {NoStop}%
\bibitem [{\citenamefont {Leung}(2003)}]{kraus2003}%
  \BibitemOpen
  \bibfield  {author} {\bibinfo {author} {\bibfnamefont {D.~W.}\ \bibnamefont {Leung}},\ }\href {\doibase 10.1063/1.1518554} {\bibfield  {journal} {\bibinfo  {journal} {J. Math. Phys.}\ }\textbf {\bibinfo {volume} {44}},\ \bibinfo {pages} {528--533} (\bibinfo {year} {2003})}\BibitemShut {NoStop}%
\bibitem [{\citenamefont {Yuan}\ \emph {et~al.}()\citenamefont {Yuan}, \citenamefont {Endo}, \citenamefont {Zhao}, \citenamefont {Li},\ and\ \citenamefont {Benjamin}}]{varqite_theory2019}%
  \BibitemOpen
  \bibfield  {author} {\bibinfo {author} {\bibfnamefont {X.}~\bibnamefont {Yuan}}, \bibinfo {author} {\bibfnamefont {S.}~\bibnamefont {Endo}}, \bibinfo {author} {\bibfnamefont {Q.}~\bibnamefont {Zhao}}, \bibinfo {author} {\bibfnamefont {Y.}~\bibnamefont {Li}}, \ and\ \bibinfo {author} {\bibfnamefont {S.~C.}\ \bibnamefont {Benjamin}},\ }\href {\doibase 10.22331/q-2019-10-07-191} {\bibfield  {journal} {\bibinfo  {journal} {{Quantum}}\ }\textbf {\bibinfo {volume} {3}},\ \bibinfo {pages} {191}}\BibitemShut {NoStop}%
\bibitem [{\citenamefont {McLachlan}(1964)}]{mac1964}%
  \BibitemOpen
  \bibfield  {author} {\bibinfo {author} {\bibfnamefont {A.}~\bibnamefont {McLachlan}},\ }\href {\doibase 10.1080/00268976400100041} {\bibfield  {journal} {\bibinfo  {journal} {Mol. Phys.}\ }\textbf {\bibinfo {volume} {8}},\ \bibinfo {pages} {39--44} (\bibinfo {year} {1964})}\BibitemShut {NoStop}%
\bibitem [{\citenamefont {Zoufal}, \citenamefont {Sutter},\ and\ \citenamefont {Woerner}(2023)}]{varqite_theory2023}%
  \BibitemOpen
  \bibfield  {author} {\bibinfo {author} {\bibfnamefont {C.}~\bibnamefont {Zoufal}}, \bibinfo {author} {\bibfnamefont {D.}~\bibnamefont {Sutter}}, \ and\ \bibinfo {author} {\bibfnamefont {S.}~\bibnamefont {Woerner}},\ }\href {\doibase 10.1103/PhysRevApplied.20.044059} {\bibfield  {journal} {\bibinfo  {journal} {Phys. Rev. Appl.}\ }\textbf {\bibinfo {volume} {20}},\ \bibinfo {pages} {044059} (\bibinfo {year} {2023})}\BibitemShut {NoStop}%
\bibitem [{\citenamefont {Tang}\ \emph {et~al.}(2021{\natexlab{b}})\citenamefont {Tang}, \citenamefont {Shkolnikov}, \citenamefont {Barron}, \citenamefont {Grimsley}, \citenamefont {Mayhall}, \citenamefont {Barnes},\ and\ \citenamefont {Economou}}]{HEA2021}%
  \BibitemOpen
  \bibfield  {author} {\bibinfo {author} {\bibfnamefont {H.~L.}\ \bibnamefont {Tang}}, \bibinfo {author} {\bibfnamefont {V.}~\bibnamefont {Shkolnikov}}, \bibinfo {author} {\bibfnamefont {G.~S.}\ \bibnamefont {Barron}}, \bibinfo {author} {\bibfnamefont {H.~R.}\ \bibnamefont {Grimsley}}, \bibinfo {author} {\bibfnamefont {N.~J.}\ \bibnamefont {Mayhall}}, \bibinfo {author} {\bibfnamefont {E.}~\bibnamefont {Barnes}}, \ and\ \bibinfo {author} {\bibfnamefont {S.~E.}\ \bibnamefont {Economou}},\ }\href {\doibase 10.1103/PRXQuantum.2.020310} {\bibfield  {journal} {\bibinfo  {journal} {PRX Quantum}\ }\textbf {\bibinfo {volume} {2}},\ \bibinfo {pages} {020310} (\bibinfo {year} {2021}{\natexlab{b}})}\BibitemShut {NoStop}%
\end{thebibliography}%

\end{document}